# The Production of Policy Knowledge in the United States

Taegyoon Kim, [1,2] Alexander Furnas, [3,4,5,6] Dashun Wang [3,4,5,6,7,*]
[1] School of Digital Humanities and Computational Social Sciences, KAIST, Daejeon, Republic of Korea
[2] Graduate School of Data Science, KAIST, Daejeon, Republic of Korea
[3] Center for Science of Science and Innovation, Northwestern University, Evanston, IL, USA
[4] Ryan Institute on Complexity, Northwestern University, Evanston, IL, USA
[5] Northwestern Innovation Institute, Northwestern University, Evanston, IL, USA
[6] Kellogg School of Management, Northwestern University, Evanston, IL, USA
[7] McCormick School of Engineering, Northwestern University, Evanston, IL, USA

[*] Corresponding author: dashun.wang@kellogg.northwestern.edu

**Abstract**

Policymaking relies on institutions that translate expertise into politically actionable knowledge. Yet little is known about how such knowledge is created, structured, and politically polarized, or how science shapes these processes. Using 2 million U.S. policy documents and nearly 1 million scientific citations from more than 200 think tanks and 100 government organizations (1998–2021), we analyze the supply side of science in policymaking: the production of policy knowledge. We find that think tanks are the dominant suppliers of science-based policy knowledge to government, but that this production has become increasingly politically polarized, driven primarily by growing insularity among left-leaning institutions. We further find that science is linked to lower polarization, as policy documents grounded in scientific evidence, especially those citing high-impact science, are less ideologically segregated and occupy more central positions in policy knowledge networks. These results reveal the potential of scientific expertise to shape and, at times, bridge the ideological landscape of policy knowledge production. Amid rising political polarization and the growing role of science in policymaking, understanding how policy knowledge is produced and how scientific expertise relates to its ideological structure is essential to strengthening the informational foundations of democratic governance.

**Significance statement**

Public policymaking relies on a complex ecosystem that transforms research and expertise into actionable policy knowledge. Scientific knowledge rarely enters this process directly; instead, it is filtered through intermediary organizations that shape how evidence is selected, framed, and circulated. Using large-scale data on U.S. policy documents, we first show that think tanks play a dominant role in supplying science-based policy knowledge. We then provide the first systematic evidence that policy knowledge production itself has become increasingly polarized, with this trend driven primarily by growing insularity among left-leaning think tanks. At the same time, policy discussion that explicitly draws on scientific research, especially high-impact science, is systematically less ideologically insular. Together, these findings reveal how policy knowledge is produced in polarized systems and why scientific expertise remains a critical source of shared informational infrastructure.

## Introduction

Public policymaking rarely draws directly on scientific research or expert consensus. Instead, it depends on an extensive ecosystem of organizations that gather, interpret, and translate expertise into politically actionable knowledge (1–4). These intermediary institutions, ranging from government research offices to think tanks, supply the informational infrastructure through which evidence enters political debate and shapes public decisions (5–7). Understanding how they organize and circulate knowledge is therefore essential to understanding how science ultimately informs governance (8).

Among these institutions, think tanks occupy a uniquely powerful position.[1] In the U.S., they provide much of the intellectual and analytical labor that underpins policy formulation (9, 10). They staff administrations, draft legislative proposals, and shape media and elite discourse (4, 9). Their reports influence what counts as credible evidence and what problems deserve attention, steering the agenda often long before policymakers act. Through this work, think tanks act as key producers of policy knowledge, generating research, synthesizing data, and reframing scientific research for political consumption (4, 9, 12). Yet despite decades of scholarship on think tanks' political influence, the processes through which they produce policy knowledge remain poorly mapped. Prior research has largely treated their work as anecdotal or ideological, emphasizing case studies of particular institutions or policy areas (4, 10, 13, 14). We lack systematic evidence on how think tanks collectively structure the informational ecosystem of policy knowledge: how they interact with one another, how their knowledge circulates between ideological communities, and how scientific evidence figures in the production of policy knowledge. As a result, the architecture of modern policy knowledge, from its organization to polarization to its dependence on science, remains poorly understood.

Understanding this architecture is particularly important amid rising political polarization (15–19) and the politicization of science (20–24). While much attention has focused on elected officials and the public, polarization among the institutions that supply policymakers with expertise may be equally consequential. If think tanks increasingly operate within ideological echo chambers, they not only reflect political divisions but may also amplify them by shaping what each side regards as legitimate knowledge (10). Conversely, if scientific evidence retains authority across ideological boundaries, it could serve as a bridge within this fragmented landscape. Assessing these possibilities requires shifting analytical attention from the use of science in policymaking to its production as policy knowledge.

In this paper, we advance a supply-side perspective on the science-policy interface. Whereas previous work, including studies of congressional citations to science, has examined how policymakers demand scientific input to inform decisions (24–26), here we focus on how intermediary institutions supply that input by creating policy knowledge itself. This distinction between demand and supply is critical: the demand side concerns how political actors select and deploy science in pursuit of their goals. The supply side concerns how organizations produce the body of knowledge from which policymakers later draw, deciding what evidence to incorporate, how to frame it, and with whom to share it. Mapping this upstream process allows us to understand the epistemic foundations on which evidence-based governance rests.

Within this framework, science can play a dual role. On the one hand, it offers epistemic authority: the use of peer-reviewed research signals credibility and differentiates analytical expertise from political opinion (27). On the other hand, science may itself be selectively mobilized and reframed to fit ideological narratives (20, 28). Whether the use of science exacerbates or mitigates polarization in policy knowledge production is an open question with important implications for both the broader roles of science and how evidence functions in democracy.

[1] See the Methods section for the definition of think tanks and sample construction.

To address these questions, we assemble a large-scale dataset of approximately 2 million U.S. policy documents produced between 1998 and 2021 by more than 200 think tanks and 100 government organizations. We trace the links between these policy documents as well as their links to the scientific literature, by linking policy documents to nearly 1 million citations to scientific publications, allowing us to construct a comprehensive map of the policy knowledge ecosystem in the U.S. Using network analysis, we quantify how think tanks exchange policy knowledge with one another as well as with government, measure the ideological polarization of these networks, and examine how engagement with science relates to their structure.

Our analyses reveal three key findings. First, we show that think tanks are the dominant external suppliers of science-based policy knowledge to government, far surpassing intergovernmental and peer government sources. Second, we provide the first systematic evidence that policy knowledge production within the think tank community has become increasingly polarized, forming distinct ideological communities that mirror broader political divisions. Third, we demonstrate that science use is consistently linked to lower ideological polarization and greater influence: policy documents that cite scientific research, especially highly impactful studies, are less ideologically segregated and occupy more central positions in the policy knowledge network. These patterns hold across policy domains and fields of research, suggesting that science has the potential to serve as a shared epistemic resource amid growing polarization.

Taken together, these results provide the first systematic analyses of the architecture of modern policy knowledge production in the U.S. They show that while the production of policy knowledge has become more ideologically divided, science remains a powerful organizing principle that can bridge political boundaries. By mapping how scientific expertise structures policy knowledge itself, this study captures the supply side of science in policymaking, which complements research on the demand side of science use and deepens our understanding of the informational foundations of evidence-based governance.

## Results

### Think tanks as a dominant supplier of expertise in policymaking

To shed light on the critical role of think tanks in producing and supplying expertise to policymakers in government, we begin by examining the extent to which government organizations rely on external sources for policy knowledge: other government organizations, intergovernmental organizations, and think tanks. Because policymakers in government often face challenges in accessing expertise (29, 30), and science has long served as a principal source of evidence-based insights relevant to complex policy problems (8, 24, 27, 31, 32), we pay particular attention to what we call "science-based policy knowledge"—policy knowledge grounded in scientific evidence, approximated by policy documents that cite at least one scientific publication (33). We then track how government organizations engage with this knowledge by examining the citations in their own policy documents to policy documents produced by other government organizations, intergovernmental organizations, and think tanks (n = 81,163). To assess the scientific basis of the cited documents, we classify them according to whether they cite at least one scientific publication, referring to these as "science-based," and to those that do not as "non-science-based" for simplicity.

Our analysis reveals that think tanks are a crucial source of science-based policy knowledge for policymakers in government (4, 10) (see Fig. S1 for the direct use of science by organization type). As shown in Fig. 1A, when government organizations seek policy knowledge that does not cite science, they predominantly engage with other government organizations. By contrast, when government organizations seek science-based policy knowledge, they most frequently engage with think tanks, followed by government organizations and then intergovernmental organizations. In Fig. S2–S4, we show that this contrast is highly consistent over time, policy domains, and fields of research (the Methods section details the measurement of policy domains and fields of research). This consistency extends across levels of government and is especially pronounced in Congress (Fig. S5). In addition, we find that government organizations rely more heavily on think tanks when policy discussion draws on higher-impact science (Fig. S7 and Table S2; see the Methods section for details on how impact is measured). Together, these findings consistently suggest that think tanks are a key component in the policy process, especially critical in integrating scientific expertise into the policy process, highlighting the important role of think tanks in policymaking (9, 34–36).

Given their central role as suppliers of policy knowledge, we next ask whether think tanks' engagement with government varies along partisan lines. Here we examine policy documents produced by the House committees and Senate committees, which together account for approximately 40% of policy documents produced by U.S. governmental organizations in our data (see Fig. S6). For both chambers, we estimate three regression models, where the outcome variables are whether the policy document cites at least one policy document from a left-leaning, non-ideological, and right-leaning think tanks, respectively. The key predictor in common is a binary measure of partisan control in the respective chamber (Democratic: 1, Republican: 0). For all models, we further incorporate policy domains and linear and quadratic year trend terms (see Table S1 for regression tables). As shown in Fig. 1B (House) and Fig. 1C (Senate), Congressional committees cite ideologically aligned think tanks, drawing more heavily on left-leaning sources under Democratic control and, conversely, on right-leaning ones under Republican control. This pattern is particularly pronounced in the House, where Democratic majorities are also less likely to cite right-leaning think tanks, indicating a partisan swing in sourcing policy knowledge. Together, these results demonstrate that policymakers preferentially draw on ideologically aligned think tanks, revealing how the channels through which expertise enters government are structured by partisan boundaries.

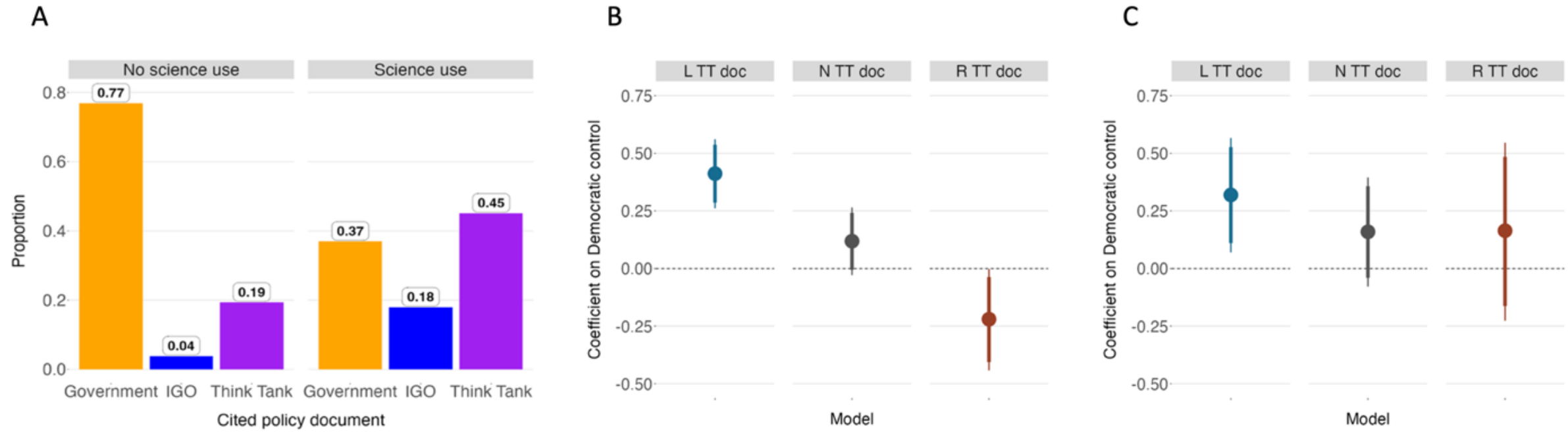

Figure 1: Thinks tanks as a dominant supplier of expertise in policymaking. (A) The sources of policy documents cited by government organizations. We subset these cited documents into those that do not include any citations to any scientific publication and those that cite at least one scientific publication. Here we include 105 unique government organizations (30 federal, 35 state, and 40 city). We examine 48,944 policy documents cited by government organizations that do not include any citations to science ("no science use") and 32,219 policy documents cited by government organizations that include at least one citation to a scientific publication ("science use"). (B) Logistic regression coefficients for political alignment between House committees and think tanks (n = 33,336). The dots represent point estimates, and the thick and thin lines indicate the 90% and 95% confidence intervals, respectively. (C) Logistic regression coefficients for political alignment between Senate committees and think tanks (n = 15,816). The dots represent point estimates, and the thick and thin lines indicate the 90% and 95% confidence intervals, respectively.

**Increasing ideological polarization of policy knowledge production**

The observed partisan alignment between Congress and think tanks suggests that polarization may extend beyond the transfer of knowledge to the institutions that produce it. To investigate this possibility, we analyze how think tanks exchange policy knowledge with one another and whether these exchanges have become increasingly organized along ideological lines. While political polarization is a multifaceted phenomenon with a range of approaches to defining and studying the topic (15, 18, 37, 38), here we focus on one critical dimension: the concentration of knowledge exchange within ideologically homogeneous communities (39, 40, 40–42). We assess this by analyzing citations among think-tank policy documents. In this context, a higher share of co-ideological citations (those linking think tanks that share the same ideological orientation) indicates greater political polarization, as policy discussion becomes increasingly confined to like-minded groups.

Fig. 2A visualizes the citation network among think tanks, revealing strong ideological segregation: policy knowledge circulates primarily within, rather than across, ideological communities. In Fig. 2B, we take a granular look at the degree of polarization at the document level, describing the policy citations (n = 51,704) in terms of the ideology of the think tank citing a policy document and that of the policy document that is cited. We find that cross-cutting citations are rare. Left-leaning think tanks cite one another roughly ten times more often than they cite right-leaning institutions, whereas right-leaning think tanks cite their peers more than twice as frequently as they cite those on the left. Across all ideological institutions, co-ideological citations are about five times more common than cross-ideological ones, and even when non-ideological think tanks are included, more than half of all citations occur within the same ideological camp. And these patterns are robust across policy areas (Fig. S8A). Taken together, these findings suggest that while policy discussion is divided along ideological lines, the insularity appears particularly pronounced among left-leaning institutions.

Furthermore, this polarization has intensified over the past two decades (Fig. 2C), driven mainly by increasing insularity among left-leaning think tanks (Fig. 2D–F).[2] Fig. 2D shows that the proportion of left-leaning think tanks' co-ideological citations steadily increases over time, whereas the proportions of right-leaning and non-ideological think tanks' co-ideological citations do not exhibit any clear trends. Fig. 2E–F corroborate this trend by examining the distribution of the citing think tank's ideology conditional on that of the cited think tank, accounting for differences in engagement across the ideological divide. Interestingly, right-leaning think tanks' cross-ideological engagement is in fact increasing, although cross-ideological citations remain rare (the red line in Fig. 2E vs. the blue line in Fig. 2F). This asymmetry suggests that polarization in policy knowledge production is not merely the product of two sides drifting apart, but of a structural consolidation on one side of the ideological spectrum. This trend holds across the major policy domains (Fig. S8B–C), and we find a similar overall increase at the think tank level (Fig. S10). Taken together, these results show that policy knowledge exchange increasingly occurs within partisan enclaves, mirroring—and potentially reinforcing—the broader polarization of American politics (39, 40, 40–43), suggesting that ideological divisions now reach deep into the institutions that generate the evidence base for policymaking.

[2] One potential concern is that the observed increase in polarization simply reflects the expansion of the Overton database over time, with newly observed organizations entering the database at different rates. Our analyses provide little support for this possibility. First, the share of policy documents produced by newly observed organizations declines over time, indicating that database expansion becomes progressively less important as coverage matures. Second, excluding all citations involving newly observed organizations yields nearly identical polarization trends to those obtained using the full network. Together, these findings, reported in Table S3 and Figure S9, suggest that the observed increase in polarization is not driven by the entry of newly observed organizations into the database, but instead reflects a broader structural change in policy knowledge production.

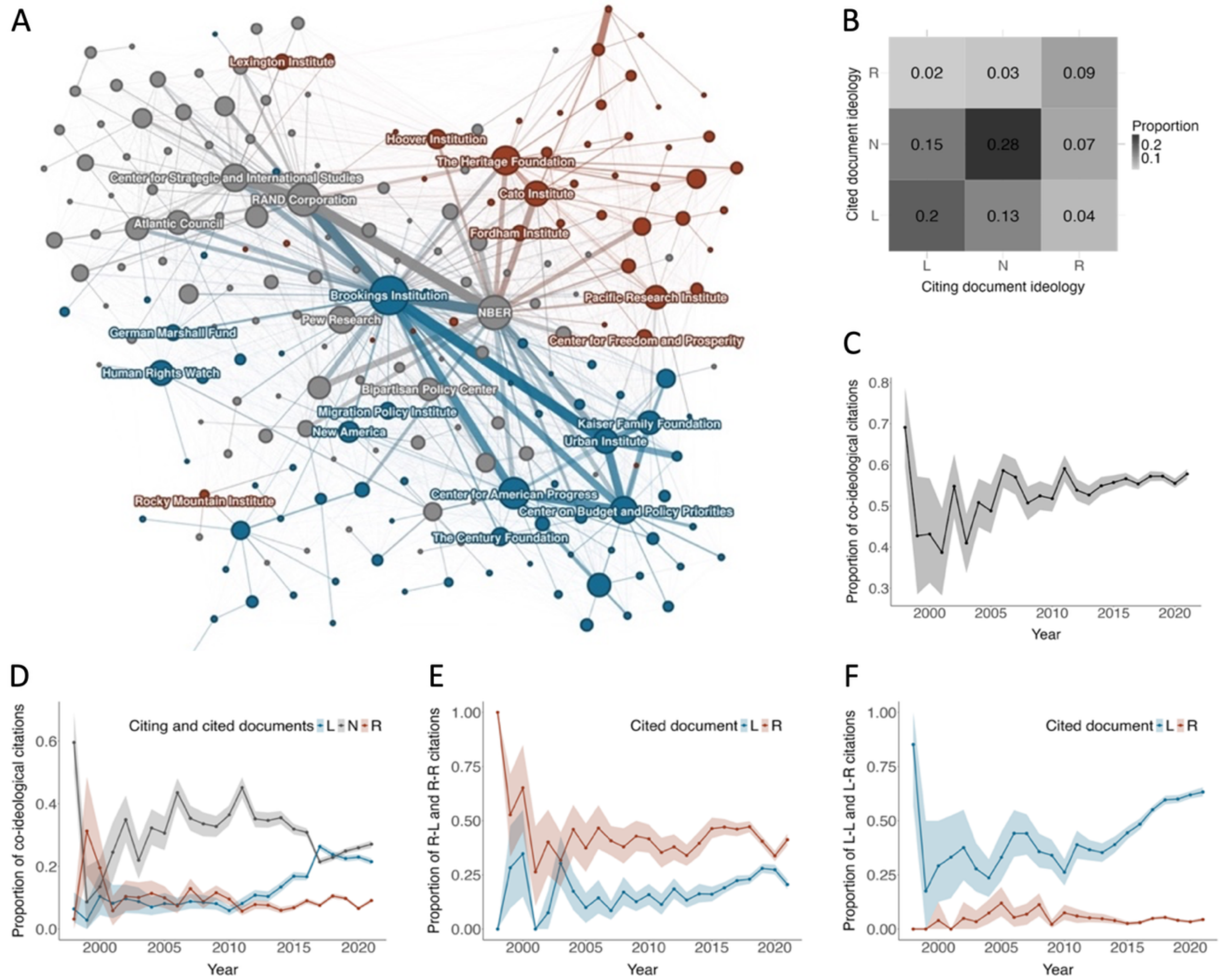

Figure 2: Increasing ideological polarization of policy knowledge production. (A) The policy citation network. Nodes represent think tanks, and weighted directed edges from nodes $i$ to $j$ represent the number of times policy documents produced by $i$ cite policy documents produced by $j$. Node color indicates the ideology of the think tank, with blue, red, and gray representing left-leaning, right-leaning, and non-ideological think tanks, respectively. Node size is proportionate to indegree. Edge color is the unweighted midpoint of its two endpoint colors, blending the source and target think tanks' ideology colors equally. Edge width is proportionate to edge weight. (B) The joint distribution of the ideology of think tanks producing the citing and cited policy documents. (C) Proportion of co-ideological citations over time. (D) Proportion of co-ideological citations over time, broken down by ideology. (E) Proportion of co- and cross-ideological citations over time, conditioning on citing policy documents from left-leaning think tanks. (F) Proportion of co- and cross-ideological citations over time, conditioning on citing policy documents from right- leaning think tanks. For (E)–(F), the colors reflect the ideology of the think tank producing the cited document. For (C)–(F), the area around the line depicts the 95% C.I.

**The use of science and policy influence**

What is the role of science in policy knowledge production, especially for ideologically different think tanks? Using the think tank policy citation network, we first measure think tanks' influence with network centrality.[3] We rely on two complementary network centrality measures: indegree and eigenvector centrality, which have been widely used to measure influence in network settings (46–52). While the former provides an intuitive metric of influence, measuring the number of policy citations think tanks receive, the latter also takes into account the importance of those who cite them (53). We employ these measures as outcome variables in regression models. The key predictor here is science use by each think tank, measured as the proportion of policy documents from the think tank that cite at least one scientific publication (see Fig. S12–S14 for descriptive statistics on science use and network centrality, and Fig. S17 for a visualization of the policy citation network illustrating the relationship between science use and indegree centrality). We use OLS regression for eigenvector centrality as it is a continuous variable, and we use negative binomial regression for indegree centrality as it is a count variable with over-dispersion (54).

We also incorporate key covariates: a think tank's ideology, the number of years since its founding, revenue, geographic location, and its number of policy documents. We classify think tanks' ideology using a three-step procedure combining IGScore estimates (43), membership in the State Policy Network, and keyword-based labeling of mission statements (see the Methods section for details). For years since founding, revenue, and geographic location, we rely on a think tank's IRS Form 990, an official tax document that contains financial information about non-profit organizations. We include the number of policy documents per think tank because think tanks producing more policy documents are more likely to be cited and therefore more central (see Fig. S11 for more details about these measures).

As shown in Fig. 3A–B, think tanks producing policy documents that frequently rely on scientific publications tend to be central in the network, in terms of both eigenvector centrality and indegree centrality (see Tables S4 and S5 for regression tables). To assess whether this association varies by ideology, we estimate regression models with multiplicative terms (55) between science use (continuous) and ideology (categorical). Fig. 3C–D show no significant ideological differences in this relationship: greater use of science predicts higher centrality for left-leaning, right-leaning, and non-ideological think tanks alike, suggesting that, even amid ideological division, science appears to function as a shared marker of authority (see Tables S6 and S7 for regression tables).

Robustness checks confirm that this positive association persists across policy domains, fields of research, and at the document level (Tables S11–S17). Also, to address the interdependence among think tanks in eigenvector centrality and potential bias in uncertainty estimation, we also report regression results based on p-values using permutation tests, arriving at consistent conclusions (Fig. S15). In addition, we find that the role of science as a source of influence in policy knowledge production is universal, regardless of the impact of the cited scientific publication within the scientific community (Tables S8–S10 and Fig. S16). This same pattern is corroborated at the document level: policy documents that cite a greater number of scientific publications receive more policy citations, a relationship that persists even when these citations are differentiated by the impact of the cited science (Table S11).

[3] Citation centrality captures a document's or organization's prominence within the policy knowledge network, as citations indicate the uptake and circulation of knowledge across organizations. However, influence can also operate through channels beyond citation networks, including personnel movement, lobbying, media visibility, private briefings, and elite social networks. We therefore interpret citation centrality as capturing influence specifically within the observable structure of policy knowledge production, rather than as a comprehensive measure of policy influence more broadly.

Together, these results suggest that engagement with science is associated with a think tank's visibility and credibility within the policy knowledge network, regardless of ideology. This raises the next important question: does science also help bridge the ideological divides that structure policy knowledge production?

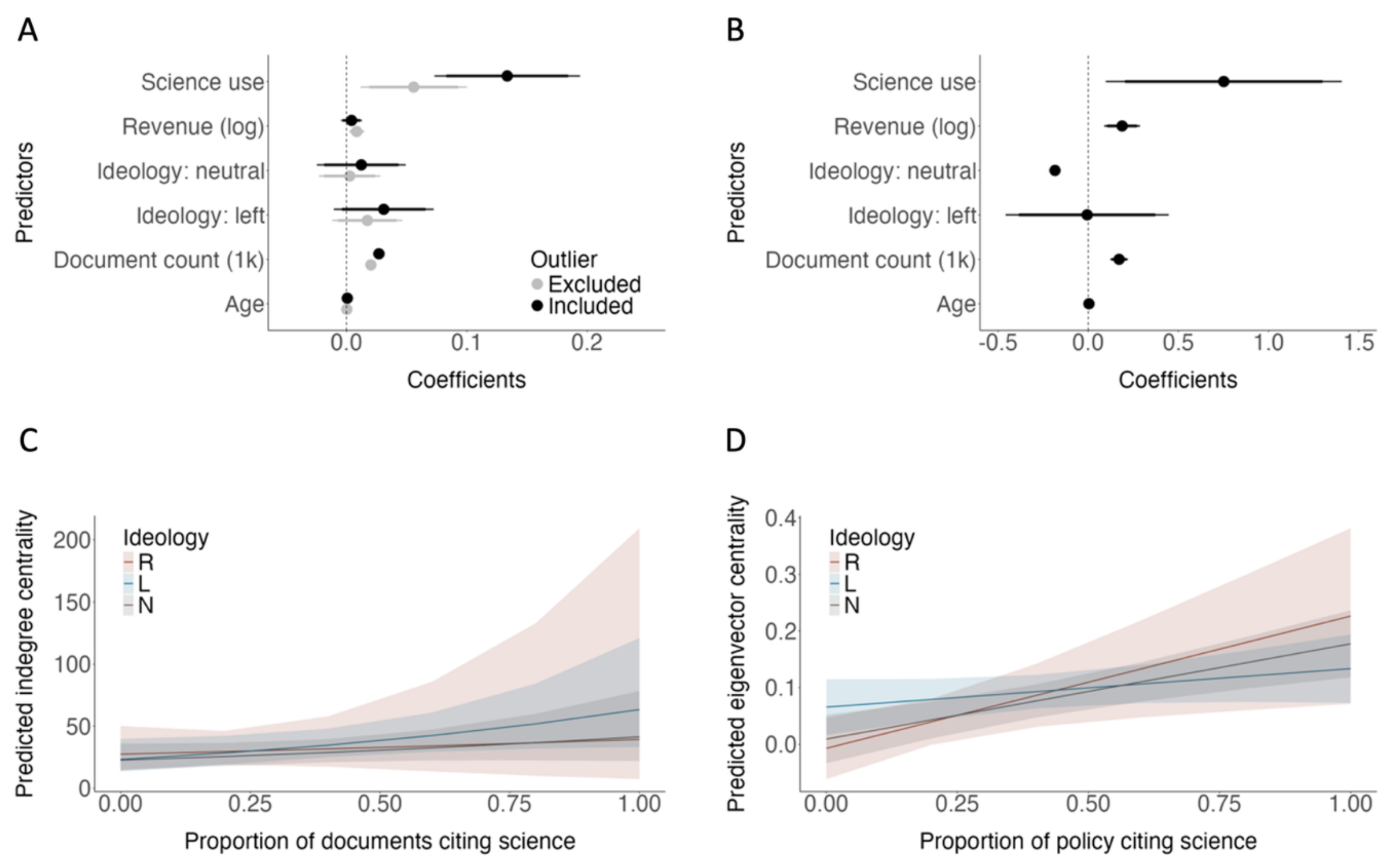

Figure 3: The use of science and policy influence. (A) Coefficients from the ordinary least squares models for eigenvector centrality. (B) Coefficients from the negative binomial regression models indegree centrality. For both, the dots depict point estimates, and the thick and thin lines around the dots depict the 90% and 95% confidence intervals. Revenue is in U.S. dollars and log-transformed. The number of policy documents is denominated in thousands. Ideology is a categorical variable with the right-leaning as the baseline. (A) shows the coefficients from two models including and excluding influential observations, identified based on Cook's distance (56). The threshold is set at 4 / the number of observations, the widely used threshold. (C) Predicted centrality values by science use for ideologically different think tanks (eigenvector centrality). (D) Predicted centrality values by science use for ideologically different think tanks (indegree centrality). For both, the areas around the line indicate the 95% confidence intervals.

**Science use and ideological polarization in policy knowledge production**

To further illuminate the relationship between science and policy knowledge production, we dissect the policy citation network in terms of science use, allowing us to measure the extent of science use and analyze its relationship with the degree of ideological polarization. As before, we begin by classifying the citations between think tanks' policy documents into those that use science and those that do not. For a policy citation to be classified as using science, both the citing policy document and the cited policy document in the citation pair must cite at least one scientific publication. We classify a policy citation as not using science when neither of the policy documents cites any scientific publication. We then assess the degree of ideological polarization by examining the proportion of co-ideological citations within the networks (we also evaluate cases in which only the citing or cited document uses science; both cases fall between those in which both or neither document uses science, consistent with the overall pattern we identify; see Fig. S18).

In doing so, we further investigate whether the relationship between science use and polarization is conditional on the impact of the cited science. Specifically, we examine whether co-ideological citations are less likely when the cited scientific research is highly influential—operationalized as citation counts relative to other publications in the same year and field. "Hit 10," "Hit 5," and "Hit 1" publications refer to papers in the top 10%, 5%, and 1%, respectively, of citations within their field and year (see the Methods section for details on how impact is measured).

We find that, when science is used, co-ideological policy discussion is less common. This association implies that policy discussion that is, in some minimal sense at least, grounded in science by directly citing scientific publications, is less polarized than policy discussion that is not grounded in science. As illustrated in Fig. 4A–B, although the decrease is larger for right-leaning think tanks than for left-leaning think tanks, the pattern holds for both (Fig. 4A for left-leaning and Fig. 4B for right-leaning). Moreover, the pattern is more prominent when the cited science is high-impact, revealing a monotonic decrease in co-ideological citation as scientific impact increases. While these results are associational rather than causal, they suggest that the quality and credibility associated with high-impact science may reduce reliance on ideologically congenial policy source. We further find that this reduction in co-ideological citation in science-based policy discussion appears across a wide range of policy domains and fields of research (Fig. S19).

Importantly, the decrease in the co-ideological policy discussion does not necessarily mean an increase in cross-ideological policy discussion. Rather, left-leaning and right-leaning think tanks interact with more technocratic, non-ideological think tanks like RAND or the NBER (the gray lines in Fig. 4A–B) to a greater degree when policy discussion is based on science. This finding accords with our previous conclusion that the degree of assortativity within the policy citation network is much lower when we incorporate non-ideological think tanks into the network. Together, these findings illustrate that non-ideological think tanks are key actors that not only incorporate scientific expertise into their own policy research (see Fig. S13), but also function as a bridge between left-leaning and right-leaning think tanks in policy discussion, encouraging science-based policy discussion beyond the ideological echo chambers of policy knowledge.

Next, to assess whether the association reflects a systematic relationship between science use and polarization, rather than differences in document- or organization-level characteristics, we also conduct regression analyses (Table S18) in which we model whether a given policy-to-policy citation is co-ideological as a function of science use, measured by either the (log-transformed) count of scientific publications cited or a binary indicator of whether science is cited. Because citations between policy documents are dyads, there are multiple ways to operationalize whether science is used in the pair. We report models specified for the citing policy document in the policy-policy dyad cites science, the cited document cites science, and both documents cite science. All models include think tank fixed effects, year fixed effects, and policy domain and field of research controls. Due to the think tank fixed effects, the estimated association is identified from variation in science use across citations from the same think tank; the issue and field controls similarly hold constant policy domains and fields of research. We find that science use remains negatively associated with co-ideological citation in these more demanding specifications.[4]

To further illuminate the relationship between science use and ideological polarization, we examine whether the same pattern of lower polarization is observed at the think tank level. Specifically, we construct policy citation networks of think tanks from citations that differ in whether scientific evidence is used in policy documents and in the impact of the cited science, following the classification framework introduced above. For each network, we compute ideological assortativity. Consistent with the document-level analysis, we find that the use of science is associated with lower levels of homophily—indicating reduced ideological polarization—and that assortativity decreases even further when policy citations draw on high-impact scientific research (Fig. 4C). This pattern is robust to how network edges are defined: the reduction in ideological assortativity holds across alternative specifications of edge directedness and weight (Fig. S20A–B) and generalizes across a wide range of policy domains and scientific fields (Fig. S20C–E, Fig. S21). Fig. 4D–E visualize this pattern of lower polarization.

Collectively, these findings indicate that policy discussion is less polarized when it is grounded in science—a relationship that is monotonic, with more impactful science associated with lower levels of polarization, suggesting that scientific expertise may serve as an alternative to ideological alignment as an organizing principle of policy knowledge production.

[4] This substantially narrows alternative explanations based on stable differences across organizations, temporal changes, and characteristics of policy domains and fields of research that may jointly shape science use and ideological polarization. As with any observational analysis, these models cannot rule out all alternative mechanisms. Nevertheless, they provide a more stringent test of the association than comparisons that do not account for organizational, temporal, policy domain, and research field differences.

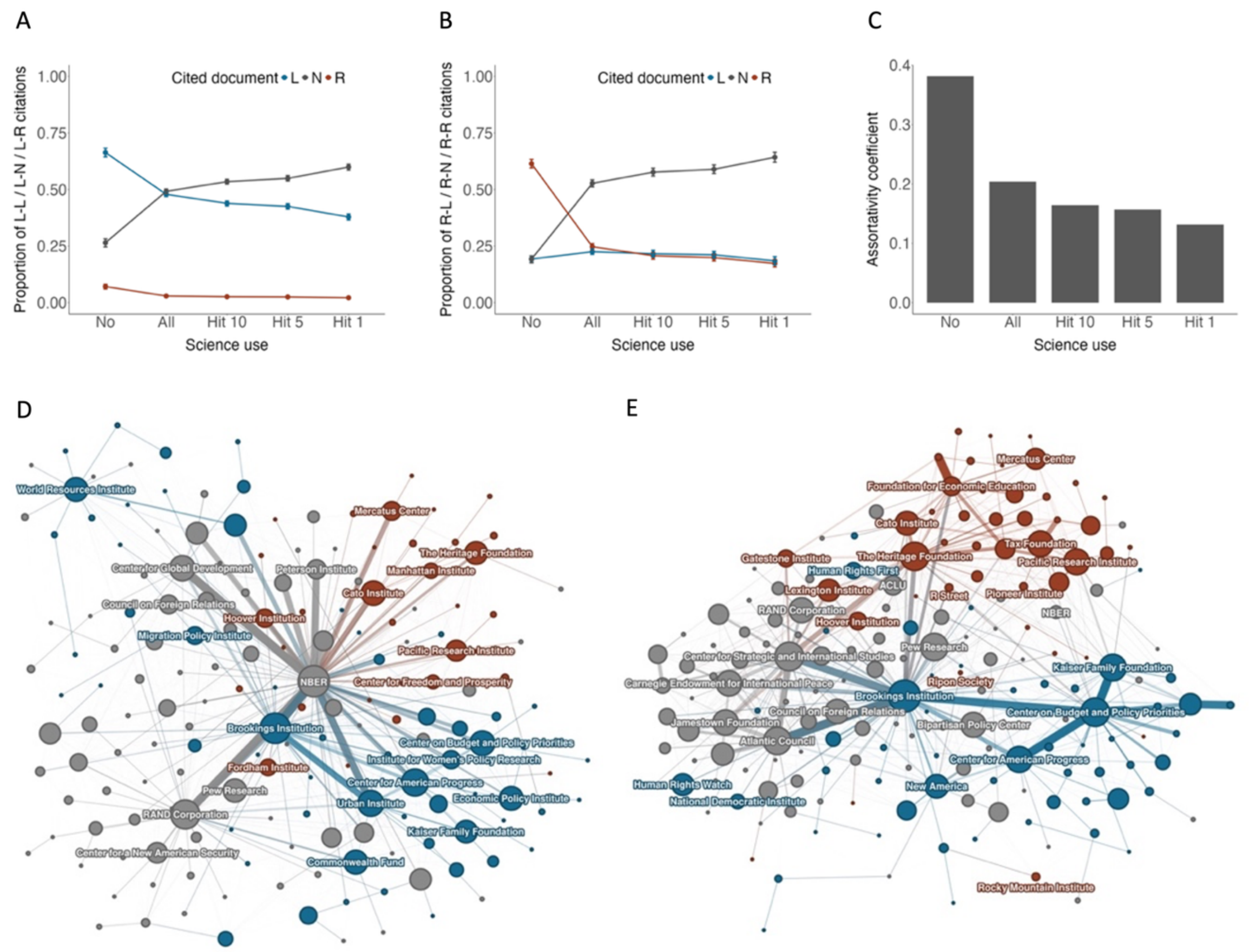

Figure 4: Science use and ideological polarization in policy knowledge production. (A) The distributions of ideology in policy citations by science use and impact: conditioning on the citing document from a left-leaning think. (B) The distributions of ideology in policy citations by science use and impact: conditioning on the citing document from a right-leaning think tank. (C) Assortativity coefficients by science use and impact. (D) The policy citation network built on policy citations that use science. (E) The policy citation network built on policy citations that do not use science. For (D)–(E), Nodes represent think tanks, and weighted directed edges from nodes $i$ to $j$ represent the number of times policy documents produced by $i$ cite policy documents produced by $j$. Node color indicates the ideology of the think tank, with blue, red, and gray representing left-leaning, right-leaning, and non-ideological think tanks, respectively. Node size is proportionate to indegree. Edge color is the unweighted midpoint of its two endpoint colors, blending the source and target think tanks' ideology colors equally. Edge width is proportionate to edge weight.

## Discussion

Taken together, this study examines the supply side of science in policymaking, offering the first large-scale evidence on the institutional and ideological dynamics underlying the production of policy knowledge. By mapping millions of policy documents and their connections to other policy documents as well as scientific publications, we show that think tanks play a crucial role in supplying science-based knowledge to government, that their networks have grown increasingly polarized, and that greater engagement with science is associated with lower levels of polarization. Together, these results reveal how science is embedded the informational foundations of modern policymaking, not only as an object of political debate but as a key feature of the architecture of policy knowledge production.

A central contribution of this work is to conceptualize and empirically map policy knowledge production as an organized system. Policy knowledge does not arise solely within government, nor does it flow linearly from academia into policy (57). Instead, it is generated within a distributed network of institutions—what scholars have called knowledge regimes (4, 13)—that translate, curate, and legitimate expertise for political use (3). Within this regime, think tanks are the principal producers of policy knowledge grounded in science. When policymakers in government seek scientific evidence, they overwhelmingly draw from think tanks rather than peer government organizations or intergovernmental organizations. This finding highlights an institutional asymmetry: much of what appears as scientific evidence in policy originates outside the state, within organizations that blend analytical expertise with ideological missions.

In doing so, our analysis uncovers deep and growing polarization in the production of policy knowledge. Over the past two decades, think tanks have become increasingly clustered along ideological lines, exchanging knowledge primarily within like-minded communities. This pattern mirrors the broader polarization observed in the U.S. politics (17, 19, 43–45), suggesting that political divisions now extend into the epistemic infrastructure of policymaking itself. The fact that polarization has intensified at the very sites where evidence is produced indicates that partisan dynamics shape not only how science is used in policy, but likely how policy knowledge itself is constructed.

Our analyses also reveal that the recent intensification of polarization is asymmetric. The growing insularity of left-leaning think tanks—contrasted with relatively stable or even modestly cross-cutting engagement among right-leaning think tanks—suggests that polarization in policy knowledge production is driven less by mutual estrangement than by consolidation on one side of the ideological spectrum. This pattern aligns with historical shifts in the institutional landscape, where new advocacy-oriented and issue-specific organizations have proliferated on the political left, tightening internal knowledge networks (4). Conceptually, this finding reframes polarization in policy knowledge production not simply as a symmetric process of divergence, but as an uneven restructuring of the epistemic ecosystem that shapes the flow of expertise into government.

Yet amid this polarization, we find consistent evidence that science use is associated with lower levels of polarization. Policy documents that engage with scientific research, especially those citing high- impact studies, are less ideologically segregated. This association appears rather universal across policy domains and fields of research, suggesting that science has the potential to function as a shared epistemic currency even within divided policy communities. This pattern is consistent with the hypothesis that science introduces a common reference system, including a set of methods, standards, and evidentiary norms, that transcends partisan frames.

In this sense, science may not dissolve ideological divisions, but may be linked to policy discourse that is more closely anchored to shared empirical ground (76). It thus further suggests that the integration of scientific expertise into policy knowledge may help sustain limited yet meaningful zones of epistemic overlap, amid increasingly polarized political environments.

Our findings complement and extend emerging evidence on how political actors use science in policymaking. Whereas previous studies have examined the demand side—how policymakers differ in the science they cite to advance partisan goals (24), we focus on the supply side—how organizations create and circulate policy knowledge before it reaches policymakers. Together, these perspectives reveal two interdependent stages of a continuous epistemic cycle linking the creation, transmission, and application of expertise in democratic governance. Overall, this paper paints a fuller picture of the overall informational environment from which policymakers draw their knowledge. By revealing how that environment is structured—who produces policy knowledge, what science it incorporates, and how it is polarized—this analysis illuminates the upstream processes that shape the evidence available for political decision-making.

Conceptually, our findings point to a network-based understanding of policy knowledge production. Rather than treating policy knowledge as a collection of individual ideas or organizational outputs, it may be understood as an institutionalized system of exchange in which organizations produce, circulate, and legitimate expertise (3, 9, 35, 58–60). Within this system, polarization manifests structurally, through the concentration of knowledge flows within ideologically aligned communities, rather than solely through rhetorical positioning or policy preferences (43, 61–63). At the same time, scientific research operates as an epistemic resource that can potentially confer authority and shape organizational interaction patterns. These empirical regularities therefore open promising avenues for future theoretical work. In particular, the patterns we uncover suggest that policy knowledge may be fruitfully modeled as a networked system of exchange among organizations, in which polarization arises from the organization of knowledge flows rather than from individual preferences alone, and in which scientific research functions as a source of epistemic authority. Developing theory that formalizes these dynamics represents an important direction for future research.

The findings also carry broader implications for debates about evidence-based policymaking. Calls for more scientific input in policy often assume that providing high-quality evidence will lead to better decisions. Our results suggest a more complex reality. Even when science enters the policy process, its effects likely depend on the institutional channels through which it flows (11). If those channels are polarized, scientific evidence may be unevenly incorporated. Strengthening evidence-based governance, therefore, may be not only about improving the quality of research or communication, but about cultivating institutional environments that support cross-ideological exchange of scientific expertise. This insight is particularly relevant in an era when new technologies, from social media to generative AI, are transforming how knowledge is produced and disseminated (64–67). As information ecosystems become more fragmented and algorithmically mediated, ensuring the integrity and connectivity of science-based policy knowledge will be critical to maintaining the epistemic foundations of future democratic governance.

Several limitations warrant caution and point toward future research. First, our analyses are associative rather than causal. While we find robust relationships between science use and reduced polarization, which hold consistently across policy topics and fields of research, we cannot determine whether science itself causes depolarization or whether less polarized topics attract more scientific engagement. Future work could leverage exogenous shocks or natural experiments to identify causal mechanisms. Second, our analysis focuses on the U.S. whereas policy knowledge regimes vary widely across countries. The United States has a particularly privately oriented network of policy research organizations—with think tanks playing a central role in generating and communicating policy ideas (4)—whereas in some other advanced democracies these functions can be more directly embedded in state research institutions and coordinated systems of expertise provision (13). Comparative analyses could reveal whether science exerts a similar moderating influence in multiparty systems or in countries with stronger state-led research infrastructures. Third, our approach captures structural relationships, not semantic fidelity. We analyze who cites what, but not how scientific findings are represented or interpreted. Future work combining network analysis with semantic or embedding-based approaches could assess both the structure and content of policy knowledge, tracing how knowledge evolves as it circulates through policy networks.

Overall, the findings presented here reveal the central role of science in the architecture of modern policymaking. Think tanks are key producers of policy knowledge that both reflect and likely reproduce political polarization, yet they also remain important conduits of scientific expertise upon which the government relies. By integrating science into policy knowledge, they not only shape the informational landscape from which policymakers draw, but also sustain potential bridges across ideological divides. Hence, even in polarized times, the pursuit of evidence endures as one of the few shared languages through which societies can understand and govern themselves.

## Methods

**Data sources for policy documents and scientific publications.** For policy documents, we leverage Overton, one of the world's largest databases of policy documents (68). Overton defines policy documents broadly as documents written primarily for or by policymakers. The database includes policy documents from think tanks, government organizations, and intergovernmental organizations. At the time of our data collection (September 13, 2021), the Overton database had approximately 4.5 million policy documents from 187 countries and from more than 1,500 different sources worldwide. For this study, we use the full set of more than 1.8 million policy documents published in the United States from January 1, 1998, to September 13, 2021. We use an API to obtain policy documents from each policy source separately. Overton extracts scientific references in documents and maps them onto DOIs, one of the most commonly used identifiers for scientific publications. We then link the scientific references cited within the policy documents to Dimensions (69), one of the largest scientific publications databases, which includes more than 100 million publications accessible from journals, conference proceedings, books and chapters, and pre-print servers. We identify scientific publications from the Dimensions API using DOI information. We find that a vast majority (approximately 94.3%) of the more than 1 million unique DOIs found in the policy documents (which comprise a total of 2.2 million citations) can be matched to Dimensions records. See Tables S19–21 for the lists of think tanks, government organizations, and intergovernmental organizations.

**Defining and identifying think tanks.** We adopt the definition developed by the Think Tanks and Civil Societies Program (TTCSP), the largest research initiative on think tanks worldwide: "Think tanks are public-policy research, analysis and engagement organizations that generate policy-oriented research, analysis and advice on domestic and international issues, thereby enabling policymakers and the public to make informed decisions about public policy. Think tanks may be affiliated or independent institutions that are structured as permanent bodies, not ad-hoc commissions" (71). This definition identifies think tanks as free-standing organizational units whose institutional purpose is to produce and disseminate policy research; it distinguishes them from universities, which are general-purpose, degree-granting institutions. We study a total of 231 think tanks, rendering our study the most comprehensive study of think tanks in the U.S. We cover 80% of the top 110 think tanks rated by the TTCSP. In addition, we incorporate 143 additional think tanks in our study that are not studied in (70–71) (See Table S19)

**Think tank ideology.** We use a three-step approach to measure a think tank's ideology. First, we leverage a previously validated measure called IGScore (43). The measure employs Bayesian Item Response Theory to infer the ideal points of various organizations, including think tanks, interest groups, and business firms. It makes use of a dataset on organizations' legislative positions on bills filed during the 109th Congress to the 114th Congress (2005–2016). If the think tank has an IGScore that is greater than 0.888, we label the organization right-leaning. If the score is less than -0.629, we label the think tank left-leaning. Otherwise, we label the think tanks non-ideological. The cutoffs center the most bound of the IQR of Democratic (Republican) members of the 114th Congress as scored by Crosson et al. (2020). Second, if the think tank is a member of the State Policy Network (SPN), we label the think tank right-leaning. Founded in 1992, the SPN is recognized as one of the most influential organizations in the conservative movement (72). It is a nonprofit organization that supports and coordinates a network of right-of-center think tanks and advocacy groups across the United States that focus on a range of issues, including limited government, free markets, and individual liberty. Third, we examine think tanks' publicly available mission statements to determine if a set of keywords that signal ideology are mentioned. The inter-coder reliability score for the assessment of ideology using this method is approximately 0.86 in terms of Cohen's $\kappa$ and Krippendorf's $\alpha$. Using this approach, we identified 104 non-ideological think tanks, 72 left-leaning think tanks, and 49 right-leaning think tanks (see Table S18 for the ideology classifications of the think tanks under study).

**Impact and field of research.** To measure impact, we calculate citation percentiles for scientific papers published in the same year and field. In so doing, we take into account heterogeneities in citations across years and fields. Following previous research (73–75), we define Hit 10, Hit 5, and Hit 1 publications as those that are ranked in the top 10%, 5%, and 1% of citations received, respectively. For the field of research, we leverage Dimensions' Fields of Research (FOR) classification for scientific papers. The FOR is based on the Australian and New Zealand Standard Research Classification (ANZSRC) system, which covers a broad set of research fields. We follow the system and classify scientific publications into 22 top-level fields: "Agricultural and Veterinary Sciences," "Biological Sciences," "Built Environment and Design," "Chemical Sciences," "Commerce, Management, Tourism and Services," "Earth Sciences," "Economics," "Education," "Engineering," "Environmental Sciences," "History and Archaeology," "Information and Computing Sciences," "Language, Communication and Culture," "Law and Legal Studies," "Mathematical Sciences," "Medical and Health Sciences," "Philosophy and Religious Studies," "Physical Sciences," "Psychology and Cognitive Sciences," "Studies in Creative Arts and Writing," "Studies in Human Society," and "Technology."

**Policy documents' issue areas.** We leverage Overton's policy issue classifications. Overton uses machine learning approaches to assign fields and topics to policy documents. The Overton policy issue classification is based primarily on the International Press Telecommunications Council (IPTC) Subject Codes taxonomy, the global standards body for the news media. Overton follows the taxonomy to classify policy documents into 18 top-level policy domains: "politics," "conflicts, war and peace," "economy, business and finance," "science and technology," "arts, culture and entertainment," "disaster, accident and emergency incident," "health," "crime, law and justice," "labour," "environment," "sport," "lifestyle and leisure," "society," "prices," "education," "religion and belief," "weather," and "human interest." A citation's issue area is defined based on the issue area of the citing policy document (and not that of the cited policy document). Also, note that policy documents are assigned multiple issue areas; the same policy documents, therefore, count toward multiple issue areas.

# Supplementary Information

# The Production of Policy Knowledge in the United States

Taegyoon Kim, [1,2] Alexander Furnas, [3,4,5,6] Dashun Wang [3,4,5,6,7,*]
[1] School of Digital Humanities and Computational Social Sciences, KAIST, Daejeon, Republic of Korea
[2] Graduate School of Data Science, KAIST, Daejeon, Republic of Korea
[3] Center for Science of Science and Innovation, Northwestern University, Evanston, IL, USA
[4] Ryan Institute on Complexity, Northwestern University, Evanston, IL, USA
[5] Northwestern Innovation Institute, Northwestern University, Evanston, IL, USA
[6] Kellogg School of Management, Northwestern University, Evanston, IL, USA
[7] McCormick School of Engineering, Northwestern University, Evanston, IL, USA

[*] Corresponding author: dashun.wang@kellogg.northwestern.edu

## S1. Thinks tanks as a dominant supplier of expertise in policymaking

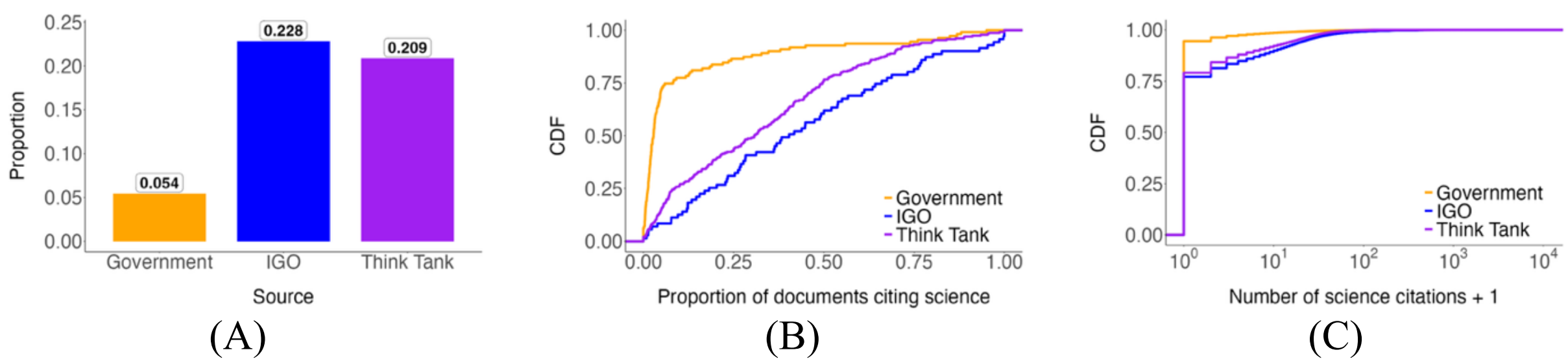

Figure S1: (A) The proportion of policy documents citing at least one scientific publication by institution type. (B) The cumulative distribution function of the proportion of policy documents citing at least one scientific publication by institution type. (C) The cumulative distribution function of the number of citations to scientific publications (per policy document) by institution type.

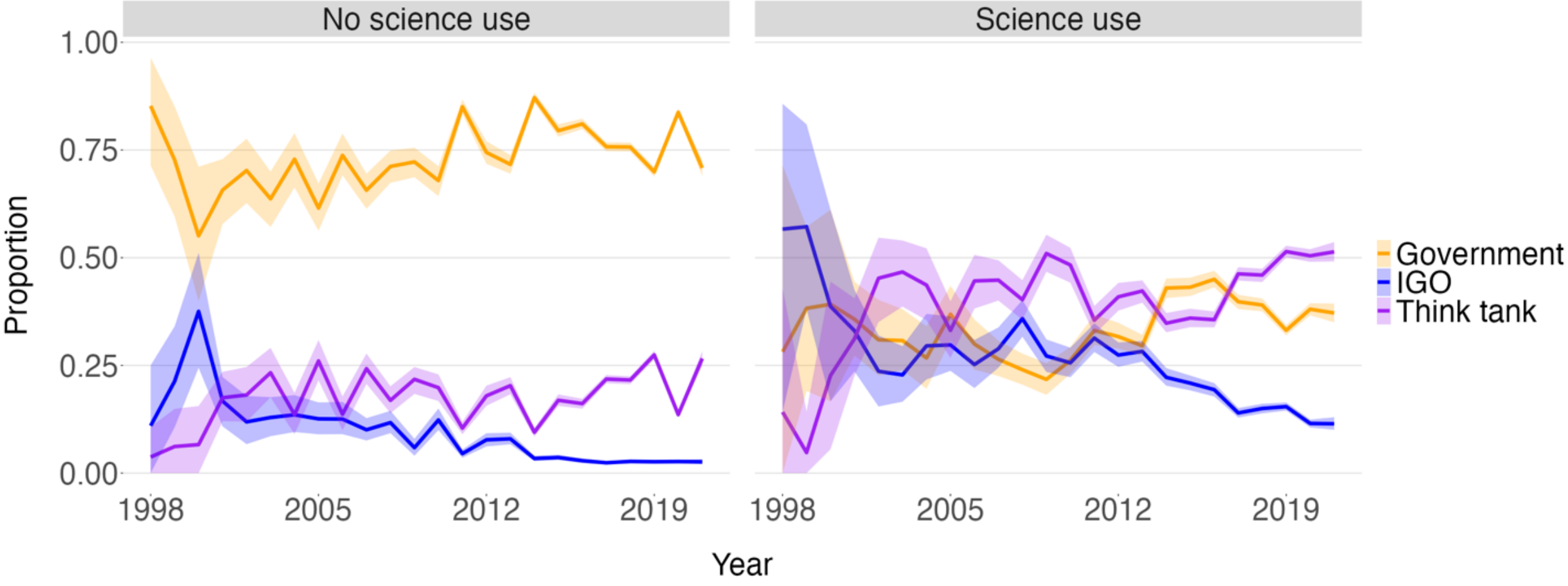

Figure S2. The timelines of the sources of the policy documents cited by governments depending on the document's use of science. The sources of policy documents cited by government organizations over time, for policy documents that do not include any citations to scientific publications and those that cite at least one scientific publication. Consistent with the time-aggregated pattern in the main text, there is a noticeable difference between the top sources of policy documents cited by government organizations that have no scientific references and those that include at least one citation to a scientific publication. Peer government organizations are the top source for the former, while think tanks are the top source for the latter for most of the years analyzed). The area around the line depicts the 95% bootstrapped confidence interval.

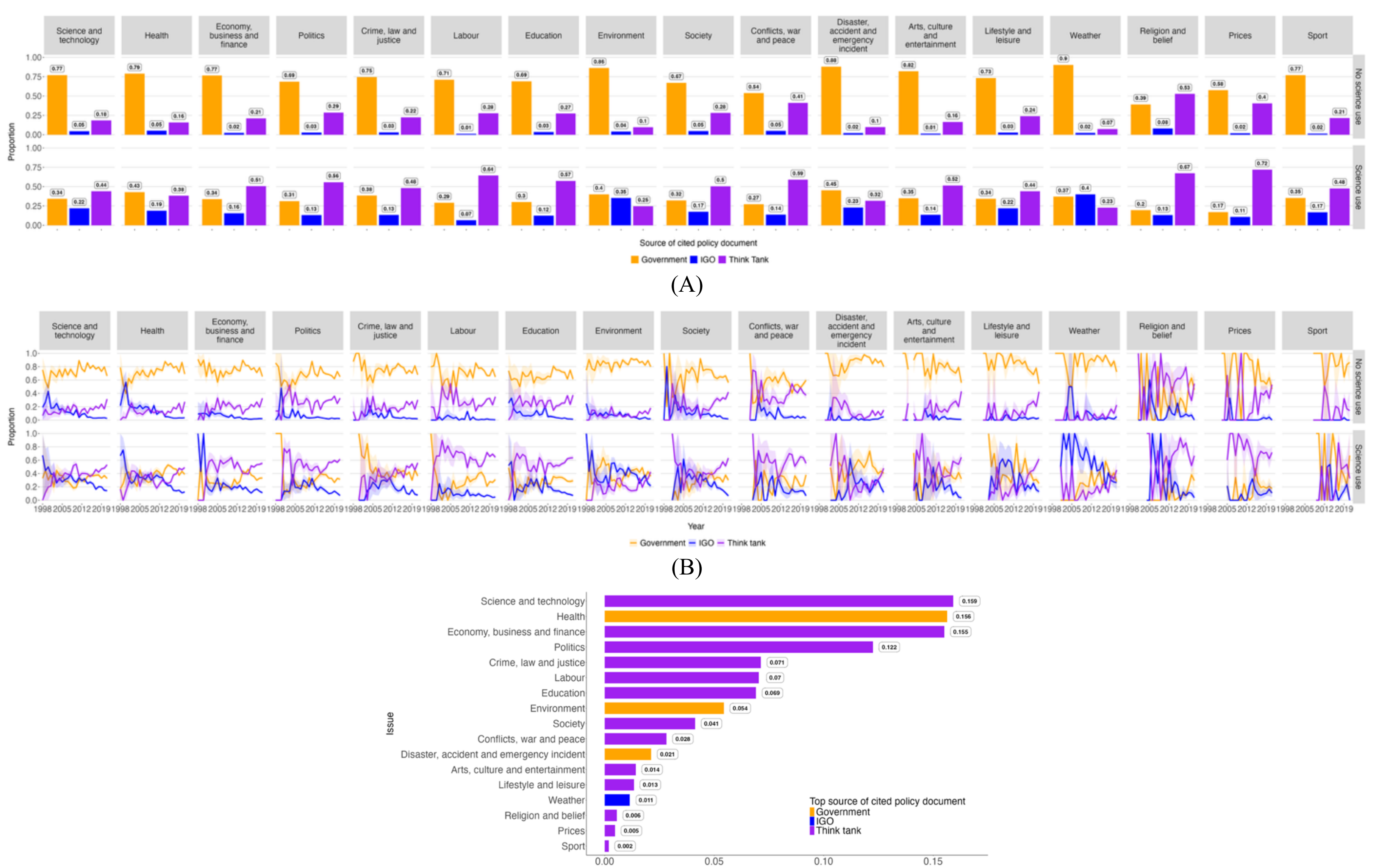

(A)

(B)

(C)

Figure S3. Breakdowns by policy issue area. Together, (A)—(C) illustrate the differences in the sources of the policy documents cited by government organizations depending on their science use, for each policy issue area. For (A)—(B), panes are ordered by the frequency with which governments' policy documents (the citing policy documents) belong to that area. Specifically, (A) depicts the time-aggregated pattern, and (B) depicts the timeline (with the 95% bootstrapped confidence interval). Finally, (C) shows the top source by policy issue area when science is used in the cited document. We can see that for 13 out of 17 issue areas, think tanks are the top source cited by governments for science-based policy information, accounting for approximately 76% of all citations by government organization

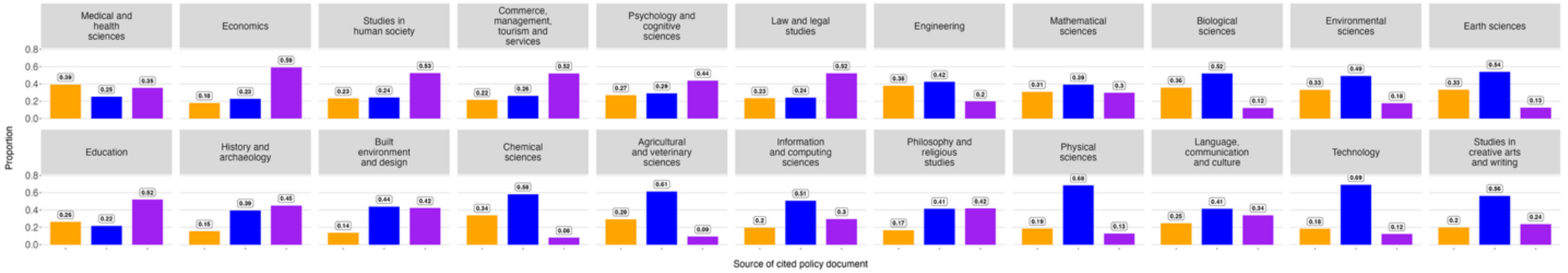

Medical and health sciences
Economics
Studies in human society
Commerce, management, tourism and services
Psychology and cognitive sciences
Law and legal studies
Engineering
Mathematical sciences
Biological sciences
Environmental sciences
Earth sciences
Education
History and archaeology
Built environment and design
Chemical sciences
Agricultural and veterinary sciences
Information and computing sciences
Philosophy and religious studies
Physical sciences
Language, communication and culture
Technology
Studies in creative arts and writing
Proportion
Source of cited policy document
Government IGO Think Tank

(A)

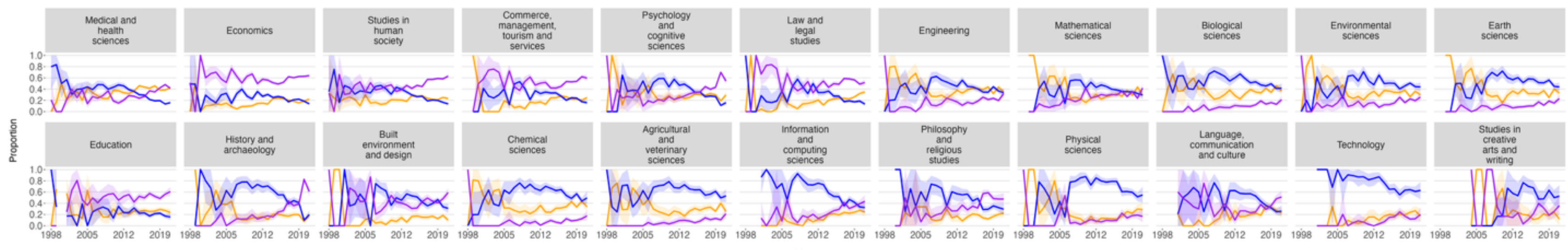

Proportion
Year

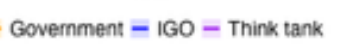

Government IGO Think tank

(B)

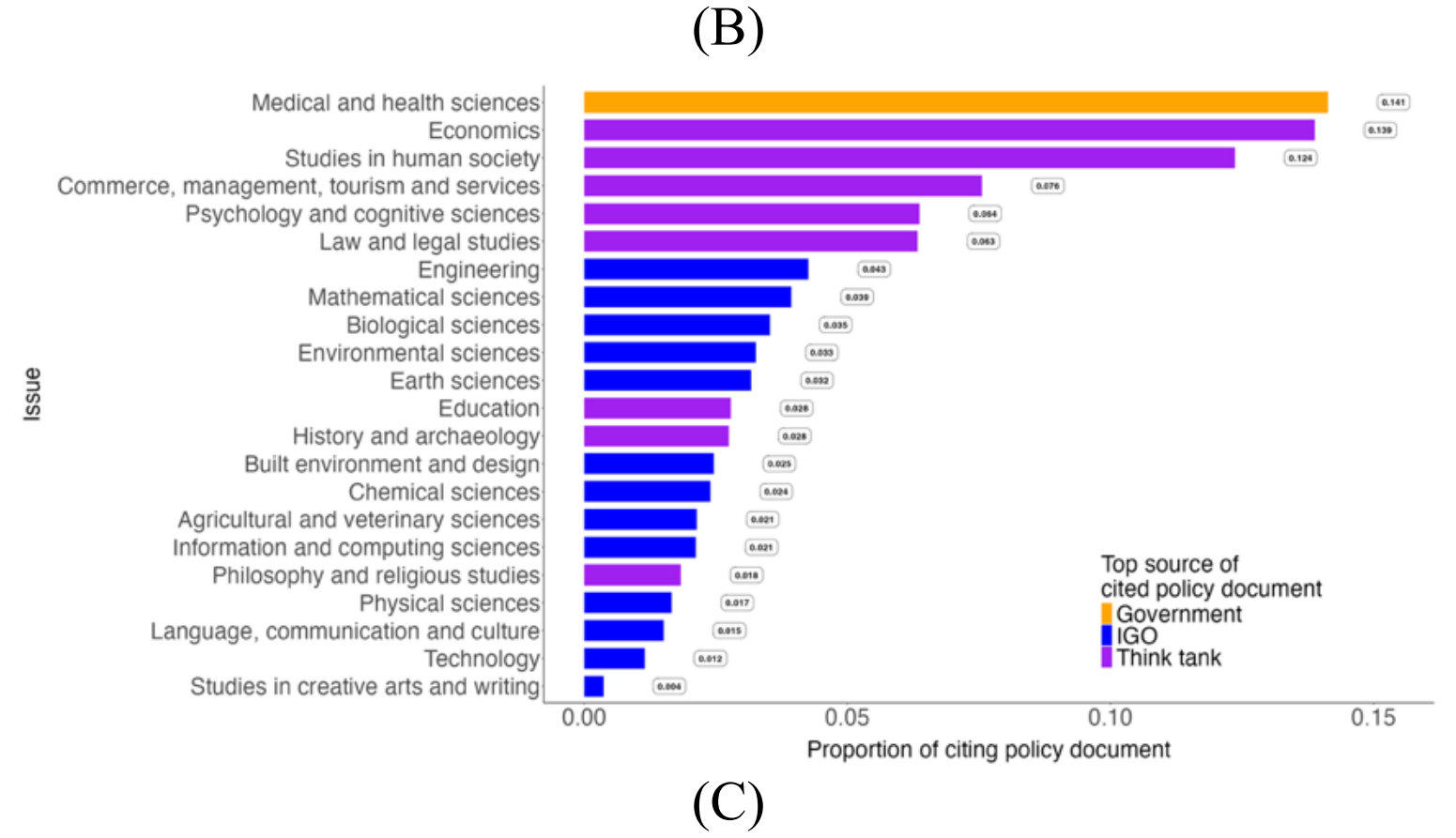

Issue
Proportion of citing policy document
Top source of cited policy document
Government
IGO
Think tank

(C)

Figure S4. Breakdowns by field of research. Together, (A)—(C) illustrate the differences in the sources of the policy documents cited by government organizations depending on their science use, for each field of research. For (A)—(B), panes are ordered by the frequency with which governments' policy documents (the citing policy documents) belong to that field. Specifically, (A) depicts the time-aggregated pattern, and (B) depicts the timeline (with the 95% bootstrapped confidence interval). Finally, (C) shows the top source by field of research when science is used in the cited document. We see that for 8 out of 22 fields, think tanks are the top source cited by governments for science-based policy information, accounting for approximately 56% of all citations by government organizations.

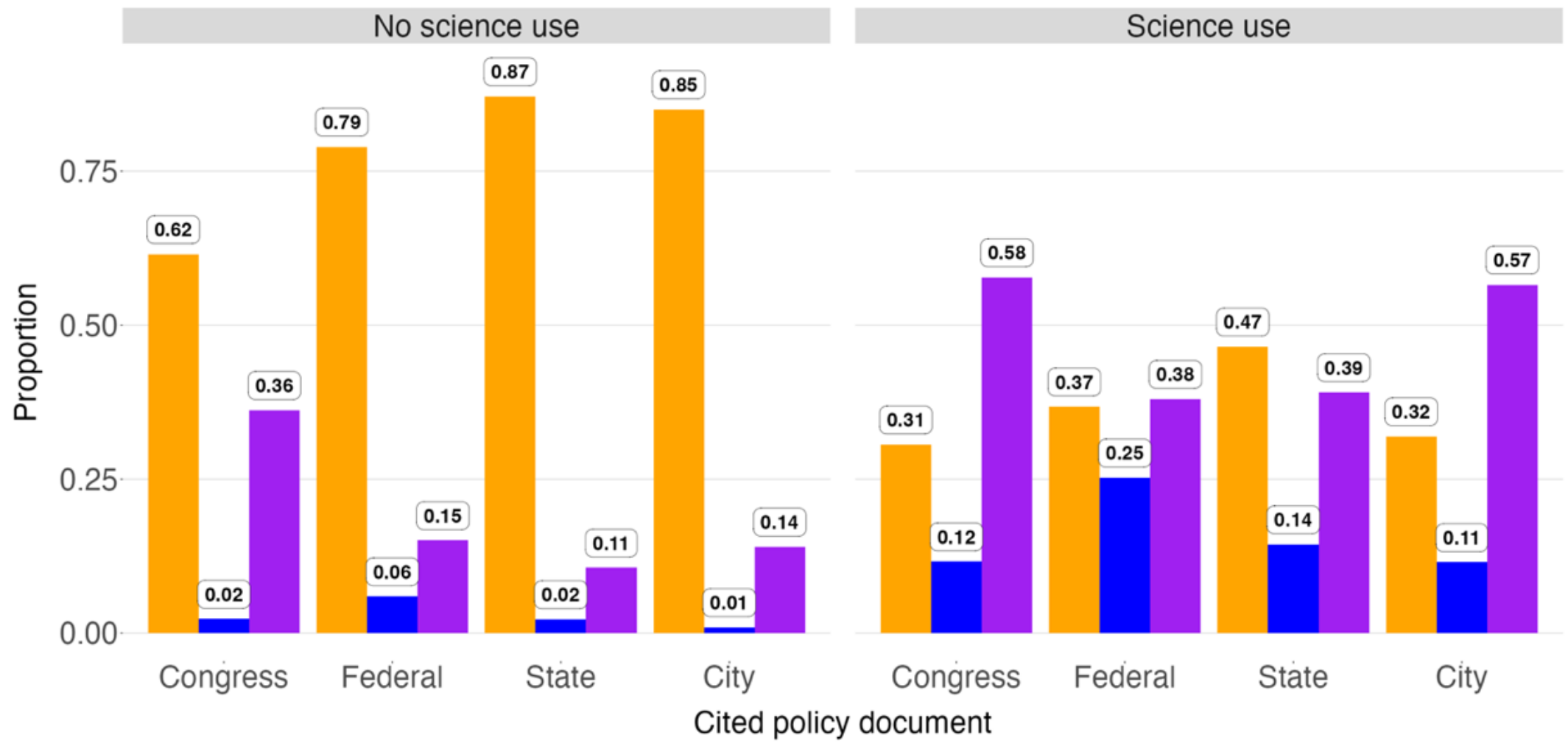

Figure S5. The sources of the policy documents cited by government organizations at different levels of government depending on the document's use of science. Government organizations across different levels of government rely most heavily upon think tanks for science-based policy knowledge. Notable, the Congress (left bar plot on each panel) cites think tanks for policy knowledge to a greater degree than other federal government organizations (right bar plot on each panel), and this is particularly pronounced for science-based policy knowledge (close to 58%)

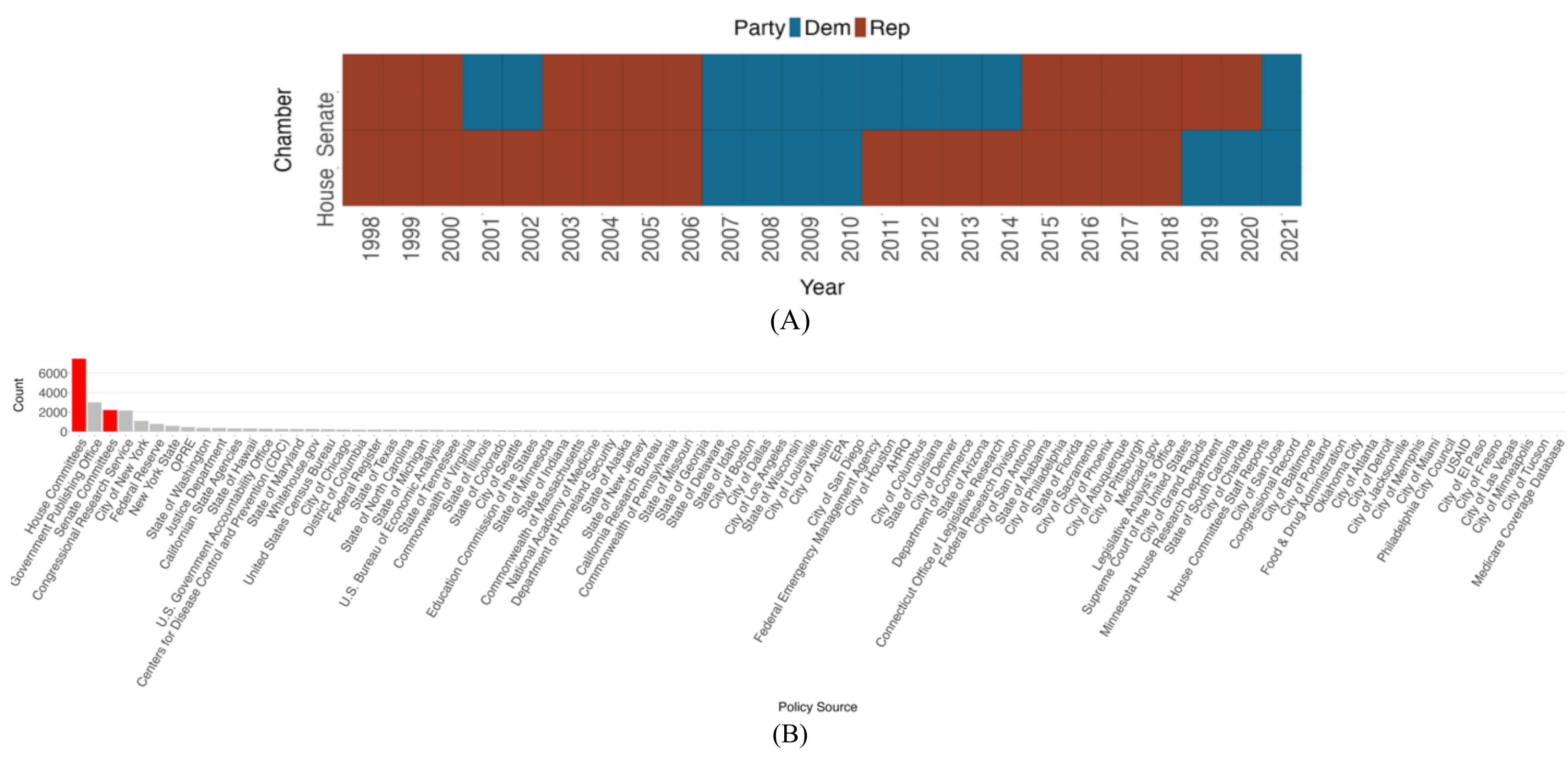

Figure S6. Political alignment between Congressional committees and think tanks in policymaking. (A) Partisan control of the House and Senate. (B) The number of policy documents produced by government organizations. It shows Congressional committees, from both the House (n = 7,479) and the Senate (n = 2,203), account for a substantive fraction of government policy documents (approximately 40%).

| | House | | | Senate | | |
|---|---|---|---|---|---|---|
| | (1) L | (2) N | (3) R | (4) L | (5) N | (6) R |
| Dem control | 0.412*** | 0.119 | -0.220** | 0.319** | 0.159 | 0.164 |
| | (0.076) | (0.075) | (0.112) | (0.127) | (0.121) | (0.197) |
| Num.Obs. | 33366 | 33366 | 33366 | 15816 | 15816 | 15816 |
| AIC | 8540.2 | 9256.1 | 5157.3 | 3185.1 | 3749.1 | 1557.8 |
| BIC | 8716.9 | 9432.9 | 5334.0 | 3346.1 | 3910.1 | 1718.9 |
| Log.Lik. | -4249.089 | -4607.071 | -2557.644 | -1571.543 | -1853.549 | -757.924 |
| RMSE | 0.18 | 0.19 | 0.13 | 0.15 | 0.17 | 0.10 |

* $p < 0.1$, ** $p < 0.05$, *** $p < 0.01$

Table S1. Logistic regression results: political alignment between Congressional committees and think tanks in policymaking. There are six models, and the outcome variable common across the models is whether a congressional committee policy document (House, Senate) cites a think tank policy document (left-leaning, non-ideological, right-leaning) or not. The key predictor is whether the Democratic Party controls a given chamber (the House or the Senate) in the year the committee policy document was published. We also incorporate binary variables for policy domains (not mutually exclusive) and linear and quadratic time trend terms (year). Standard errors are in parentheses.

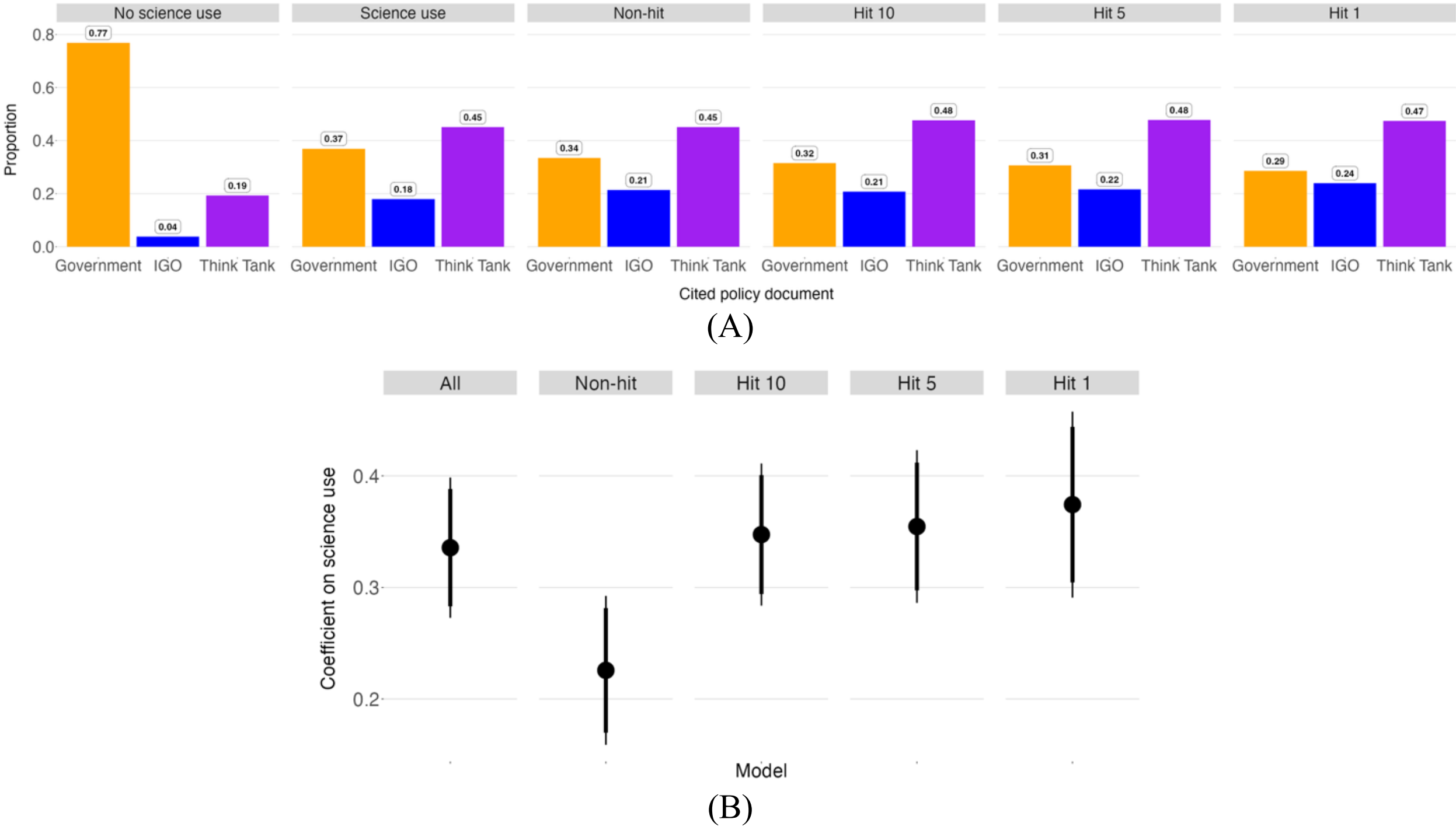

Figure S7. Heterogeneity by impact. (A) The source of the policy document cited by government organizations depending on science use at varying levels of impact. (B) Logistic regression coefficients. We estimate four parallel logistic regression models on the policy citations made by government organizations. The outcome variable in common is whether the source of the cited policy document is a think tank (as opposed to a government organization or an intergovernmental organization). The key predictor in each model is the logged count of scientific citations in the cited document, and it varies in impact. By comparing the coefficients on science use, we examine whether government organizations' reliance on think tanks for science is particularly prominent when they seek high-impact science. While the use of science in the cited policy document positively correlates with the outcome variable across all four of the models, we find a noticeable difference between the coefficient on the non-hit model and those on the hit models (Hit 10, Hit 5, and Hit 1). We conduct a linear hypothesis test using **systemfit** R package by estimating two linear probabilities models simultaneously (non-hit vs. Hit 10). We find a statistically significant difference (Chi square = 153.18, p-value = 2.2e-16). This finding means that when government organizations cite a policy document that uses science, it is more likely to have come from a think tank, especially if that policy document uses high-impact science. The difference implies that government organizations are particularly reliant on think tanks for policy information that is based on high-impact science.

| | (1) | (2) | (3) | (4) | (5) |
|---|---|---|---|---|---|
| All | 0.336*** | | | | |
| | (0.006) | | | | |
| Non-hit | | 0.226*** | | | |
| | | (0.009) | | | |
| Hit 10 | | | 0.347*** | | |
| | | | (0.007) | | |
| Hit 5 | | | | 0.355*** | |
| | | | | (0.008) | |
| Hit 1 | | | | | 0.374*** |
| | | | | | (0.009) |
| Government FE | ✓ | ✓ | ✓ | ✓ | ✓ |
| Year FE | ✓ | ✓ | ✓ | ✓ | ✓ |
| Issue FE | ✓ | ✓ | ✓ | ✓ | ✓ |
| Num.Obs. | 80335 | 80335 | 80335 | 80335 | 80335 |
| AIC | 76062.6 | 78144.3 | 76457.7 | 76632.7 | 77117.4 |
| BIC | 77428.8 | 79510.6 | 77823.9 | 77998.9 | 78483.6 |
| Log.Lik. | −37884.300 | −38925.169 | −38081.865 | −38169.358 | −38411.687 |
| RMSE | 0.39 | 0.40 | 0.39 | 0.39 | 0.40 |

* $p < 0.1$, ** $p < 0.05$, *** $p < 0.01$

Table S2. Logistic regression results: think tanks as a source of science-based policy knowledge. For all models, the outcome variable is whether the source of the document cited by government organizations is a think tank or not. The other categories in the outcome variables include government organizations and intergovernmental organizations. The predictor is the logged count of scientific citations in the cited document. Each model estimates the association between the outcome variable and science use at one impact level (Non-hit, Hit 10, Hit 5, Hit 1). We also account for unobserved heterogeneities across think tanks, years, and policy domains using fixed effects. For, policy domains, we use 18 binary variables reflecting each policy issue, each of which records whether the policy document is classified into the issue. The results for the fixed effects are omitted from the table. Standard errors are clustered for government organizations (in parentheses).

## S2. Increasing ideological polarization of policy knowledge production

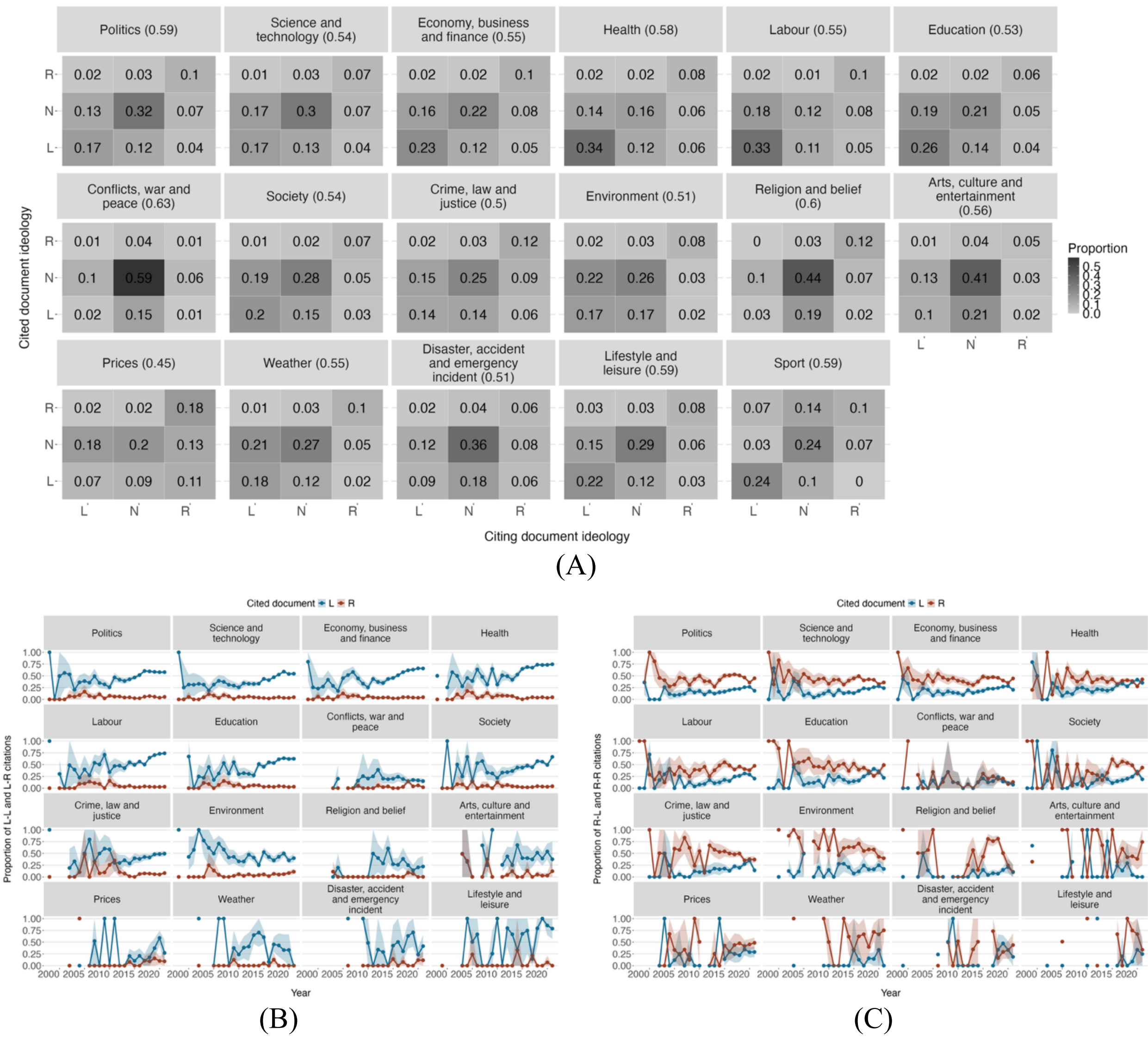

Figure S8: Breakdowns by policy issue area: (A) The joint distribution of the ideology of think tanks producing the citing and cited policy documents. The numbers next to each policy is- sue area indicates the proportion of co-ideological citations (L-L, N-N, R-R). (B) The timeline of co- and cross-ideological citations conditioning on the citing policy documents from left- leaning think tanks (C) The timeline of co- and cross-ideological citations conditioning on the citing policy documents from right-leaning think tanks. “Sport” and “Human interest” are excluded due to extreme sparsity. For both (B) and (C), the area around the line depicts the 95% bootstrapped confidence interval.

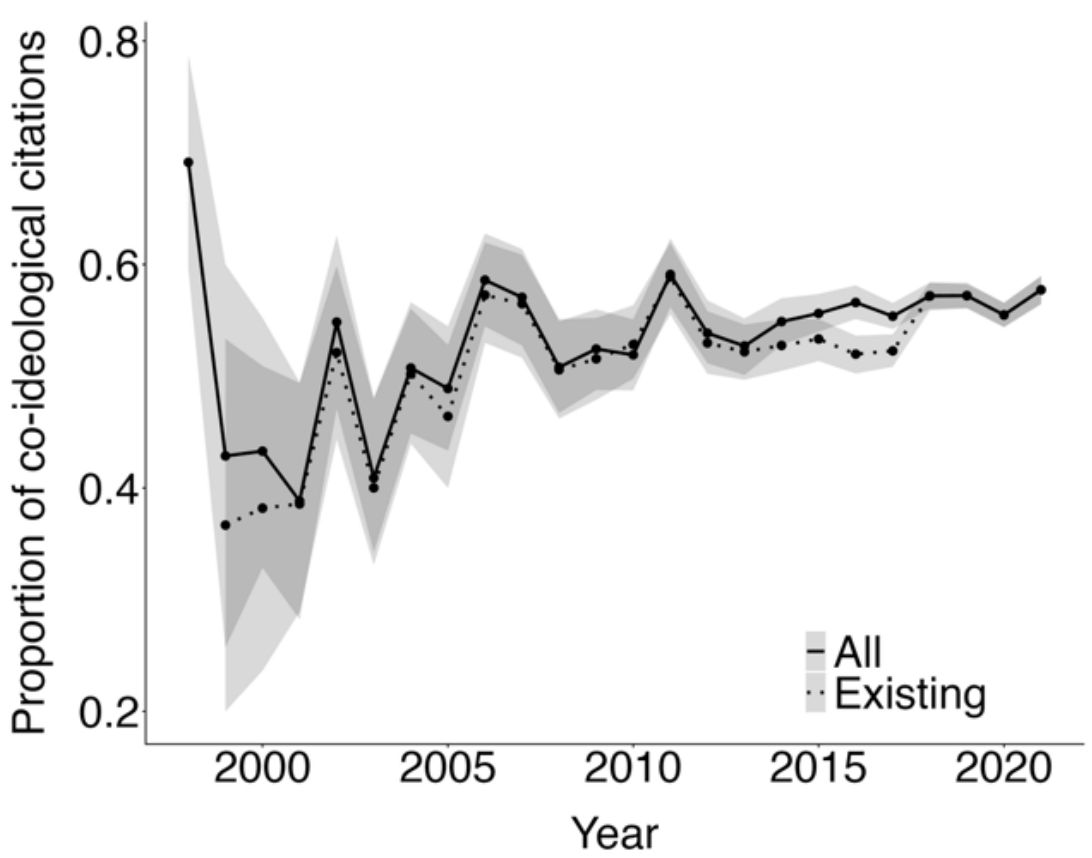

Figure S9. Ideological polarization of policy knowledge production by coverage. The solid line shows estimates based on all policy citations, whereas the dotted line excludes citations involving organizations first observed in the focal year. The nearly identical trends indicate that the observed increase in ideological polarization is unlikely to be an artifact of expanding database coverage over time.

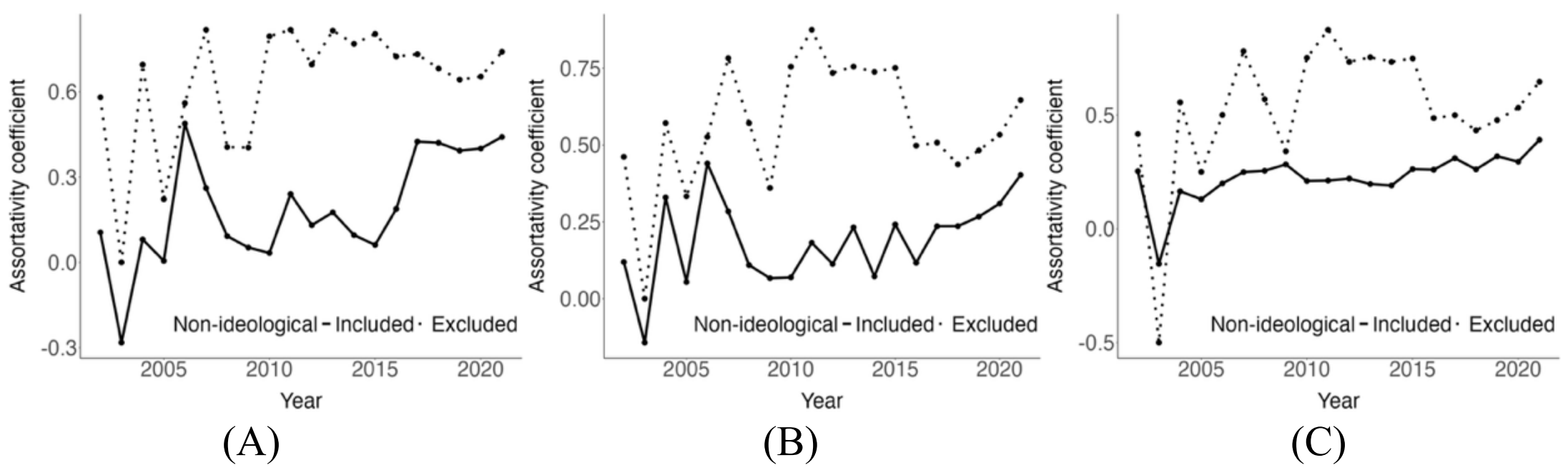

Figure S10. Ideological polarization of policy knowledge production at the think tank level. (A) The timeline of assortativity coefficients based on the directed/weighted version of the policy citation network. (B) The timeline of assortativity coefficients based on the directed/unweighted version of the policy citation network. (C) The timeline of assortativity coefficients based on the undirected/unweighted version of the policy citation network. We measure the degree to which think tanks' policy knowledge production is polarized along the ideological divide using the assortativity coefficient—a widely-used network measure for homophily. We find that policy knowledge exchange is ideologically homophilous and that it is increasingly so, consistent across alternative approaches to conceptualizing the edges in the network in terms of directedness and weight. Importantly, the degree of assortativity within the network is much lower when we incorporate non-ideological think tanks into the network (the regular line versus. the dotted line). This means that, although policy knowledge production is becoming more and more polarized, non-ideological think tanks function as bridges between the ideological think tanks.

| Year | All | Existing | New | Share new |
|---|---|---|---|---|
| 1,998 | 43 | 0 | 43 | 1.000 |
| 1,999 | 52 | 37 | 15 | 0.288 |
| 2,000 | 60 | 50 | 10 | 0.167 |
| 2,001 | 69 | 60 | 9 | 0.130 |
| 2,002 | 80 | 72 | 8 | 0.100 |
| 2,003 | 88 | 75 | 13 | 0.148 |
| 2,004 | 97 | 84 | 13 | 0.134 |
| 2,005 | 103 | 94 | 9 | 0.087 |
| 2,006 | 114 | 105 | 9 | 0.079 |
| 2,007 | 115 | 110 | 5 | 0.043 |
| 2,008 | 132 | 121 | 11 | 0.083 |
| 2,009 | 147 | 132 | 15 | 0.102 |
| 2,010 | 145 | 141 | 4 | 0.028 |
| 2,011 | 158 | 154 | 4 | 0.025 |
| 2,012 | 164 | 154 | 10 | 0.061 |
| 2,013 | 178 | 165 | 13 | 0.073 |
| 2,014 | 193 | 183 | 10 | 0.052 |
| 2,015 | 205 | 194 | 11 | 0.054 |
| 2,016 | 215 | 206 | 9 | 0.042 |
| 2,017 | 214 | 207 | 7 | 0.033 |
| 2,018 | 220 | 214 | 6 | 0.027 |
| 2,019 | 221 | 220 | 1 | 0.005 |
| 2,020 | 208 | 208 | 0 | 0.000 |
| 2,021 | 172 | 171 | 1 | 0.006 |

Table S3: The number and share of newly observed think tanks by year. Existing organizations are organizations that had already appeared in the database before the focal year. New organizations are organizations first observed in the focal year. The share of new organizations is calculated among organizations publishing at least one policy document in that year.

## S3. Science as a source of influence in the policy process

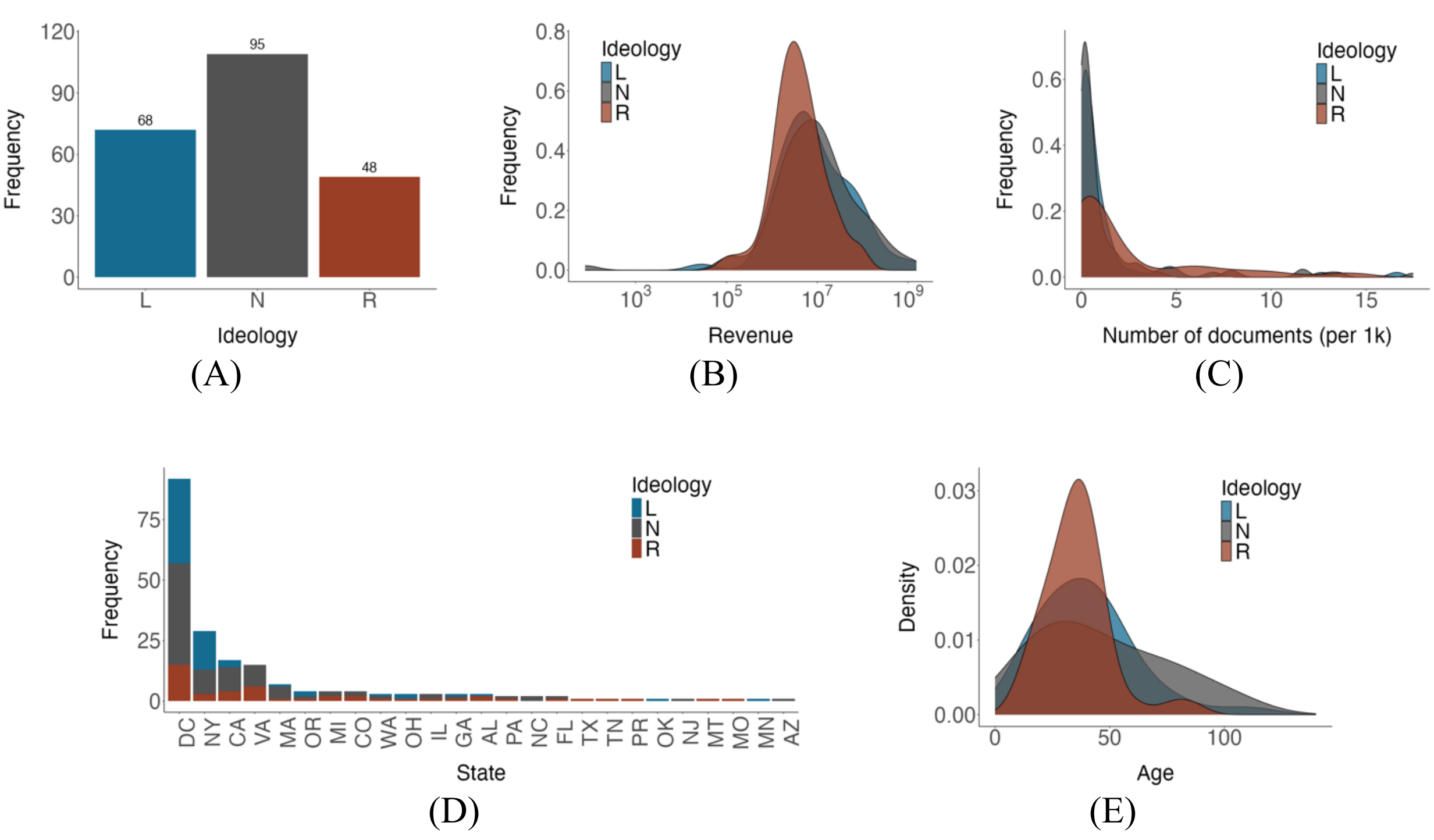

Figure S11: Measures for think tanks. (A) Ideology. (B) Revenue. (C) Age. (D) Location (state). (E) Number of policy documents.

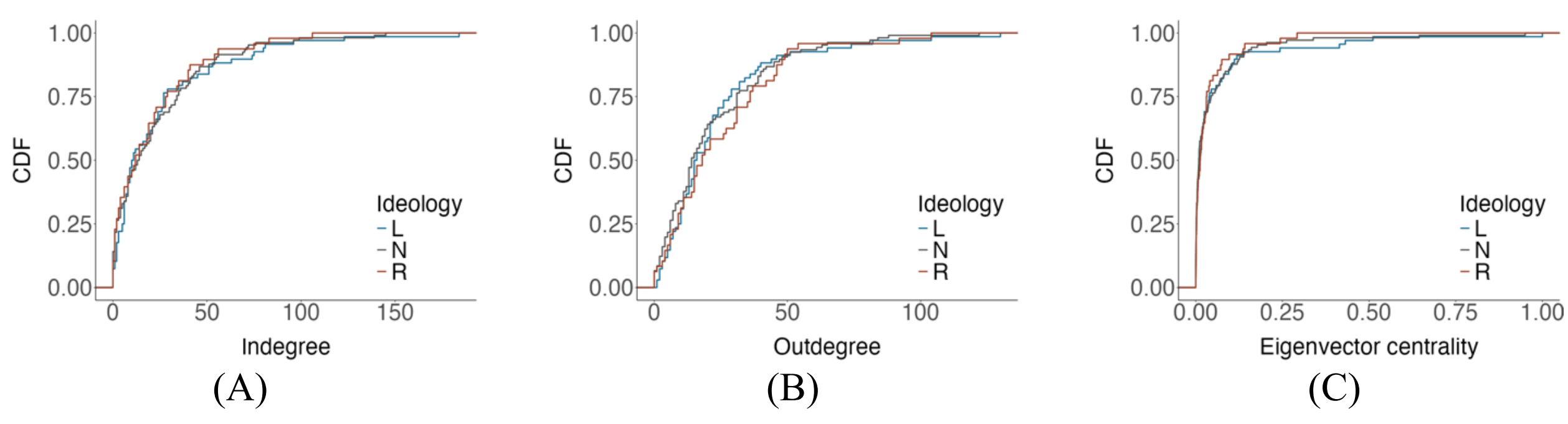

Figure S12: Degree distribution by ideology. (A) Cumulative density function for indegree centrality. (B) Cumulative density function for outdegree centrality. (C) Cumulative density function for eigenvector centrality.

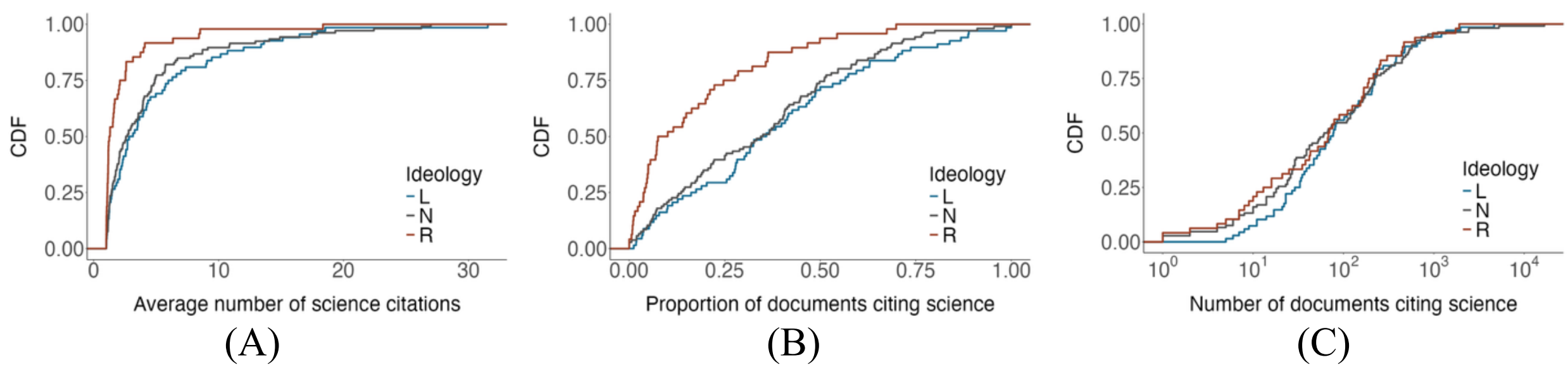

Figure S13: Science use by ideology. (A)–(C) depict the cumulative distribution functions of science use measures based on alternative approaches. (A) presents the proportion of policy documents citing at least one scientific publication. (B) illustrates the number of policy documents citing any scientific publication. (C) shows the average number of citations to scientific publications. Across the measures, we see that right-leaning think tanks are less likely to use science than left-leaning and non-ideological think tanks.

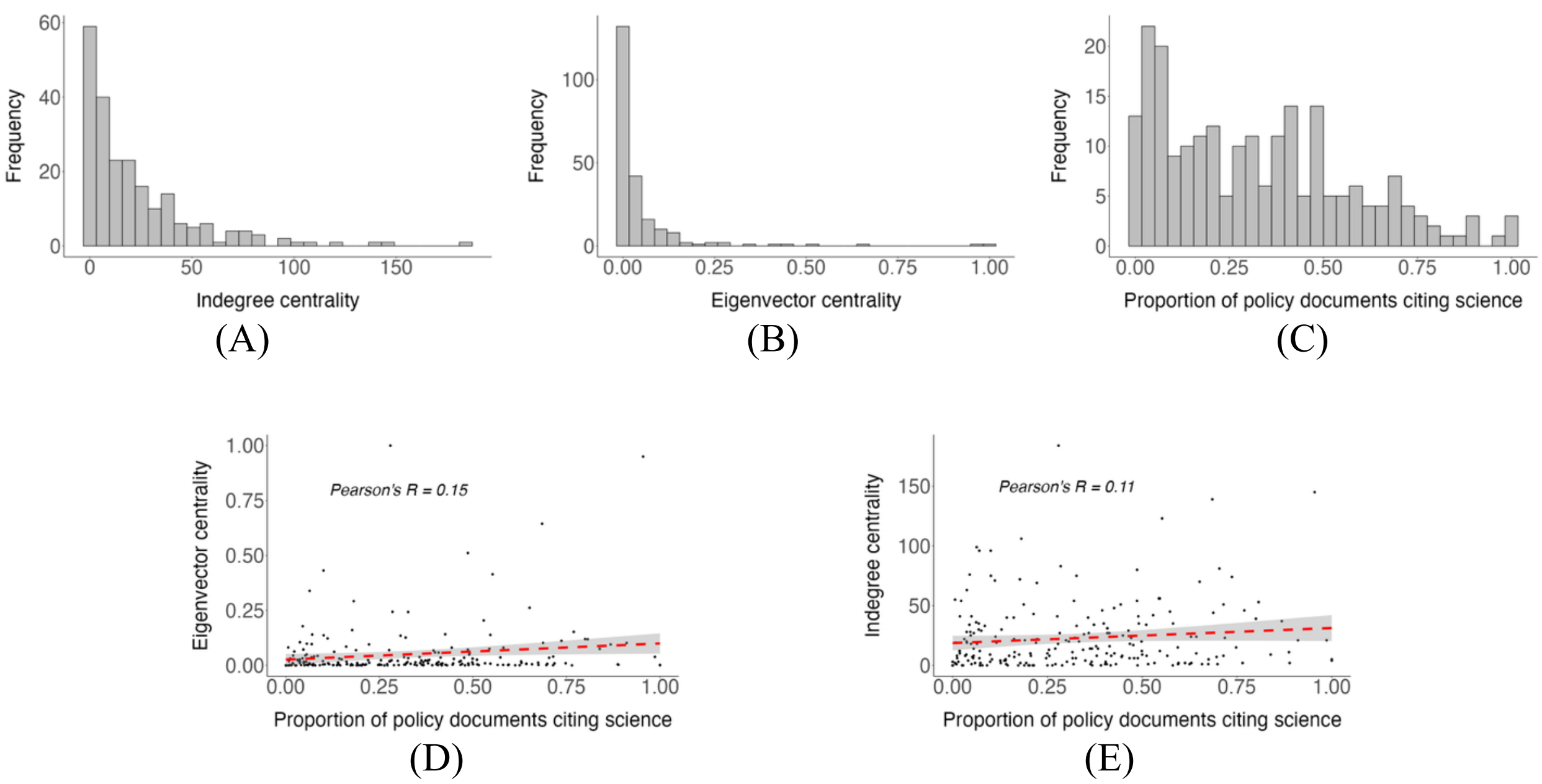

Figure S14: Descriptive statistics related to regression analysis. (A) depicts the distribution of eigenvector centrality. (B) presents the distribution of indegree centrality. (C) shows the distribution of science use (the proportion of policy documents citing any scientific publication). (D) is a scatter plot for science use and eigenvector centrality. (E) is a scatter plot for science use and indegree centrality.

| | (1) | (2) |
|---|---|---|
| Science use | 0.134*** | 0.056** |
| | (0.031) | (0.023) |
| Ideology: L | 0.031 | 0.017 |
| | (0.021) | (0.015) |
| Ideology: R | 0.012 | 0.003 |
| | (0.019) | (0.013) |
| Age | 0.001** | 0.000 |
| | (0.000) | (0.000) |
| Revenue (logged) | 0.004 | 0.008*** |
| | (0.004) | (0.003) |
| Policy document count | 0.027*** | 0.020*** |
| | (0.002) | (0.002) |
| Num.Obs. | 202 | 194 |
| R2 | 0.558 | 0.445 |
| R2 Adj. | 0.481 | 0.343 |
| AIC | -363.7 | -498.5 |
| BIC | -257.8 | -393.9 |
| Log.Lik. | 213.843 | 281.249 |
| F | 7.199 | 4.364 |
| RMSE | 0.08 | 0.06 |

* p < 0.1, ** p < 0.05, *** p < 0.01

Table S4: Main OLS regression results (eigenvector centrality). Model (1) is run on the full sample, and Model (2) is run without influential observations. We identify influential observations based on Cook's distance (59). The threshold is set at 4 / the number of observations, the widely used threshold. For both models, the outcome variable is think tanks' eigenvector centrality. Science use is measured using the proportion of policy documents citing any scientific publication per think tank. For ideology, the baseline category is the right. We calculate years since founding by subtracting the year founded from 2023. We calculate revenue by taking the average of the years for which data are available. The results for geographic location (state) are omitted. Standard errors are in parentheses.

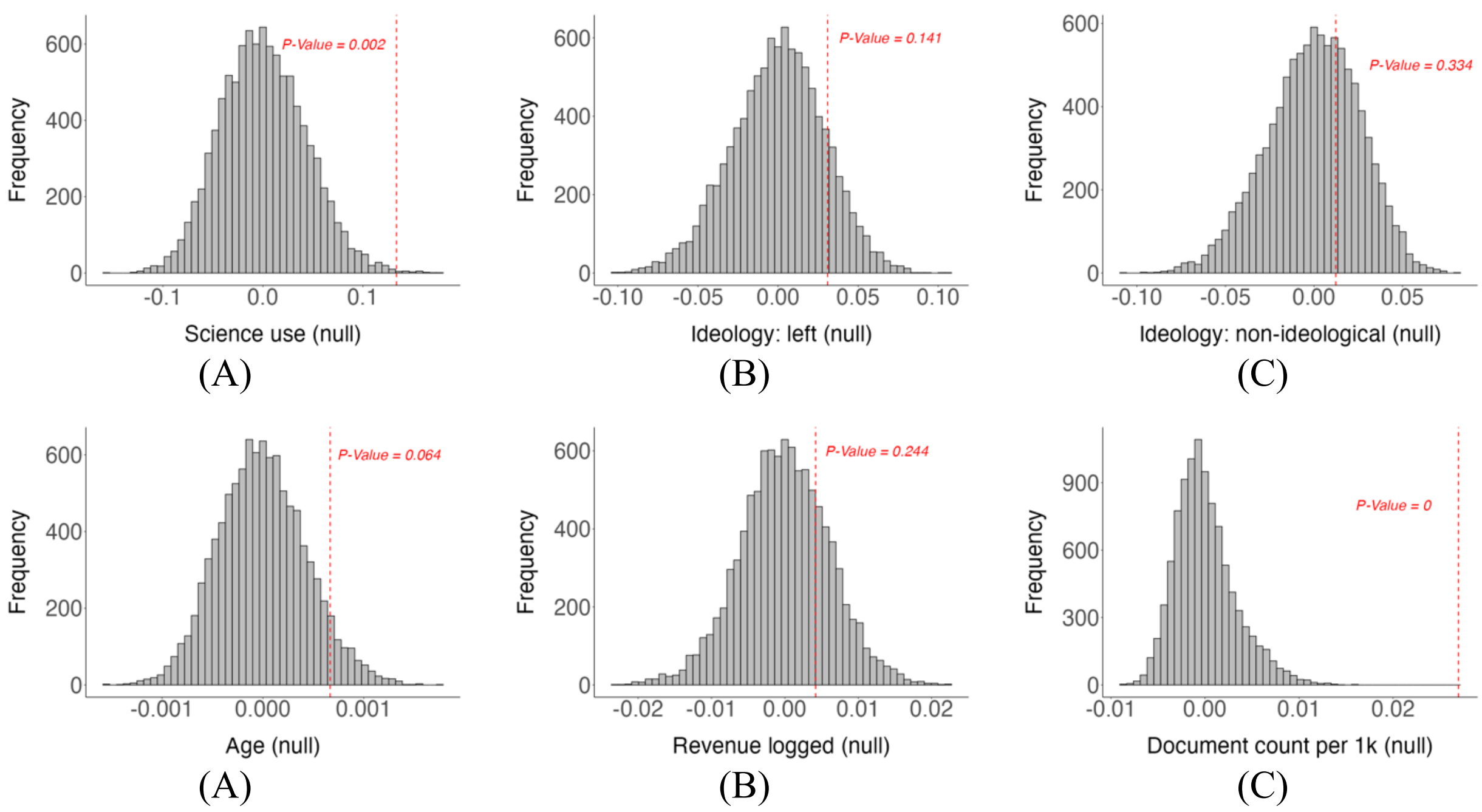

Figure S15: Results for permutation regression for eigenvector centrality. (A) science use (B) ideology (left) (relative to right) (C) ideology (non-ideological) (relative to right) (D) age (E) revenue (F) document count. Through shuffling the think tanks' order while maintaining centrality distribution, we conduct the regression model repeatedly (10,000 times) to generate null distributions and calculate p-values for the observed regression coefficients. The results are highly consistent with the main results. The only difference is that revenue is no longer statistically significant. Note that science use is still positively associated with eigenvector centrality.

| | (1) |
|---|---|
| Science use | 0.751** |
| | (0.334) |
| Ideology: L | -0.006 |
| | (0.231) |
| Ideology: N | -0.184 |
| | (0.206) |
| Age | 0.004 |
| | (0.003) |
| Revenue (logged) | 0.188*** |
| | (0.051) |
| Policy paper count | 0.171*** |
| | (0.025) |
| Num.Obs. | 202 |
| AIC | 1595.3 |
| BIC | 1701.2 |
| Log.Lik. | -765.657 |
| RMSE | 36.85 |

* $p < 0.1$, ** $p < 0.05$, *** $p < 0.01$

Table S5: Main negative binomial regression results (indegree centrality). The outcome variable is think tanks' indegree centrality. Science use is measured using the proportion of policy documents citing any scientific publication per think tank. We calculate years since founding by subtracting the year founded from 2023. We calculate revenue by taking the average of the years for which data are available. The results for geographic location (state) are omitted. Standard errors are in parentheses.

| | (1) |
|---|---|
| Science use | 0.233** |
| | (0.094) |
| Ideology: L | 0.073** |
| | (0.030) |
| Ideology: N | 0.016 |
| | (0.028) |
| Age | 0.001** |
| | (0.000) |
| Revenue (logged) | 0.004 |
| | (0.004) |
| Policy paper count | 0.027*** |
| | (0.002) |
| Science use * Ideology: L | -0.166 |
| | (0.102) |
| Science use * Ideology: N | -0.065 |
| | (0.102) |
| Num.Obs. | 202 |
| R2 | 0.569 |
| R2 Adj. | 0.487 |
| AIC | -364.5 |
| BIC | -252.1 |
| Log.Lik. | 216.268 |
| F | 6.960 |
| RMSE | 0.08 |

* $p < 0.1$, ** $p < 0.05$, *** $p < 0.01$

Table S6: Interactive OLS regression results (eigenvector centrality): science use * ideology. The outcome variable is think tanks' eigenvector centrality. Science use is measured using the proportion of policy documents citing any scientific publication per think tank. For ideology, the baseline category is the right. We calculate years since founding by subtracting the year founded from 2023. We calculate revenue by taking the average of the years for which data are available. The results for geographic location (state) are omitted. Standard errors are clustered for government (in parentheses).

| | (1) |
|---|---|
| Science use | 0.359 |
| | (1.026) |
| Ideology: L | -0.167 |
| | (0.335) |
| Ideology: N | -0.193 |
| | (0.306) |
| Age | 0.004 |
| | (0.003) |
| Revenue (logged) | 0.187*** |
| | (0.051) |
| Policy paper count | 0.171*** |
| | (0.025) |
| Science use * Ideology: L | 0.643 |
| | (1.109) |
| Science use * Ideology: N | 0.243 |
| | (1.111) |
| Num.Obs. | 202 |
| AIC | 1598.9 |
| BIC | 1711.4 |
| Log.Lik. | -765.447 |
| RMSE | 35.54 |

* $p < 0.1$, ** $p < 0.05$, *** $p < 0.01$

Table S7: Interactive negative binomial regression results (indegree centrality): science use x ideology. The outcome variable is think tanks' indegree centrality. Science use is measured using the proportion of policy documents citing any scientific publication per think tank. For ideology, the baseline category is the right. We calculate years since founding by subtracting the year founded from 2023. We calculate revenue by taking the average of the years for which data are available. The results for geographic location (state) are omitted. Standard errors are clustered for government (in parentheses).

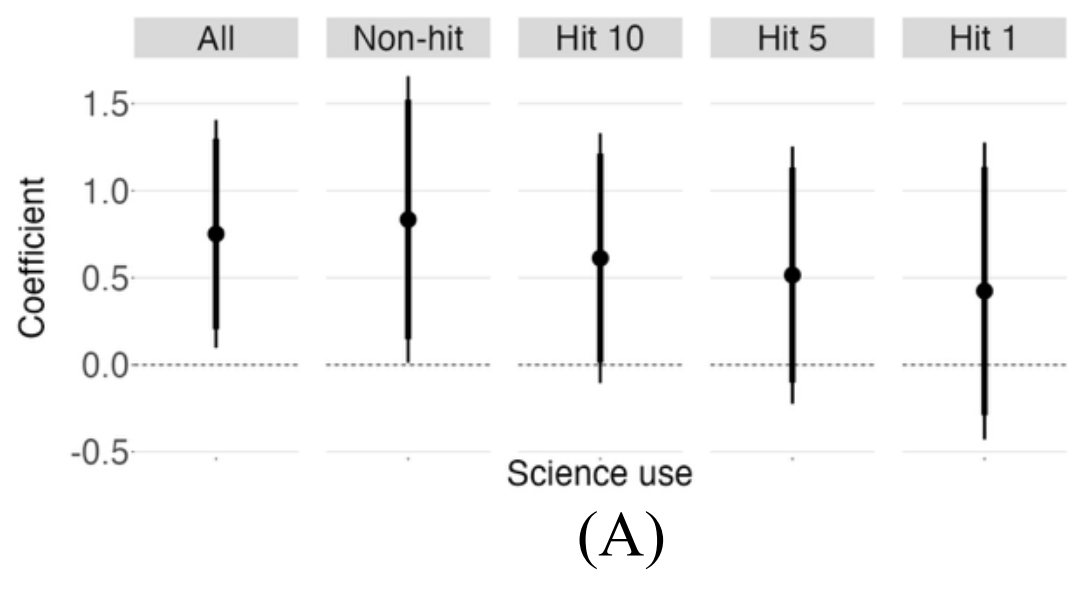

(A)

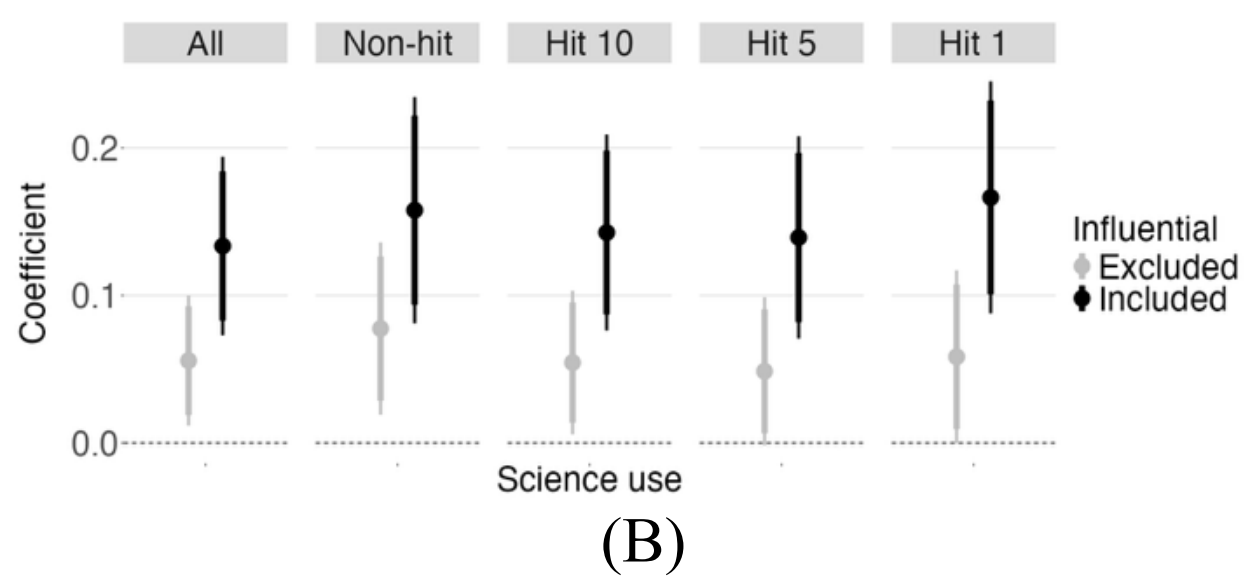

(B)

Figure S16: The role of the impact of science. (A) Coefficients from regression analyses by impact of science where the outcome variable is eigenvector centrality. (B) Coefficients from regression analyses by impact of science where the outcome variable is indegree centrality. For (A)–(B), the dots depict point estimates, and the thick and thin lines around the dots depict the 90% and 95% confidence intervals. For (A), outliers are detected using Cook's distance. The threshold is set at the widely-used threshold: 4 / the number of observations. Although the coefficients for Hit 5 and Hit 1 are not statistically significant at conventional levels in (B), they are still positively signed and of similar magnitude to the all science, non-hit science, and Hit 10 science models. Also, note that we have low power (n = 202). The results indicate that think tanks that frequently use science in producing policy documents are more central in the policy citation network, regardless of the impact of the science used. That is, using science is associated with higher centrality, but the impact of the science used makes little difference. See Tables S8, S9, S10 for regression results.

| | (1) | (2) | (3) | (4) | (5) |
|---|---|---|---|---|---|
| Science use (all) | 0.134*** | | | | |
| | (0.031) | | | | |
| Science use (non-hit) | | 0.158*** | | | |
| | | (0.039) | | | |
| Science use (hit 10) | | | 0.143*** | | |
| | | | (0.034) | | |
| Science use (hit 5) | | | | 0.139*** | |
| | | | | (0.035) | |
| Science use (hit 1) | | | | | 0.167*** |
| | | | | | (0.040) |
| Ideology: L | 0.031 | 0.035* | 0.035 | 0.036* | 0.033 |
| | (0.021) | (0.021) | (0.021) | (0.021) | (0.021) |
| Ideology: N | 0.012 | 0.016 | 0.016 | 0.017 | 0.017 |
| | (0.019) | (0.019) | (0.019) | (0.019) | (0.019) |
| Age | 0.001** | 0.001* | 0.001** | 0.001** | 0.001** |
| | (0.000) | (0.000) | (0.000) | (0.000) | (0.000) |
| Revenue (logged) | 0.004 | 0.005 | 0.005 | 0.005 | 0.005 |
| | (0.004) | (0.004) | (0.004) | (0.004) | (0.004) |
| Policy paper count | 0.027*** | 0.026*** | 0.026*** | 0.026*** | 0.025*** |
| | (0.002) | (0.002) | (0.002) | (0.002) | (0.002) |
| Num.Obs. | 202 | 202 | 202 | 202 | 202 |
| R2 | 0.558 | 0.552 | 0.556 | 0.551 | 0.555 |
| R2 Adj. | 0.481 | 0.474 | 0.478 | 0.473 | 0.476 |
| AIC | −363.7 | −361.1 | −362.6 | −360.6 | −362.1 |
| BIC | −257.8 | −255.2 | −256.8 | −254.7 | −256.2 |
| Log.Lik. | 213.843 | 212.529 | 213.311 | 212.295 | 213.043 |
| F | 7.199 | 7.032 | 7.131 | 7.003 | 7.097 |
| RMSE | 0.08 | 0.08 | 0.08 | 0.08 | 0.08 |

* $p < 0.1$, ** $p < 0.05$, *** $p < 0.01$

Table S8: OLS regression results (eigenvector centrality): impact of science (influential observations included). The outcome variable in common is think tanks' eigenvector centrality. Science use is measured using the proportion of policy documents citing any scientific publication per think tank. Different models focus on different levels of impact. Note that Hit 10 includes Hit 5 and Hit 1, and Hit 5 includes Hit 1. For ideology, the baseline category is the right. We calculate years since founding by subtracting the year founded from 2023. We calculate revenue by taking the average of the years for which data are available. The results for geographic location (state) are omitted. Standard errors are clustered for government (in parentheses).

| | (1) | (2) | (3) | (4) | (5) |
|---|---|---|---|---|---|
| Science use (all) | 0.056** | | | | |
| | (0.023) | | | | |
| Science use (non-hit) | | 0.078** | | | |
| | | (0.030) | | | |
| Science use (hit 10) | | | 0.055** | | |
| | | | (0.025) | | |
| Science use (hit 5) | | | | 0.049* | |
| | | | | (0.026) | |
| Science use (hit 1) | | | | | 0.058* |
| | | | | | (0.030) |
| Ideology: L | 0.017 | 0.017 | 0.019 | 0.020 | 0.019 |
| | (0.015) | (0.015) | (0.015) | (0.015) | (0.015) |
| Ideology: N | 0.003 | 0.003 | 0.005 | 0.005 | 0.005 |
| | (0.013) | (0.013) | (0.013) | (0.013) | (0.013) |
| Age | 0.000 | 0.000 | 0.000 | 0.000 | 0.000 |
| | (0.000) | (0.000) | (0.000) | (0.000) | (0.000) |
| Revenue (logged) | 0.008*** | 0.009*** | 0.009*** | 0.009*** | 0.009*** |
| | (0.003) | (0.003) | (0.003) | (0.003) | (0.003) |
| Policy paper count | 0.020*** | 0.020*** | 0.020*** | 0.020*** | 0.019*** |
| | (0.002) | (0.002) | (0.002) | (0.002) | (0.002) |
| Num.Obs. | 194 | 193 | 194 | 194 | 193 |
| R2 | 0.445 | 0.449 | 0.441 | 0.437 | 0.438 |
| R2 Adj. | 0.343 | 0.347 | 0.338 | 0.333 | 0.334 |
| AIC | -498.5 | -496.1 | -497.0 | -495.6 | -492.1 |
| BIC | -393.9 | -391.7 | -392.4 | -391.0 | -387.7 |
| Log.Lik. | 281.249 | 280.074 | 280.483 | 279.779 | 278.072 |
| F | | 4.403 | | | 4.207 |
| RMSE | 0.06 | 0.06 | 0.06 | 0.06 | 0.06 |

* p < 0.1, ** p < 0.05, *** p < 0.01

Table S9: OLS regression results (eigenvector centrality): impact of science (influential observations excluded). The outcome variable in common is think tanks' eigenvector centrality. Science use is measured using the proportion of policy documents citing any scientific publication per think tank. Different models focus on different levels of impact. Note that Hit 10 includes Hit 5 and Hit 1, and Hit 5 includes Hit 1. For ideology, the baseline category is the right. We calculate years since founding by subtracting the year founded from 2023. We calculate revenue by taking the average of the years for which data are available. The results for geographic location (state) are omitted. Standard errors are clustered for government (in parentheses).

| | (1) | (2) | (3) | (4) | (5) |
|---|---|---|---|---|---|
| Science use (all) | 0.751** | | | | |
| | (0.334) | | | | |
| Science use (non-hit) | | 0.834** | | | |
| | | (0.420) | | | |
| Science use (hit 10) | | | 0.613* | | |
| | | | (0.366) | | |
| Science use (hit 5) | | | | 0.515 | |
| | | | | (0.377) | |
| Science use (hit 1) | | | | | 0.424 |
| | | | | | (0.435) |
| Ideology: L | -0.006 | 0.038 | 0.037 | 0.060 | 0.089 |
| | (0.231) | (0.229) | (0.230) | (0.230) | (0.232) |
| Ideology: N | -0.184 | -0.164 | -0.151 | -0.137 | -0.124 |
| | (0.206) | (0.205) | (0.205) | (0.205) | (0.205) |
| Age | 0.004 | 0.004 | 0.004 | 0.004 | 0.004 |
| | (0.003) | (0.003) | (0.003) | (0.003) | (0.003) |
| Revenue (logged) | 0.188*** | 0.195*** | 0.198*** | 0.203*** | 0.210*** |
| | (0.051) | (0.051) | (0.051) | (0.051) | (0.051) |
| Policy paper count | 0.171*** | 0.165*** | 0.162*** | 0.160*** | 0.156*** |
| | (0.025) | (0.025) | (0.025) | (0.025) | (0.024) |
| Num.Obs. | 202 | 202 | 202 | 202 | 202 |
| AIC | 1595.3 | 1596.4 | 1597.2 | 1597.9 | 1598.6 |
| BIC | 1701.2 | 1702.3 | 1703.1 | 1703.8 | 1704.4 |
| Log.Lik. | -765.657 | -766.221 | -766.599 | -766.953 | -767.282 |
| RMSE | 36.85 | 34.64 | 34.48 | 33.93 | 33.30 |

* p < 0.1, ** p < 0.05, *** p < 0.01

Table S10: Negative binomial regression results (indegree centrality): impact of science. The outcome variable in common is think tanks' indegree centrality. Science use is measured using the proportion of policy documents citing any scientific publication per think tank. Different models focus on different levels of impact. Note that Hit 10 includes Hit 5 and Hit 1, and Hit 5 includes Hit 1. For ideology, the baseline category is the right. We calculate years since founding by subtracting the year founded from 2023. We calculate revenue by taking the average of the years for which data are available. The results for geographic location (state) are omitted. Standard errors are clustered for government (in parentheses).

| | (1) | (2) | (3) | (4) | (5) |
|---|---|---|---|---|---|
| Science use (all) | 0.362*** | | | | |
| | (0.076) | | | | |
| Science use (non-hit) | | 0.359*** | | | |
| | | (0.111) | | | |
| Science use (hit 10) | | | 0.311*** | | |
| | | | (0.090) | | |
| Science use (hit 5) | | | | 0.299*** | |
| | | | | (0.096) | |
| Science use (hit 1) | | | | | 0.263** |
| | | | | | (0.114) |
| Think Tank FE | ✓ | ✓ | ✓ | ✓ | ✓ |
| Year FE | ✓ | ✓ | ✓ | ✓ | ✓ |
| Issue FE | ✓ | ✓ | ✓ | ✓ | ✓ |
| Num.Obs. | 347179 | 347179 | 347179 | 347179 | 347179 |
| AIC | 154752.6 | 155713.8 | 155570.1 | 155760.1 | 156174.1 |
| BIC | 157743.2 | 158704.4 | 158560.8 | 158750.7 | 159164.7 |
| Log.Lik. | −77098.283 | −77578.910 | −77507.072 | −77602.063 | −77809.051 |
| F | 83.380 | 80.272 | 80.827 | 80.202 | 78.751 |
| RMSE | 0.25 | 0.25 | 0.25 | 0.25 | 0.25 |

* $p < 0.1$, ** $p < 0.05$, *** $p < 0.01$

Table S11: Policy document-level logistic regression results. We test whether the positive association between centrality and science use is also visible at the document level. Specifically, we examine 353,403 policy documents produced by think tanks by estimating five parallel logistic regression models. The outcome variable common across the models is whether a given policy document receives any citation from other think tanks. The key predictor, science use, is measured with the number of scientific publications per policy document and varies by impact across the models (all, non-hit, Hit 10, Hit 5, Hit 1). We also account for unobserved heterogeneities across think tanks tanks, years, and policy domains using fixed effects. For, policy domains, we use 18 binary variables reflecting each policy issue, each of which records whether the policy document is classified into the issue. The results for the fixed effects are omitted from the table. Standard errors are clustered for think tanks (in parentheses)

| | Politics | Science and technology | Economy, business, and finance | Health | Labour | Education | Conflicts, war, and peace | Society | Crime, law, and justice | Environment | Religion and belief | Arts, culture, and entertainment | Prices | Weather | Disaster, accident, and emergency incident | Lifestyle and leisure | Sport | Human interest |
|---|---|---|---|---|---|---|---|---|---|---|---|---|---|---|---|---|---|---|
| Science use | 0.070** | 0.155*** | 0.134*** | 0.108** | 0.123*** | 0.133*** | 0.018 | 0.114*** | 0.068* | 0.194*** | -0.010 | 0.068** | 0.154*** | 0.213*** | 0.102*** | 0.132*** | 0.062* | -0.027 |
| | (0.032) | (0.029) | (0.031) | (0.047) | (0.038) | (0.036) | (0.040) | (0.033) | (0.035) | (0.043) | (0.027) | (0.034) | (0.027) | (0.043) | (0.039) | (0.039) | (0.034) | (0.035) |
| Ideology: L | 0.032 | 0.020 | 0.028 | 0.071** | 0.054** | 0.047* | 0.028 | 0.037* | 0.012 | 0.038 | -0.017 | 0.040* | -0.017 | 0.025 | 0.006 | 0.015 | 0.036 | -0.026 |
| | (0.022) | (0.020) | (0.022) | (0.033) | (0.026) | (0.024) | (0.028) | (0.022) | (0.024) | (0.030) | (0.019) | (0.023) | (0.019) | (0.029) | (0.027) | (0.027) | (0.024) | (0.024) |
| Ideology: N | 0.043** | 0.009 | -0.006 | -0.010 | -0.006 | 0.011 | 0.089*** | 0.032 | 0.018 | 0.024 | -0.007 | 0.066*** | -0.029* | 0.009 | 0.024 | 0.033 | 0.004 | -0.003 |
| | (0.020) | (0.018) | (0.019) | (0.029) | (0.023) | (0.022) | (0.025) | (0.020) | (0.021) | (0.026) | (0.017) | (0.020) | (0.017) | (0.026) | (0.024) | (0.024) | (0.021) | (0.021) |
| Age | 0.000 | 0.001** | 0.001** | 0.001** | 0.001* | 0.001* | 0.000 | 0.000 | 0.000 | 0.000 | 0.000 | 0.000 | 0.001*** | 0.001 | 0.000 | 0.000 | 0.000 | -0.001* |
| | (0.000) | (0.000) | (0.000) | (0.000) | (0.000) | (0.000) | (0.000) | (0.000) | (0.000) | (0.000) | (0.000) | (0.000) | (0.000) | (0.000) | (0.000) | (0.000) | (0.000) | (0.000) |
| Revenue (logged) | 0.006 | 0.003 | 0.003 | 0.005 | 0.003 | 0.004 | 0.005 | 0.007 | 0.010* | 0.004 | 0.006 | 0.005 | -0.003 | 0.002 | 0.007 | 0.002 | -0.002 | 0.005 |
| | (0.005) | (0.004) | (0.005) | (0.007) | (0.005) | (0.005) | (0.006) | (0.005) | (0.005) | (0.006) | (0.004) | (0.005) | (0.004) | (0.006) | (0.006) | (0.006) | (0.005) | (0.005) |
| Policy paper count | 0.029*** | 0.024*** | 0.024*** | 0.023*** | 0.023*** | 0.024*** | 0.022*** | 0.027*** | 0.032*** | 0.023*** | 0.018*** | 0.024*** | 0.019*** | 0.021*** | 0.023*** | 0.026*** | 0.012*** | 0.005* |
| | (0.002) | (0.002) | (0.002) | (0.004) | (0.003) | (0.003) | (0.003) | (0.002) | (0.003) | (0.003) | (0.002) | (0.003) | (0.002) | (0.003) | (0.003) | (0.003) | (0.003) | (0.003) |
| Num.Obs. | 202 | 202 | 202 | 202 | 202 | 202 | 202 | 202 | 202 | 202 | 202 | 202 | 202 | 202 | 202 | 202 | 202 | 202 |
| R2 | 0.552 | 0.536 | 0.507 | 0.349 | 0.410 | 0.440 | 0.376 | 0.528 | 0.573 | 0.364 | 0.537 | 0.449 | 0.475 | 0.328 | 0.384 | 0.404 | 0.169 | 0.074 |
| R2 Adj. | 0.474 | 0.455 | 0.421 | 0.235 | 0.306 | 0.342 | 0.266 | 0.446 | 0.498 | 0.252 | 0.456 | 0.353 | 0.383 | 0.210 | 0.276 | 0.299 | 0.024 | -0.088 |
| AIC | -343.3 | -390.0 | -355.9 | -189.8 | -284.7 | -305.1 | -255.4 | -342.1 | -317.0 | -229.0 | -416.6 | -330.3 | -416.4 | -231.0 | -269.2 | -272.3 | -320.2 | -312.6 |
| BIC | -237.5 | -284.1 | -250.0 | -83.9 | -178.9 | -199.2 | -149.5 | -236.3 | -211.2 | -123.2 | -310.7 | -224.5 | -310.5 | -125.2 | -163.3 | -166.4 | -214.4 | -206.7 |
| Log.Lik. | 203.667 | 226.983 | 209.949 | 126.884 | 174.360 | 184.529 | 159.681 | 203.064 | 190.510 | 146.515 | 240.292 | 197.167 | 240.190 | 147.517 | 166.578 | 168.141 | 192.120 | 188.276 |
| F | 7.037 | 6.584 | 5.868 | 3.053 | 3.957 | 4.476 | 3.429 | 6.387 | 7.648 | 3.261 | 6.613 | 4.649 | 5.162 | 2.783 | 3.560 | 3.864 | 1.163 | 0.456 |
| RMSE | 0.09 | 0.08 | 0.09 | 0.13 | 0.10 | 0.10 | 0.11 | 0.09 | 0.09 | 0.12 | 0.07 | 0.09 | 0.07 | 0.12 | 0.11 | 0.11 | 0.09 | 0.10 |

* p < 0.1, ** p < 0.05, *** p < 0.01

Table S12: OLS regression results by issue (eigenvector centrality). All models are run on the full sample. For each issue, we build a policy citation network based on a set of edges corresponding to the issue, and we compute indegree centrality. Science use is measured using the proportion of policy documents citing any scientific publication per think tank. For ideology, the baseline category is the right. We calculate years since founding by subtracting the year founded from 2023. We calculate revenue by taking the average of the years for which data are available. The results for geographic location (state) are omitted. Standard errors are in parentheses.

| | Politics | Science and technology | Economy, business, and finance | Health | Labour | Education | Conflicts, war, and peace | Society | Crime, law, and justice | Environment | Religion and belief | Arts, culture, and entertainment | Prices | Weather | Disaster, accident, and emergency incident | Lifestyle and leisure | Sport | Human interest |
|---|---|---|---|---|---|---|---|---|---|---|---|---|---|---|---|---|---|---|
| Science use | 14.856** | 21.668*** | 18.082*** | 16.245*** | 15.460*** | 16.660*** | 4.113 | 12.639*** | 7.560** | 10.091*** | -0.035 | 2.182* | 3.894*** | 3.733*** | 2.707*** | 2.103*** | 0.459*** | -0.026 |
| | (5.870) | (5.847) | (5.464) | (4.603) | (4.557) | (4.262) | (3.913) | (3.974) | (3.296) | (2.727) | (1.747) | (1.254) | (1.014) | (0.889) | (0.958) | (0.478) | (0.166) | (0.025) |
| Ideology: L | 2.908 | 3.002 | 3.025 | 2.504 | 3.136 | 2.247 | 2.488 | 4.371 | 2.657 | 0.417 | 1.577 | 1.564* | -0.075 | -0.278 | 0.557 | 0.171 | -0.028 | 0.006 |
| | (4.030) | (4.014) | (3.751) | (3.160) | (3.128) | (2.926) | (2.686) | (2.728) | (2.263) | (1.872) | (1.199) | (0.861) | (0.696) | (0.611) | (0.658) | (0.328) | (0.114) | (0.017) |
| Ideology: N | 3.094 | 0.456 | -0.131 | -3.350 | -4.145 | -2.357 | 9.285*** | 1.857 | 2.452 | 0.197 | 3.458*** | 1.594** | -1.303** | 0.205 | 0.629 | 0.100 | -0.129 | 0.021 |
| | (3.584) | (3.571) | (3.337) | (2.811) | (2.783) | (2.602) | (2.390) | (2.427) | (2.013) | (1.665) | (1.067) | (0.766) | (0.619) | (0.543) | (0.585) | (0.292) | (0.101) | (0.015) |
| Age | 0.082 | 0.104* | 0.107* | 0.116** | 0.088* | 0.095** | 0.014 | 0.056 | 0.040 | 0.025 | 0.004 | 0.005 | 0.029*** | 0.005 | 0.012 | 0.004 | 0.003* | 0.000* |
| | (0.059) | (0.059) | (0.055) | (0.047) | (0.046) | (0.043) | (0.040) | (0.040) | (0.033) | (0.028) | (0.018) | (0.013) | (0.010) | (0.009) | (0.010) | (0.005) | (0.002) | (0.000) |
| Revenue (logged) | 1.773** | 1.970** | 1.654** | 1.254* | 0.749 | 1.053* | 0.940* | 1.119* | 0.938** | 0.737* | 0.240 | 0.325* | -0.001 | 0.223* | 0.137 | 0.092 | -0.016 | 0.004 |
| | (0.841) | (0.838) | (0.783) | (0.660) | (0.653) | (0.611) | (0.561) | (0.569) | (0.472) | (0.391) | (0.250) | (0.180) | (0.145) | (0.127) | (0.137) | (0.068) | (0.024) | (0.004) |
| Policy paper count | 4.960*** | 5.075*** | 4.738*** | 3.516*** | 3.201*** | 3.514*** | 2.679*** | 3.327*** | 3.276*** | 2.007*** | 1.366*** | 1.031*** | 0.815*** | 0.573*** | 0.682*** | 0.346*** | 0.090*** | 0.003* |
| | (0.444) | (0.442) | (0.413) | (0.348) | (0.344) | (0.322) | (0.296) | (0.300) | (0.249) | (0.206) | (0.132) | (0.095) | (0.077) | (0.067) | (0.072) | (0.036) | (0.013) | (0.002) |
| Num.Obs. | 202 | 202 | 202 | 202 | 202 | 202 | 202 | 202 | 202 | 202 | 202 | 202 | 202 | 202 | 202 | 202 | 202 | 202 |
| R2 | 0.561 | 0.581 | 0.577 | 0.535 | 0.483 | 0.554 | 0.475 | 0.552 | 0.612 | 0.491 | 0.493 | 0.515 | 0.530 | 0.442 | 0.454 | 0.472 | 0.329 | 0.059 |
| R2 Adj. | 0.484 | 0.507 | 0.503 | 0.453 | 0.392 | 0.475 | 0.383 | 0.474 | 0.544 | 0.402 | 0.404 | 0.430 | 0.447 | 0.345 | 0.358 | 0.379 | 0.212 | -0.106 |
| AIC | 1756.5 | 1755.0 | 1727.6 | 1658.3 | 1654.3 | 1627.2 | 1592.8 | 1598.9 | 1523.4 | 1446.9 | 1267.0 | 1133.0 | 1047.2 | 994.2 | 1024.1 | 743.0 | 315.2 | -448.2 |
| BIC | 1862.4 | 1860.9 | 1833.5 | 1764.2 | 1760.1 | 1733.1 | 1698.6 | 1704.8 | 1629.3 | 1552.7 | 1372.9 | 1238.9 | 1153.1 | 1100.1 | 1130.0 | 848.8 | 421.1 | -342.4 |
| Log.Lik. | -846.274 | -845.504 | -831.815 | -797.162 | -795.140 | -781.603 | -764.377 | -767.465 | -729.702 | -691.439 | -601.496 | -534.508 | -491.597 | -465.106 | -480.073 | -339.483 | -125.598 | 256.120 |
| F | 7.286 | 7.899 | 7.781 | 6.548 | 5.325 | 7.068 | 5.165 | 7.029 | 8.982 | 5.501 | 5.541 | 6.052 | 6.419 | 4.523 | 4.739 | 5.097 | 2.798 | 0.355 |
| RMSE | 15.97 | 15.91 | 14.86 | 12.52 | 12.40 | 11.59 | 10.65 | 10.81 | 8.97 | 7.42 | 4.75 | 3.41 | 2.76 | 2.42 | 2.61 | 1.30 | 0.45 | 0.07 |

p < 0.1, ** p < 0.05, *** p < 0.01

Table S13: OLS regression results by issue (indegree centrality). All models are run on the full sample. For each issue, we build a policy citation network based on a set of edges corresponding to the issue, and we compute indegree centrality. Science use is measured using the proportion of policy documents citing any scientific publication per think tank. For ideology, the baseline category is the right. We calculate years since founding by subtracting the year founded from 2023. We calculate revenue by taking the average of the years for which data are available. The results for geographic location (state) are omitted. Standard errors are in parentheses.

| | Politics | Science and technology | Economy, business, and finance | Health | Labour | Education | Conflicts, war, and peace | Society | Crime, law, and justice | Environment | Religion and belief | Arts, culture, and entertainment | Prices | Weather | Disaster, accident, and emergency incident | Lifestyle and leisure |
|---|---|---|---|---|---|---|---|---|---|---|---|---|---|---|---|---|
| Science use | 0.598 | 0.989*** | 0.788** | 0.832** | 1.030** | 1.192*** | 0.541 | 1.018** | 0.395 | 1.417*** | -0.240 | 0.394 | 1.281** | 1.736*** | 1.007* | 2.530*** |
| | (0.387) | (0.347) | (0.373) | (0.394) | (0.470) | (0.394) | (0.569) | (0.399) | (0.427) | (0.473) | (0.668) | (0.497) | (0.609) | (0.642) | (0.582) | (0.675) |
| Ideology: L | -0.095 | -0.159 | -0.034 | -0.112 | 0.000 | -0.332 | 0.175 | 0.126 | 0.030 | -0.423 | -0.592 | -0.166 | -0.564 | -0.797 | -0.238 | -0.464 |
| | (0.268) | (0.241) | (0.258) | (0.272) | (0.323) | (0.273) | (0.412) | (0.281) | (0.301) | (0.339) | (0.479) | (0.364) | (0.429) | (0.498) | (0.425) | (0.502) |
| Ideology: N | -0.127 | -0.279 | -0.242 | -0.736*** | -0.944*** | -0.844*** | 1.235*** | -0.260 | -0.088 | -0.222 | 0.497 | 0.052 | -1.098*** | -0.120 | 0.314 | -0.221 |
| | (0.239) | (0.214) | (0.230) | (0.245) | (0.293) | (0.245) | (0.362) | (0.255) | (0.270) | (0.298) | (0.417) | (0.318) | (0.382) | (0.426) | (0.373) | (0.449) |
| Age | 0.004 | 0.003 | 0.004 | 0.007 | 0.006 | 0.006 | 0.002 | 0.005 | 0.003 | 0.002 | 0.003 | -0.004 | 0.006 | -0.003 | 0.007 | 0.001 |
| | (0.004) | (0.004) | (0.004) | (0.004) | (0.005) | (0.004) | (0.006) | (0.004) | (0.004) | (0.005) | (0.007) | (0.005) | (0.006) | (0.007) | (0.006) | (0.007) |
| Revenue (logged) | 0.185*** | 0.230*** | 0.215*** | 0.257*** | 0.214*** | 0.284*** | 0.266*** | 0.240*** | 0.273*** | 0.314*** | 0.357*** | 0.393*** | 0.271*** | 0.402*** | 0.224** | 0.403*** |
| | (0.059) | (0.053) | (0.057) | (0.061) | (0.073) | (0.062) | (0.088) | (0.062) | (0.067) | (0.076) | (0.106) | (0.081) | (0.100) | (0.110) | (0.092) | (0.113) |
| Policy paper count | 0.180*** | 0.176*** | 0.192*** | 0.183*** | 0.212*** | 0.186*** | 0.224*** | 0.194*** | 0.202*** | 0.189*** | 0.220*** | 0.181*** | 0.244*** | 0.181*** | 0.213*** | 0.215*** |
| | (0.029) | (0.026) | (0.028) | (0.029) | (0.035) | (0.029) | (0.042) | (0.029) | (0.031) | (0.035) | (0.047) | (0.034) | (0.039) | (0.045) | (0.040) | (0.043) |
| Num.Obs. | 202 | 202 | 202 | 202 | 202 | 202 | 202 | 202 | 202 | 202 | 202 | 202 | 202 | 202 | 202 | 202 |
| AIC | 1460.2 | 1461.6 | 1418.5 | 1251.1 | 1161.1 | 1213.6 | 993.9 | 1180.9 | 1130.3 | 955.5 | 667.7 | 681.0 | 526.7 | 505.5 | 554.9 | 384.4 |
| BIC | 1566.1 | 1567.5 | 1524.3 | 1357.0 | 1267.0 | 1319.5 | 1099.8 | 1286.8 | 1236.2 | 1061.3 | 773.6 | 786.9 | 632.6 | 611.4 | 660.8 | 490.3 |
| Log.Lik. | -698.111 | -698.804 | -677.241 | -593.552 | -548.572 | -574.800 | -464.955 | -558.463 | -533.175 | -445.733 | -301.843 | -308.513 | -231.349 | -220.761 | -245.451 | -160.196 |
| RMSE | 28.95 | 28.91 | 34.17 | 25.58 | 42.19 | 24.58 | 23.36 | 23.44 | 19.31 | 12.35 | 12.19 | 3.05 | 15.25 | 8.13 | 2.62 | 1.37 |

* p < 0.1, ** p < 0.05, *** p < 0.01

Table S14: Negative binomial regression results by issue (indegree centrality). All models are run on the full sample. For each issue, we build a policy citation network based on a set of edges corresponding to the issue, and we compute indegree centrality. “Sport” and “human interest” are excluded as there are not enough edges. Science use is measured using the proportion of policy documents citing any scientific publication per think tank. For ideology, the baseline category is the right. We calculate age by subtracting the year founded from 2023. We calculate revenue by taking the average of the years for which data are available. The results for geographic location (state) are omitted. Standard errors are in parentheses.

| | (1) | (2) | (3) | (4) | (5) | (6) | (7) | (8) | (9) | (10) | (11) | (12) | (13) | (14) | (15) | (16) | (17) | (18) | (19) | (20) | (21) | (22) |
|---|---|---|---|---|---|---|---|---|---|---|---|---|---|---|---|---|---|---|---|---|---|---|
| Science use | 0.178*** | -0.018 | -0.002 | 0.215*** | -0.030 | 0.317* | 0.330*** | 0.256*** | 0.360*** | 0.064 | 0.245 | 0.415* | 0.132* | 0.761*** | 0.116 | 0.416** | -0.285 | -0.150 | -0.061 | 0.331 | 0.100 | -0.484 |
| | (0.051) | (0.134) | (0.129) | (0.040) | (0.104) | (0.170) | (0.075) | (0.079) | (0.127) | (0.082) | (0.223) | (0.238) | (0.069) | (0.190) | (0.150) | (0.185) | (0.364) | (0.164) | (0.117) | (0.397) | (0.331) | (0.575) |
| Ideology: L | 0.042** | 0.055** | 0.055** | 0.043** | 0.055** | 0.053** | 0.037* | 0.041* | 0.044** | 0.053** | 0.050** | 0.048** | 0.050** | 0.048** | 0.053** | 0.049** | 0.057*** | 0.058*** | 0.056** | 0.054** | 0.054** | 0.056*** |
| | (0.021) | (0.022) | (0.022) | (0.020) | (0.022) | (0.021) | (0.021) | (0.021) | (0.021) | (0.022) | (0.022) | (0.022) | (0.021) | (0.021) | (0.022) | (0.021) | (0.022) | (0.022) | (0.022) | (0.022) | (0.022) | (0.022) |
| Ideology: N | 0.015 | 0.031 | 0.031 | 0.024 | 0.031 | 0.024 | 0.021 | 0.027 | 0.020 | 0.029 | 0.027 | 0.027 | 0.029 | 0.022 | 0.028 | 0.027 | 0.031 | 0.031 | 0.031 | 0.027 | 0.030 | 0.033* |
| | (0.019) | (0.019) | (0.020) | (0.018) | (0.019) | (0.019) | (0.018) | (0.019) | (0.019) | (0.019) | (0.020) | (0.019) | (0.019) | (0.019) | (0.020) | (0.019) | (0.019) | (0.019) | (0.019) | (0.020) | (0.019) | (0.020) |
| Age | 0.001** | 0.001* | 0.001* | 0.001* | 0.001* | 0.001** | 0.001** | 0.000 | 0.001** | 0.001* | 0.001** | 0.001* | 0.001* | 0.001** | 0.001* | 0.001** | 0.001* | 0.001* | 0.001* | 0.001* | 0.001* | 0.001* |
| | (0.000) | (0.000) | (0.000) | (0.000) | (0.000) | (0.000) | (0.000) | (0.000) | (0.000) | (0.000) | (0.000) | (0.000) | (0.000) | (0.000) | (0.000) | (0.000) | (0.000) | (0.000) | (0.000) | (0.000) | (0.000) | (0.000) |
| Revenue (logged) | 0.006 | 0.009* | 0.009* | 0.006 | 0.009* | 0.009* | 0.007 | 0.006 | 0.006 | 0.008* | 0.008* | 0.008* | 0.008* | 0.006 | 0.008* | 0.007 | 0.009** | 0.009** | 0.009* | 0.009* | 0.009* | 0.009** |
| | (0.004) | (0.005) | (0.005) | (0.004) | (0.005) | (0.004) | (0.004) | (0.004) | (0.005) | (0.005) | (0.005) | (0.004) | (0.004) | (0.004) | (0.005) | (0.005) | (0.005) | (0.005) | (0.005) | (0.005) | (0.005) | (0.005) |
| Policy paper count | 0.026*** | 0.024*** | 0.024*** | 0.025*** | 0.024*** | 0.025*** | 0.025*** | 0.026*** | 0.025*** | 0.024*** | 0.024*** | 0.025*** | 0.025*** | 0.024*** | 0.024*** | 0.024*** | 0.024*** | 0.024*** | 0.024*** | 0.024*** | 0.024*** | 0.024*** |
| | (0.002) | (0.002) | (0.002) | (0.002) | (0.002) | (0.002) | (0.002) | (0.002) | (0.002) | (0.002) | (0.002) | (0.002) | (0.002) | (0.002) | (0.002) | (0.002) | (0.002) | (0.002) | (0.002) | (0.002) | (0.002) | (0.002) |
| Num.Obs. | 202 | 202 | 202 | 202 | 202 | 202 | 202 | 202 | 202 | 202 | 202 | 202 | 202 | 202 | 202 | 202 | 202 | 202 | 202 | 202 | 202 | 202 |
| R2 | 0.542 | 0.510 | 0.510 | 0.580 | 0.510 | 0.520 | 0.560 | 0.538 | 0.532 | 0.511 | 0.513 | 0.518 | 0.520 | 0.552 | 0.511 | 0.524 | 0.511 | 0.512 | 0.510 | 0.512 | 0.510 | 0.512 |
| R2 Adj. | 0.462 | 0.424 | 0.424 | 0.506 | 0.424 | 0.435 | 0.483 | 0.457 | 0.450 | 0.426 | 0.428 | 0.434 | 0.436 | 0.473 | 0.426 | 0.440 | 0.426 | 0.426 | 0.425 | 0.426 | 0.424 | 0.426 |
| AIC | -356.6 | -342.7 | -342.7 | -373.9 | -342.8 | -346.8 | -364.5 | -354.8 | -352.0 | -343.4 | -344.1 | -346.2 | -347.0 | -360.8 | -343.4 | -348.5 | -343.4 | -343.7 | -343.0 | -343.5 | -342.8 | -343.5 |
| BIC | -250.7 | -236.8 | -236.8 | -268.0 | -236.9 | -240.9 | -258.7 | -248.9 | -246.1 | -237.5 | -238.2 | -240.4 | -241.1 | -254.9 | -237.5 | -242.7 | -237.5 | -237.8 | -237.1 | -237.6 | -236.9 | -237.7 |
| Log.Lik. | 210.292 | 203.353 | 203.342 | 218.950 | 203.391 | 205.386 | 214.268 | 209.400 | 207.989 | 203.704 | 204.056 | 205.112 | 205.503 | 212.381 | 203.692 | 206.269 | 203.706 | 203.833 | 203.503 | 203.752 | 203.396 | 203.760 |
| F | 6.753 | 5.926 | 5.925 | 7.868 | 5.931 | 6.163 | 7.253 | 6.644 | 6.472 | 5.967 | 6.008 | 6.131 | 6.176 | 7.013 | 5.965 | 6.267 | 5.967 | 5.982 | 5.944 | 5.972 | 5.931 | 5.973 |
| RMSE | 0.09 | 0.09 | 0.09 | 0.08 | 0.09 | 0.09 | 0.08 | 0.09 | 0.09 | 0.09 | 0.09 | 0.09 | 0.09 | 0.08 | 0.09 | 0.09 | 0.09 | 0.09 | 0.09 | 0.09 | 0.09 | 0.09 |

* $p < 0.1$, ** $p < 0.05$, *** $p < 0.01$

Table S15: OLS regression results by field (eigenvector centrality). (1) Studies in human society, (2) Earth sciences, (3) Engineering, (4) Economics, (5) Environmental sciences, (6) History and archaeology, (7) Commerce management tourism and services, (8) Medical and health sciences, (9) Psychology and cognitive sciences, (10) Education, (11) Language communication and culture, (12) Built environment and design, (13) Law and legal studies, (14) Mathematical sciences, (15) Information and computing sciences, (16) Philosophy and religious studies, (17) Chemical sciences, (18) Agricultural and veterinary sciences, (19) Biological sciences, (20) Physical sciences, (21) Studies in creative arts and writing. (22) Technology. All models are run on the full sample. For each field, we compute the proportion of policy documents citing at least one scientific publication in that field. For ideology, the baseline category is the right. We calculate years since founding by subtracting the year founded from 2023. We calculate revenue by taking the average of the years for which data are available. The results for geographic location (state) are omitted. Standard errors are in parentheses.

| | (1) | (2) | (3) | (4) | (5) | (6) | (7) | (8) | (9) | (10) | (11) | (12) | (13) | (14) | (15) | (16) | (17) | (18) | (19) | (20) | (21) | (22) |
|---|---|---|---|---|---|---|---|---|---|---|---|---|---|---|---|---|---|---|---|---|---|---|
| Science use | 40.761*** | -12.801 | -5.300 | 34.318*** | -13.656 | 44.697 | 37.262** | 44.329** | 42.157 | 2.440 | 36.280 | 78.371 | 24.647 | 60.618 | -16.190 | 149.797*** | -113.327 | -36.976 | -30.442 | 55.889 | -22.331 | -193.132 |
| | (11.298) | (29.657) | (28.629) | (9.279) | (23.052) | (37.906) | (17.273) | (17.706) | (28.677) | (18.208) | (49.555) | (53.086) | (15.332) | (43.926) | (33.399) | (40.194) | (80.427) | (36.465) | (25.848) | (88.259) | (73.595) | (127.020) |
| Ideology: L | 5.725 | 8.877* | 8.751* | 6.746 | 9.019* | 8.368* | 6.625 | 6.293 | 7.343 | 8.565* | 7.974 | 7.339 | 7.814 | 8.103* | 8.920* | 6.522 | 9.486** | 9.350* | 9.376* | 8.470* | 8.716* | 9.348* |
| | (4.686) | (4.816) | (4.824) | (4.636) | (4.825) | (4.774) | (4.816) | (4.795) | (4.839) | (4.821) | (4.867) | (4.839) | (4.780) | (4.777) | (4.819) | (4.640) | (4.798) | (4.824) | (4.809) | (4.790) | (4.793) | (4.779) |
| Ideology: N | 1.331 | 5.183 | 5.111 | 3.885 | 5.102 | 4.150 | 3.916 | 4.394 | 3.744 | 4.961 | 4.510 | 4.422 | 4.733 | 4.291 | 5.397 | 3.773 | 4.997 | 5.227 | 5.395 | 4.353 | 5.128 | 6.016 |
| | (4.266) | (4.315) | (4.335) | (4.146) | (4.296) | (4.341) | (4.270) | (4.228) | (4.356) | (4.310) | (4.344) | (4.289) | (4.269) | (4.305) | (4.370) | (4.146) | (4.273) | (4.290) | (4.293) | (4.414) | (4.316) | (4.320) |
| Age | 0.125* | 0.117 | 0.119 | 0.116 | 0.119 | 0.131* | 0.134* | 0.096 | 0.130* | 0.120 | 0.129* | 0.125* | 0.116 | 0.125* | 0.114 | 0.149** | 0.121* | 0.118 | 0.115 | 0.122* | 0.117 | 0.117 |
| | (0.070) | (0.073) | (0.073) | (0.070) | (0.073) | (0.073) | (0.072) | (0.072) | (0.073) | (0.073) | (0.074) | (0.073) | (0.073) | (0.073) | (0.074) | (0.071) | (0.073) | (0.073) | (0.073) | (0.073) | (0.074) | (0.073) |
| Revenue (logged) | 2.508** | 3.191*** | 3.200*** | 2.755*** | 3.205*** | 3.166*** | 2.953*** | 2.772*** | 2.824*** | 3.172*** | 3.085*** | 3.067*** | 3.032*** | 2.959*** | 3.318*** | 2.630*** | 3.262*** | 3.263*** | 3.252*** | 3.180*** | 3.232*** | 3.457*** |
| | (0.986) | (1.004) | (1.007) | (0.974) | (1.004) | (1.001) | (0.997) | (1.000) | (1.028) | (1.010) | (1.012) | (1.001) | (1.002) | (1.012) | (1.040) | (0.977) | (1.000) | (1.004) | (1.002) | (1.003) | (1.016) | (1.014) |
| Policy paper count | 5.887*** | 5.491*** | 5.495*** | 5.629*** | 5.475*** | 5.575*** | 5.566*** | 5.755*** | 5.591*** | 5.513*** | 5.547*** | 5.582*** | 5.601*** | 5.496*** | 5.477*** | 5.563*** | 5.436*** | 5.470*** | 5.450*** | 5.535*** | 5.492*** | 5.411*** |
| | (0.517) | (0.526) | (0.528) | (0.506) | (0.527) | (0.526) | (0.519) | (0.525) | (0.525) | (0.528) | (0.527) | (0.524) | (0.524) | (0.522) | (0.528) | (0.505) | (0.524) | (0.525) | (0.525) | (0.526) | (0.527) | (0.525) |
| Num.Obs. | 202 | 202 | 202 | 202 | 202 | 202 | 202 | 202 | 202 | 202 | 202 | 202 | 202 | 202 | 202 | 202 | 202 | 202 | 202 | 202 | 202 | 202 |
| R2 | 0.581 | 0.549 | 0.549 | 0.582 | 0.550 | 0.553 | 0.561 | 0.565 | 0.555 | 0.549 | 0.550 | 0.555 | 0.556 | 0.554 | 0.550 | 0.583 | 0.554 | 0.552 | 0.553 | 0.550 | 0.549 | 0.555 |
| R2 Adj. | 0.507 | 0.470 | 0.470 | 0.509 | 0.471 | 0.474 | 0.484 | 0.489 | 0.476 | 0.470 | 0.471 | 0.476 | 0.478 | 0.476 | 0.471 | 0.510 | 0.476 | 0.473 | 0.474 | 0.471 | 0.470 | 0.477 |
| AIC | 1825.4 | 1840.0 | 1840.2 | 1824.7 | 1839.8 | 1838.6 | 1834.8 | 1832.9 | 1837.7 | 1840.2 | 1839.6 | 1837.6 | 1837.2 | 1838.0 | 1839.9 | 1824.4 | 1837.9 | 1839.0 | 1838.6 | 1839.7 | 1840.1 | 1837.5 |
| BIC | 1931.3 | 1945.9 | 1946.0 | 1930.5 | 1945.7 | 1944.4 | 1940.6 | 1938.8 | 1943.5 | 1946.0 | 1945.4 | 1943.5 | 1943.0 | 1943.8 | 1945.8 | 1930.3 | 1943.7 | 1944.9 | 1944.4 | 1945.6 | 1946.0 | 1943.4 |
| Log.Lik. | -880.693 | -887.993 | -888.083 | -880.331 | -887.896 | -887.285 | -885.391 | -884.467 | -886.835 | -888.093 | -887.787 | -886.824 | -886.588 | -886.985 | -887.964 | -880.216 | -886.937 | -887.498 | -887.287 | -887.867 | -888.049 | -886.747 |
| F | 7.898 | 6.950 | 6.939 | 7.947 | 6.962 | 7.039 | 7.280 | 7.399 | 7.096 | 6.937 | 6.976 | 7.097 | 7.127 | 7.077 | 6.953 | 7.962 | 7.083 | 7.012 | 7.039 | 6.966 | 6.943 | 7.107 |
| RMSE | 18.93 | 19.63 | 19.64 | 18.90 | 19.62 | 19.56 | 19.38 | 19.29 | 19.52 | 19.64 | 19.61 | 19.52 | 19.49 | 19.53 | 19.63 | 18.89 | 19.53 | 19.58 | 19.56 | 19.62 | 19.64 | 19.51 |

* p < 0.1, ** p < 0.05, *** p < 0.01

Table S16: OLS regression results by field (indegree centrality). (1) Studies in human society, (2) Earth sciences, (3) Engineering, (4) Economics, (5) Environmental sciences, (6) History and archaeology, (7) Commerce management tourism and services, (8) Medical and health sciences, (9) Psychology and cognitive sciences, (10) Education, (11) Language communication and culture, (12) Built environment and design, (13) Law and legal studies, (14) Mathematical sciences, (15) Information and computing sciences, (16) Philosophy and religious studies, (17) Chemical sciences, (18) Agricultural and veterinary sciences, (19) Biological sciences, (20) Physical sciences, (21) Studies in creative arts and writing. (22) Technology. All models are run on the full sample. For each field, we compute the proportion of policy documents citing at least one scientific publication in that field. For ideology, the baseline category is the right. We calculate years since founding by subtracting the year founded from 2023. We calculate revenue by taking the average of the years for which data are available. The results for geographic location (state) are omitted. Standard errors are in parentheses.

|  | (1) | (2) | (3) | (4) | (5) | (6) | (7) | (8) | (9) | (10) | (11) | (12) | (13) | (14) | (15) | (16) | (17) | (18) | (19) | (20) | (21) | (22) |
|---|---|---|---|---|---|---|---|---|---|---|---|---|---|---|---|---|---|---|---|---|---|---|
| Science use | 2.167*** | -1.438 | -1.271 | 1.080** | -1.442 | 1.465 | 0.446 | 0.915 | 0.631 | -0.624 | 0.496 | 1.060 | 1.123 | -1.153 | -2.715* | 3.895** | -8.754** | -2.759 | -2.718** | 3.745 | -2.093 | -19.899*** |
|  | (0.528) | (1.411) | (1.344) | (0.440) | (1.080) | (1.781) | (0.813) | (0.838) | (1.347) | (0.861) | (2.332) | (2.504) | (0.708) | (2.072) | (1.603) | (1.896) | (3.891) | (1.720) | (1.239) | (4.101) | (3.529) | (6.388) |
| Ideology: L | -0.055 | 0.180 | 0.183 | 0.051 | 0.194 | 0.143 | 0.119 | 0.079 | 0.125 | 0.162 | 0.143 | 0.137 | 0.088 | 0.173 | 0.213 | 0.095 | 0.200 | 0.190 | 0.210 | 0.148 | 0.170 | 0.227 |
|  | (0.222) | (0.226) | (0.227) | (0.223) | (0.226) | (0.225) | (0.229) | (0.229) | (0.229) | (0.227) | (0.229) | (0.229) | (0.226) | (0.226) | (0.226) | (0.224) | (0.225) | (0.226) | (0.225) | (0.225) | (0.225) | (0.223) |
| Ideology: N | -0.284 | -0.059 | -0.052 | -0.143 | -0.069 | -0.122 | -0.105 | -0.092 | -0.100 | -0.087 | -0.093 | -0.095 | -0.102 | -0.068 | -0.021 | -0.111 | -0.078 | -0.075 | -0.058 | -0.140 | -0.075 | 0.006 |
|  | (0.203) | (0.203) | (0.204) | (0.201) | (0.202) | (0.206) | (0.204) | (0.202) | (0.207) | (0.203) | (0.205) | (0.204) | (0.202) | (0.204) | (0.206) | (0.201) | (0.201) | (0.202) | (0.202) | (0.208) | (0.204) | (0.202) |
| Age | 0.005 | 0.003 | 0.003 | 0.003 | 0.004 | 0.004 | 0.004 | 0.003 | 0.004 | 0.004 | 0.004 | 0.004 | 0.004 | 0.004 | 0.002 | 0.005 | 0.004 | 0.003 | 0.003 | 0.004 | 0.003 | 0.004 |
|  | (0.003) | (0.003) | (0.003) | (0.003) | (0.003) | (0.003) | (0.003) | (0.003) | (0.003) | (0.003) | (0.003) | (0.003) | (0.003) | (0.003) | (0.003) | (0.003) | (0.003) | (0.003) | (0.003) | (0.003) | (0.003) | (0.003) |
| Revenue (logged) | 0.172*** | 0.226*** | 0.229*** | 0.203*** | 0.230*** | 0.222*** | 0.219*** | 0.207*** | 0.217*** | 0.226*** | 0.222*** | 0.220*** | 0.208*** | 0.228*** | 0.238*** | 0.193*** | 0.232*** | 0.235*** | 0.235*** | 0.225*** | 0.225*** | 0.246*** |
|  | (0.049) | (0.050) | (0.050) | (0.049) | (0.050) | (0.050) | (0.050) | (0.050) | (0.051) | (0.050) | (0.050) | (0.050) | (0.050) | (0.050) | (0.051) | (0.050) | (0.049) | (0.050) | (0.049) | (0.050) | (0.050) | (0.050) |
| Policy paper count | 0.179*** | 0.150*** | 0.150*** | 0.157*** | 0.149*** | 0.156*** | 0.152*** | 0.157*** | 0.153*** | 0.150*** | 0.153*** | 0.153*** | 0.156*** | 0.155*** | 0.151*** | 0.156*** | 0.147*** | 0.149*** | 0.147*** | 0.156*** | 0.152*** | 0.145*** |
|  | (0.024) | (0.024) | (0.024) | (0.024) | (0.024) | (0.024) | (0.024) | (0.024) | (0.024) | (0.024) | (0.024) | (0.024) | (0.024) | (0.024) | (0.024) | (0.024) | (0.024) | (0.024) | (0.024) | (0.024) | (0.024) | (0.024) |
| Num.Obs. | 202 | 202 | 202 | 202 | 202 | 202 | 202 | 202 | 202 | 202 | 202 | 202 | 202 | 202 | 202 | 202 | 202 | 202 | 202 | 202 | 202 | 202 |
| AIC | 1587.4 | 1598.4 | 1598.5 | 1594.3 | 1597.8 | 1598.8 | 1599.1 | 1598.3 | 1599.2 | 1599.1 | 1599.3 | 1599.2 | 1597.8 | 1599.1 | 1596.9 | 1594.5 | 1595.5 | 1597.3 | 1595.5 | 1598.4 | 1599.0 | 1591.6 |
| BIC | 1693.3 | 1704.2 | 1704.4 | 1700.1 | 1703.7 | 1704.7 | 1704.9 | 1704.2 | 1705.0 | 1704.9 | 1705.2 | 1705.1 | 1703.7 | 1704.9 | 1702.8 | 1700.4 | 1701.4 | 1703.2 | 1701.3 | 1704.3 | 1704.9 | 1697.5 |
| Log.Lik. | -761.711 | -767.177 | -767.272 | -765.131 | -766.899 | -767.421 | -767.535 | -767.175 | -767.591 | -767.531 | -767.647 | -767.607 | -766.908 | -767.535 | -766.463 | -765.247 | -765.769 | -766.646 | -765.734 | -767.223 | -767.507 | -763.812 |
| RMSE | 41.91 | 33.84 | 34.56 | 33.71 | 34.96 | 35.14 | 32.93 | 32.23 | 33.57 | 33.76 | 34.24 | 33.85 | 33.12 | 36.30 | 34.13 | 33.72 | 34.83 | 35.21 | 34.57 | 35.60 | 33.79 | 33.08 |

* p < 0.1, ** p < 0.05, *** p < 0.01

Table S17: Negative binomial regression results by field (indegree centrality). (1) Studies in human society, (2) Earth sciences, (3) Engineering, (4) Economics, (5) Environmental sciences, (6) History and archaeology, (7) Commerce management tourism and services, (8) Medical and health sciences, (9) Psychology and cognitive sciences, (10) Education, (11) Language communication and culture, (12) Built environment and design, (13) Law and legal studies, (14) Mathematical sciences, (15) Information and computing sciences, (16) Philosophy and religious studies, (17) Chemical sciences, (18) Agricultural and veterinary sciences, (19) Biological sciences, (20) Physical sciences, (21) Studies in creative arts and writing. (22) Technology. All models are run on the full sample. For each field, we compute the proportion of policy documents citing at least one scientific publication in that field. For ideology, the baseline category is the right. We calculate years since founding by subtracting the year founded from 2023. We calculate revenue by taking the average of the years for which data are available. The results for geographic location (state) are omitted. Standard errors are in parentheses.

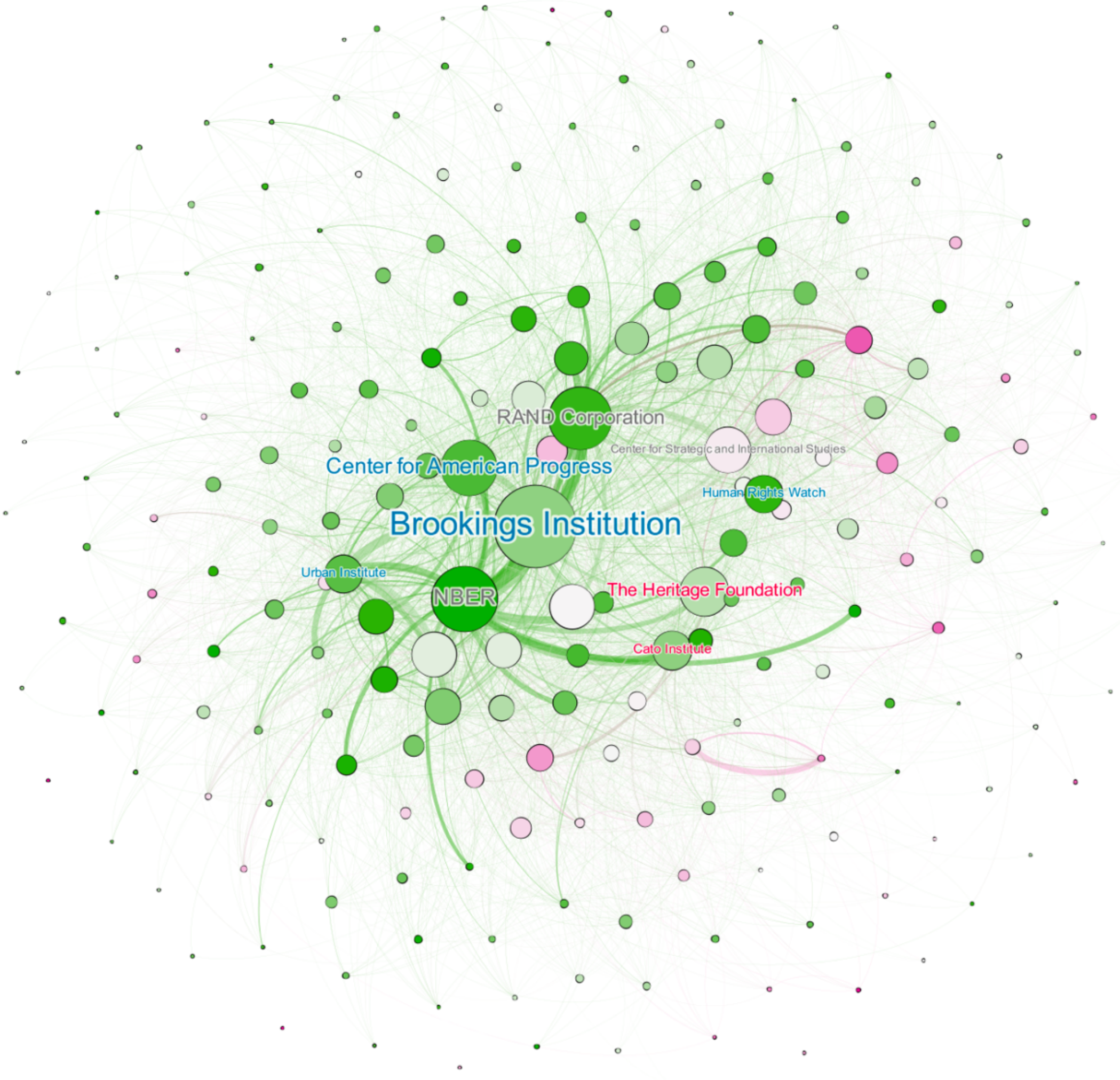

Figure S17: The policy citation network depicting the relationship between science use and indegree centrality. Node size is proportionate to the number of citations think tanks receive, and node color reflects the proportion of policy documents citing any scientific publication (pink and green indicating 0 to 1). Edge width is proportionate to indegree, and edge direction is clockwise. The network shows that more of the green nodes tend to be bigger, indicating a positive association between science and indegree centrality (think tanks with a high indegree are labeled with their names).

## S4. Science use and ideological polarization in policy knowledge production

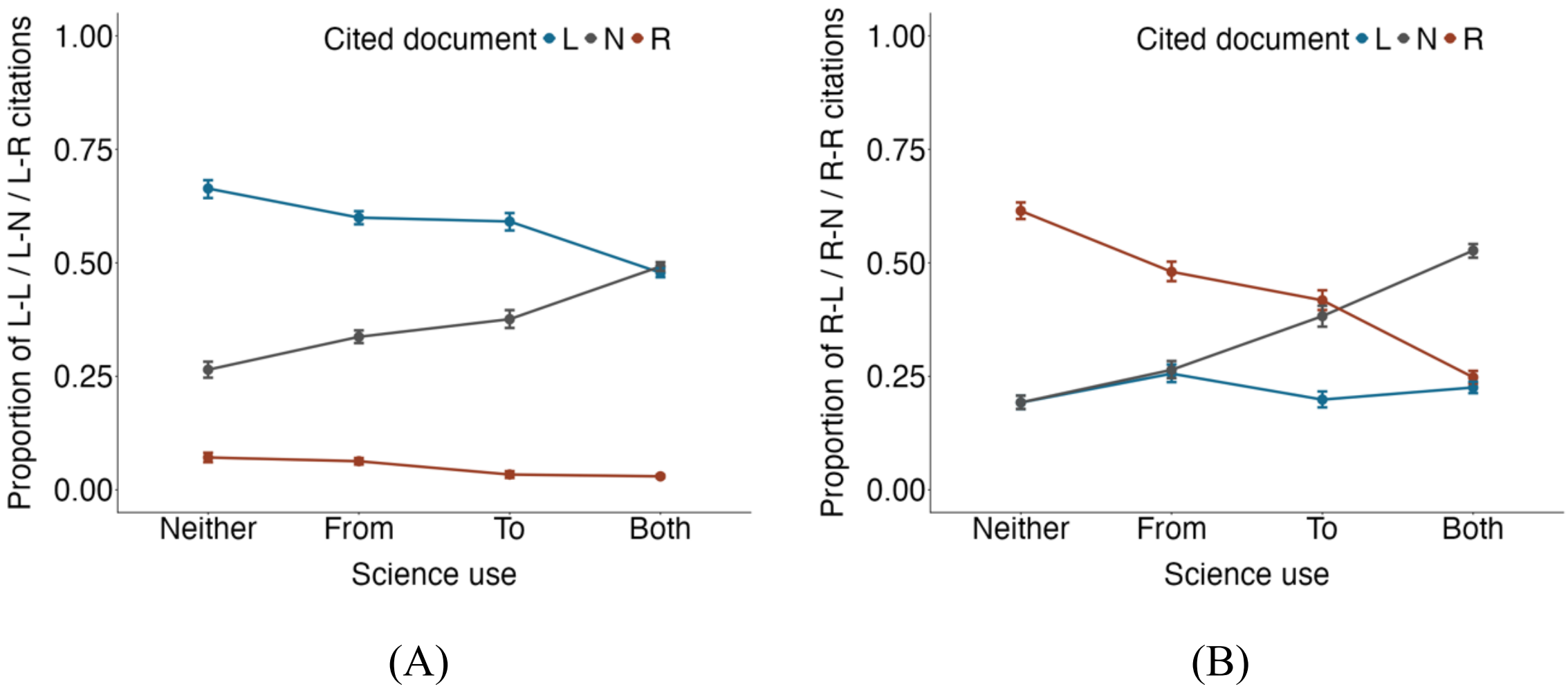

Figure S18: Science use and ideological polarization in policy knowledge production. (A) The distributions of ideology in policy citations by science use in the citing and cited documents: conditioning on the citing document from a left-leaning think. (B) The distributions of ideology in policy citations by science use in the citing and cited documents: conditioning on the citing document from a left-leaning think.

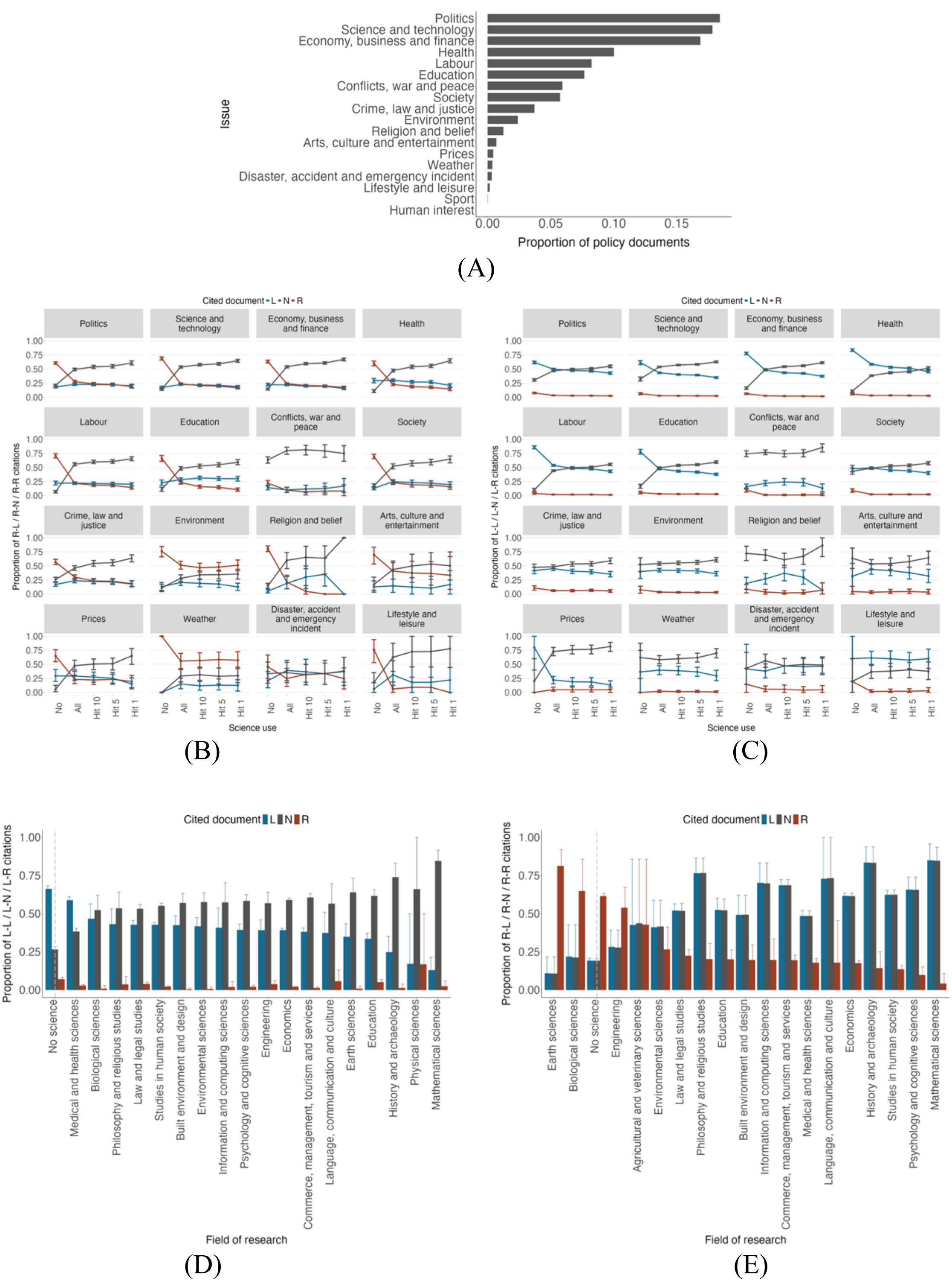

Figure S19: Co-ideological policy citations by science use and impact. (A) presents the distribution of policy domains. The issue of a policy citation is defined in terms of the issue of the citing document. As policy documents are assigned multiple issues, the same policy citations can count toward multiple issues. Here we present the results for 16 policy domains with at least one citation that uses science and at least one citation that does not use science (from a total of 18 areas). “Sport” and “Human interest” are excluded due to sparsity. (B) shows the conditional distribution of ideology in policy citations by science use and impact (the citing policy documents come from left-leaning think tanks). (C) depicts the conditional distribution of ideology in policy citations by science use and impact (the citing policy documents come from right-leaning think tanks). For (B)–(C), the error bars indicate the bootstrapped 95% confidence interval. In (B), we observe lower levels of co-ideological citations for 8 out of 16 policy domains for left-leaning think tanks when policy citations are grounded in science (“All”) compared to when they are not (“No). These 8 areas accounts for around 80% of the policy citations. In (C), we find an even more consistent pattern for right-leaning think tanks: the proportion of co-ideological citations is lower when science is used for all 16 of the policy domains. (D) shows the conditional distribution of ideology in policy citations by science use, broken down by field of research (the citing policy documents come from left-leaning think tanks). (E) shows the conditional distribution of ideology in policy citations by science use, broken down by field of research (the citing policy documents come from right-leaning think tanks). For (D)–(E), “No science” on the x-axis indicates that no scientific publication is cited in either of the policy documents. The error bar indicates the bootstrapped 95% confidence interval. Overall, (D) and (E) show that lower levels of co-ideological citations when science is used are highly consistent across fields of research. Compared to the baseline (science is not used in a given policy citation), science use is associated with lower levels of co-ideological citations both for left-leaning and for right-leaning think tanks for almost all fields of research. The only exceptions are earth sciences and biological science for right-leaning think tanks.

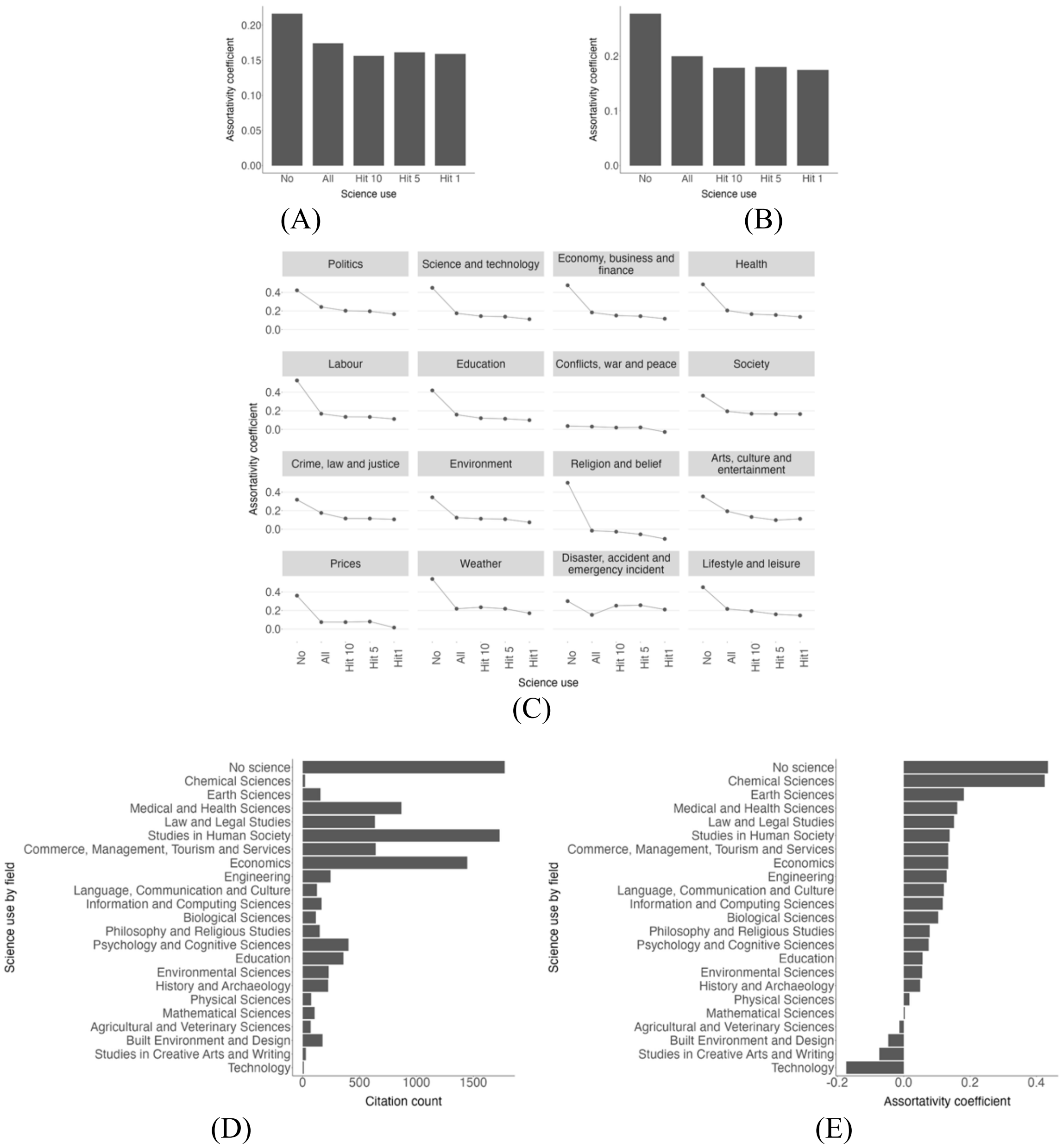

Figure S20: Assortativity coefficients by science use and impact. (A) shows assortativity coefficients based on the directed/unweighted version of the policy citation network. (B) shows assortativity coefficients based on the undirected/unweighted version of the policy citation network. With the exception of Hit 10, science use is associated with lower assortativity, and more impactful science use is associated with even lower assortativity. (C) illustrates assortativity coefficients across policy domains (issues are ordered by the number of policy citations corresponding to each issue). (D) presents the number of policy citations for which both the citing and the cited documents include at least one citation to a scientific publication in the field of research. (E) depicts assortativity coefficient by field of research. One directed, weighted policy citation network is built using the policy citations corresponding to each field.

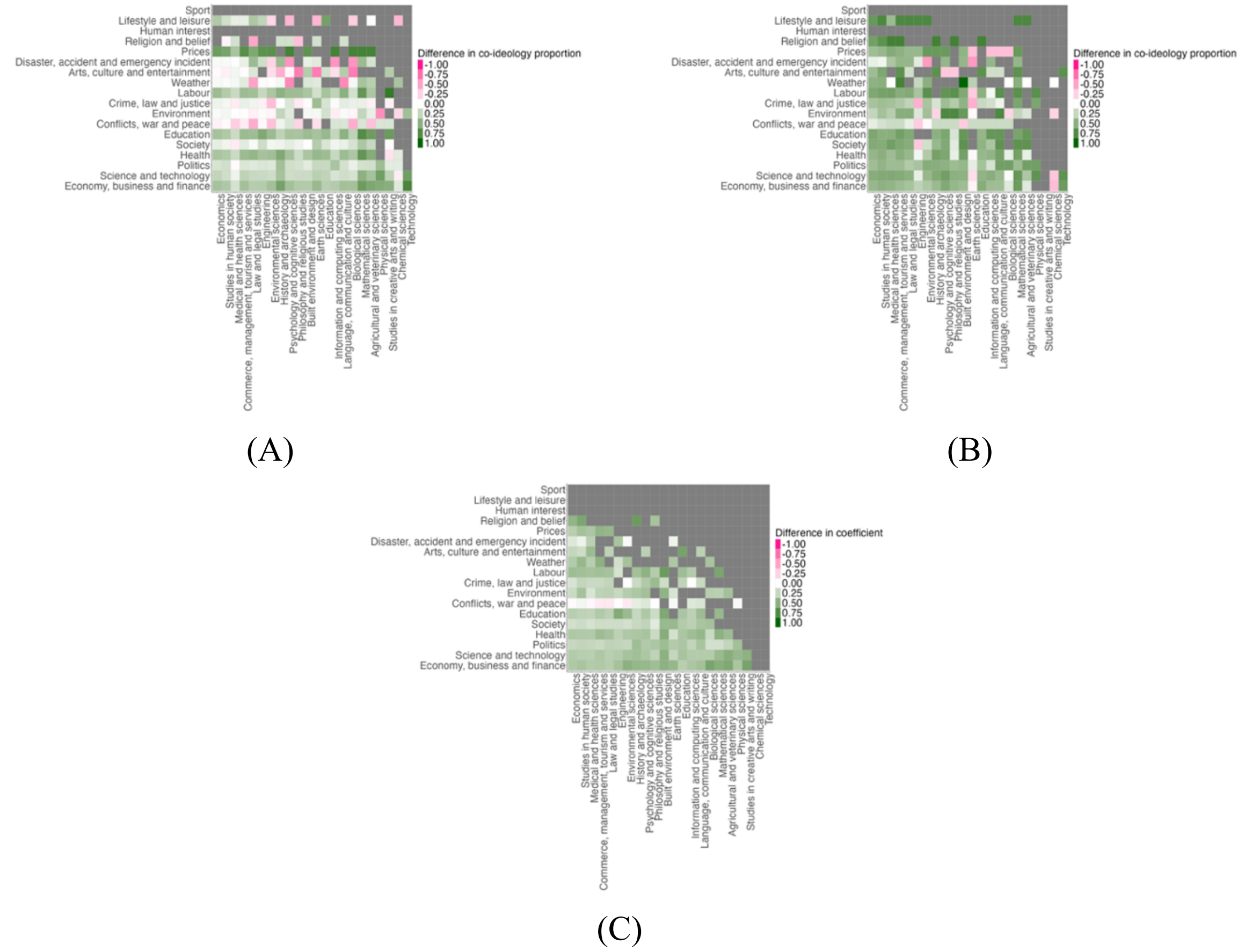

Figure S21: Assortativity coefficients by science use and impact. (A) shows the difference in the proportion of co-ideological citations when policy citations use science and when they do not use science (the citing policy documents from left-leaning think tanks). (B) presents the difference in the proportion of co-ideological citations policy citations use science and when they do not use science (the citing policy documents from right-leaning think tanks). For both (A) and (B), each cell represents one issue-field cross-section. The color of the cell depicts the proportion of the co-ideological citations when neither of the documents in the policy citation cites any scientific publication subtracted from the proportion of the co-ideological citations when both of the documents in the policy citations cite at least one scientific publication. Therefore, greener cells (or greater differences) imply less co-ideological policy discussion in the cross-section, with white (or zero) indicating that there is no difference between when science is used and when it is not used. (C) depicts the difference in the assortativity coefficient when science is not used and when science is used. This figure provides essentially the same insights as (A) and (B), but it is at the think tank level. For each cross-section, the color of the cell represents the assortativity coefficient for the network built on policy citations that do not use science subtracted from the assortativity coefficient for the network built on policy citations that use science. The assortativity coefficient is omitted for cross-sections where there are fewer than 20 policy citations. Across (A)–(C), we can see that science use is associated with lower levels of ideological segregation for a vast majority of the cross-sections, although there are some cross-sections exhibiting the opposite pattern (e.g., “Conflicts, war, and peace”).

| | Both: Count | From: Count | To: Count | Both: Any | From: Any | To: Any |
|---|---|---|---|---|---|---|
| Science use: both (count) | -0.212*** | | | | | |
| | (0.060) | | | | | |
| Science use: Citing (count) | | -0.127*** | | | | |
| | | (0.037) | | | | |
| Science use: Cited (count) | | | -0.158* | | | |
| | | | (0.064) | | | |
| Science use: Both (any) | | | | -0.243* | | |
| | | | | (0.098) | | |
| Science use: Citing (any) | | | | | -0.166* | |
| | | | | | (0.069) | |
| Science use: Cited (any) | | | | | | -0.248* |
| | | | | | | (0.104) |
| Policy domains | ✓ | ✓ | ✓ | ✓ | ✓ | ✓ |
| Fields of research | ✓ | ✓ | ✓ | ✓ | ✓ | ✓ |
| Think tank FE | ✓ | ✓ | ✓ | ✓ | ✓ | ✓ |
| Year FE | ✓ | ✓ | ✓ | ✓ | ✓ | ✓ |
| Num.Obs. | 51480 | 51480 | 51480 | 51480 | 51480 | 51480 |
| AIC | 60964.8 | 61338.2 | 61052.7 | 61320.4 | 61406.2 | 61307.9 |
| BIC | 63283.3 | 63656.6 | 63371.1 | 63638.8 | 63724.6 | 63626.4 |
| RMSE | 0.45 | 0.45 | 0.45 | 0.45 | 0.45 | 0.45 |

* $p < 0.05$, ** $p < 0.01$, *** $p < 0.001$

Table S18: Logistic regression results. We test whether the negative association between science use and co-ideological citation is also observed after accounting for document- and organization-level characteristics. Specifically, we examine policy-to-policy citations between think tanks by estimating six parallel logistic regression models. The outcome variable common across the models is whether a given policy-to-policy citation is co-ideological. The key predictor, science use, is measured either as the log-transformed number of scientific publications cited or as a binary indicator of whether science is cited, each specified for the citing document alone, the cited document alone, and both documents combined. We account for unobserved heterogeneity across think tanks and years using fixed effects and additionally control for policy domains and fields of research. The results for the fixed effects and controls are omitted from the table. Standard errors are clustered by think tank (in parentheses).

**S5. List of organizations**

| Index | Think Tank | Ideology | Rank in McGann (2020) |
|---|---|---|---|
| 1 | Peterson Institute for International Economics | N | 1 |
| 2 | Center for Strategic and International Studies | N | 2 |
| 3 | Carnegie Endowment for International Peace | N | 3 |
| 4 | Urban Institute | L | 4 |
| 5 | Center for American Progress | L | 5 |
| 6 | Heritage Foundation | R | 6 |
| 7 | Atlantic Council | N | 7 |
| 8 | Wilson Center | N | 8 |
| 9 | RAND Corporation | N | 9 |
| 10 | Stimson Center | N | 10 |
| 11 | Council on Foreign Relations | N | 11 |
| 12 | Belfer Center | N | 12 |
| 13 | Cato Institute | R | 13 |
| 14 | Hudson Institute | N | 14 |
| 15 | Baker Institute | N | 15 |
| 16 | Center for a New American Security | N | 16 |
| 17 | American Enterprise Institute | N | 17 |
| 18 | German Marshall Fund | L | 18 |
| 19 | Resources for the Future | L | 19 |
| 20 | Freedom House | N | 20 |
| 21 | Human Rights Watch | L | 21 |
| 22 | Hoover Institution | R | 22 |
| 23 | Mercatus Center | R | 23 |
| 24 | NBER | N | 25 |
| 25 | The Dialogue | N | 26 |
| 26 | World Resources Institute | L | 27 |
| 27 | Center for Global Development | N | 28 |
| 28 | Asia Society Policy Institute | N | 29 |
| 29 | Economic Policy Institute | L | 30 |
| 30 | Middle East Institute | N | 31 |
| 31 | The Chicago Council on Global Affairs | N | 32 |
| 32 | National Bureau of Asian Research | N | 33 |
| 33 | Migration Policy Institute | L | 34 |
| 34 | Acton Institute | R | 35 |
| 35 | United States Institute of Peace | N | 36 |
| 36 | Manhattan Institute | R | 37 |
| 37 | Data & Society | L | 39 |
| 38 | Carnegie Council | N | 40 |
| 39 | Pew Research | N | 41 |
| 40 | Independent Institute | R | 42 |
| 41 | Center for Climate and Energy Solutions | N | 43 |
| 42 | Center on Budget and Policy Priorities | L | 44 |
| 43 | Foreign Policy Research Institute | N | 45 |
| 44 | Joint Center for Political and Economic Studies | N | 46 |
| 45 | Bipartisan Policy Center | N | 47 |
| 46 | The Aspen Institute | N | 48 |
| 47 | EastWest Institute | N | 49 |
| 48 | New America | L | 50 |
| 49 | Atlas Network | R | 54 |
| 50 | Foundation for Economic Education | R | 55 |
| 51 | Reason Foundation | R | 56 |
| 52 | Pacific Research Institute | R | 57 |
| 53 | Center for the National Interest | R | 58 |
| 54 | Milken Institute | R | 59 |
| 55 | Third Way | N | 62 |
| 56 | Committee for Economic Development | N | 63 |
| 57 | Center for European Policy Analysis | N | 64 |
| 58 | Center for Strategic and Budgetary Assessments | N | 65 |
| 59 | Institute for Policy Studies | L | 66 |
| 60 | Open Society Foundations | L | 68 |
| 61 | Competitive Enterprise Institute | R | 69 |
| 62 | Institute for Women's Policy Research | L | 70 |
| 63 | The Arctic Institute | L | 71 |
| 64 | R Street | R | 72 |
| 65 | Center for International Security and Arms Control | N | 73 |
| 66 | Institute for New Economic Thinking | L | 75 |
| 67 | Berggruen Institute | N | 76 |
| 68 | Demos | L | 77 |
| 69 | Africa Center for Strategic Studies | N | 79 |
| 70 | Institute on Religion & Democracy | R | 81 |
| 71 | Institute for Science and International Security | N | 82 |
| 72 | Mackinac Center | R | 83 |
| 73 | Center for International Policy | L | 84 |
| 74 | Washington Center for Equitable Growth | L | 85 |
| 75 | Center for Naval Analysis | N | 86 |
| 76 | Institute for Defense Analysis | N | 87 |
| 77 | Public Policy Institute of California | N | 89 |
| 78 | Roosevelt Institute | L | 91 |
| 79 | Institute for the Study of War | N | 92 |
| 80 | Center for Immigration Studies | R | 93 |
| 81 | Center for International Development | N | 96 |
| 82 | Pacific Forum | N | 97 |
| 83 | The Beacon Hill Institute | N | 98 |
| 84 | Federation for American Immigration Reform | R | 102 |
| 85 | Employment Policies Institute | R | 103 |
| 86 | Potomac Institute for Policy Studies | N | 105 |
| 87 | Lexington Institute | R | 106 |
| 88 | Texas Public Policy Foundation | R | 107 |
| 89 | National Employment Law Project | L | No included |
| 90 | Union of Concerned Scientists | L | No included |
| 91 | IATP | L | No included |
| 92 | Oceana | L | No included |
| 93 | WWF | L | No included |
| 94 | Jobs With Justice | L | No included |
| 95 | People For the American Way | L | No included |
| 96 | National Association of Development Organizations | N | No included |
| 97 | Future of Life Institute | N | No included |
| 98 | Group of Thirty | N | No included |
| 99 | Gates Foundation | N | No included |
| 100 | Centro Para Una Nueva EconomÃa | R | No included |
| 101 | Council on Hemispheric Affairs | L | No included |
| 102 | Hewlett Foundation | L | No included |
| 103 | India China America Institute | N | No included |
| 104 | International Center for Research on Women | L | No included |
| 105 | Institute for Energy Research | R | No included |
| 106 | IFPRI | N | No included |
| 107 | Internet Society | N | No included |
| 108 | International Peace Institute | N | No included |
| 109 | Institute for Transportation and Development Policy | L | No included |
| 110 | Jamestown Foundation | N | No included |
| 111 | JINSA | N | No included |
| 112 | Kaiser Family Foundation | L | No included |
| 113 | Knight Foundation | N | No included |
| 114 | Lincoln Institute | N | No included |
| 115 | Lown Institute | N | No included |
| 116 | Macy Foundation | N | No included |
| 117 | MassINC | L | No included |
| 118 | Mercy Corps | N | No included |
| 119 | MITRE Corporation | N | No included |
| 120 | National Association of State Boards of Education | N | No included |
| 121 | National Association of Scholars | N | No included |
| 122 | National Endowment for Democracy | N | No included |
| 123 | National Democratic Institute | L | No included |
| 124 | New Teacher Center | L | No included |
| 125 | oaklandinstitute.org | N | No included |
| 126 | Open Philanthropy | N | No included |
| 127 | Paulson Institute | N | No included |
| 128 | Project 2049 Institute | N | No included |
| 129 | Partnership for Public Service | N | No included |
| 130 | Natural Resource Governance Institute | N | No included |
| 131 | Results for America | N | No included |
| 132 | Rocky Mountain Institute | R | No included |
| 133 | RTI International | N | No included |
| 134 | Santa Fe Institute | N | No included |
| 135 | Schiller Institute | N | No included |
| 136 | Secure World Foundation | N | No included |
| 137 | Seven Pillars Institute | N | No included |
| 138 | SRI | N | No included |
| 139 | Tax Foundation | R | No included |
| 140 | Tellus Insitute | N | No included |
| 141 | Center for Development and Strategy | N | No included |
| 142 | Upjohn Institute | N | No included |
| 143 | Upturn | L | No included |
| 144 | American Iranian Council | N | No included |
| 145 | Wallace Foundation | L | No included |
| 146 | The Washington Institute | N | No included |
| 147 | Water for People | N | No included |
| 148 | WEDO | L | No included |
| 149 | William T Grant Foundation | N | No included |
| 150 | The Constitution Project | L | No included |
| 151 | Public Citizen | L | No included |
| 152 | Center for Effective Government | L | No included |
| 153 | Leadership Conference on Civil and Human Rights | L | No included |
| 154 | AARP | L | No included |
| 155 | Human Rights First | L | No included |
| 156 | Southern Poverty Law Center | L | No included |
| 157 | Center for International Environmental Law | L | No included |
| 158 | Environmental Investigation Agency | L | No included |
| 159 | CLASP | L | No included |
| 160 | Natural Resources Defense Council | L | No included |
| 161 | Institute on Taxation and Economic Policy | L | No included |
| 162 | Environmental Defense Fund | L | No included |
| 163 | Environmental and Energy Study Institute | L | No included |
| 164 | ACLU | N | No included |
| 165 | Accenture | N | No included |
| 166 | ACEEE | N | No included |
| 167 | Niskanen Center | N | No included |
| 168 | American Legislative Exchange Council | R | No included |
| 169 | Center for Freedom and Prosperity | R | No included |
| 170 | American Action Forum | R | No included |
| 171 | Heartland Institute | R | No included |
| 172 | 38 North | N | No included |
| 173 | Access Now | L | No included |
| 174 | Alabama Policy Institute | R | No included |
| 175 | Allegheny Institute | R | No included |
| 176 | Americans for the Arts | N | No included |
| 177 | The Asia Foundation | N | No included |
| 178 | Beacon Center of Tennessee | R | No included |
| 179 | Berkley Center | N | No included |
| 180 | Bertelsmann Foundation | N | No included |
| 181 | Boston Consulting Group | N | No included |
| 182 | Brookings Institution | L | No included |
| 183 | Buckeye Institute | R | No included |
| 184 | C4ADS | N | No included |
| 185 | Center for Automotive Research | N | No included |
| 186 | Cascade Policy Institute | R | No included |
| 187 | CCHPCA | N | No included |
| 188 | Center for Security Policy | R | No included |
| 189 | The Century Foundation | L | No included |
| 190 | C-Fam | R | No included |
| 191 | CGR | N | No included |
| 192 | Center for International Private Enterprise | R | No included |
| 193 | Center for Climate & Security | N | No included |
| 194 | Columbia Center on Sustainable Investment | N | No included |
| 195 | Commonwealth Fund | L | No included |
| 196 | Center for Court Innovation | L | No included |
| 197 | Center for Data Innovation | N | No included |
| 198 | Data-Pop Alliance | L | No included |
| 199 | Democracy Collaborative | L | No included |
| 200 | Discovery Institute | R | No included |
| 201 | Center for the Study of the Drone | N | No included |
| 202 | Economic Innovation Group | N | No included |
| 203 | Ecotrust | L | No included |
| 204 | Education Commission of the States | N | No included |
| 205 | Educopia Institute | L | No included |
| 206 | Environmental Law Institute | L | No included |
| 207 | Federation of American Scientists | N | No included |
| 208 | Foundation for Defense of Democracies | N | No included |
| 209 | FHI 360 | N | No included |
| 210 | Ford Foundation | L | No included |
| 211 | Fordham Institute | R | No included |
| 212 | Friends of the Earth US | L | No included |
| 213 | Gatestone Institute | R | No included |
| 214 | Guttmacher Institute | L | No included |
| 215 | Independence Institute | R | No included |
| 216 | Islands Society | N | No included |
| 217 | The James Madison Institute | R | No included |
| 218 | Mises Institute | R | No included |
| 219 | National Policy Institute | R | No included |
| 220 | Oklahoma Policy Institute | L | No included |
| 221 | Economic Opportunity Institute | L | No included |
| 222 | Oregon Center for Public Policy | L | No included |
| 223 | Pacific Institute | N | No included |
| 224 | Pioneer Institute | R | No included |
| 225 | Policy Matters Ohio | L | No included |
| 226 | Population Research Institute | R | No included |
| 227 | Progressive Policy Institute | L | No included |
| 228 | Ripon Society | R | No included |
| 229 | Show Me Institute | R | No included |
| 230 | Solidarity Centre | L | No included |
| 231 | Tax Policy Center | L | No included |

Table S19. List of think tanks. We study 231 think tanks. Of the 231 think tanks, 88 think tanks appear in the list of the top 110 think tanks covered by (70). Please note that we study 143 think tanks that are not covered. Please also note that Pacific Forum was part of the Center for Strategic and International Studies until 2018. We use the latter's policy documents to study the former.

| Index | Government | Level |
|---|---|---|
| 1 | AHRQ | Federal |
| 2 | Centers for Disease Control and Prevention (CDC) | Federal |
| 3 | Congressional Record | Federal |
| 4 | Congressional Research Service | Federal |
| 5 | Department of Commerce | Federal |
| 6 | Department of Homeland Security | Federal |
| 7 | EPA | Federal |
| 8 | Education Commission of the States | Federal |
| 9 | Federal Emergency Management Agency | Federal |
| 10 | Federal Register | Federal |
| 11 | Federal Research Divison | Federal |
| 12 | Federal Reserve | Federal |
| 13 | Food & Drug Administration | Federal |
| 14 | Government Publishing Office | Federal |
| 15 | House Committees | Federal |
| 16 | House Committees Staff Reports | Federal |
| 17 | Justice Department | Federal |
| 18 | Legislative Analyst's Office | Federal |
| 19 | Medicaid.gov | Federal |
| 20 | Medicare Coverage Database | Federal |
| 21 | National Academy of Medicine | Federal |
| 22 | OPRE | Federal |
| 23 | Senate Committees | Federal |
| 24 | Supreme Court of the United States | Federal |
| 25 | U.S. Bureau of Economic Analysis | Federal |
| 26 | U.S. Government Accountability Office | Federal |
| 27 | US Preventive Services | Federal |
| 28 | USAID | Federal |
| 29 | United States Census Bureau | Federal |
| 30 | Whitehouse.gov | Federal |
| 31 | California Research Bureau | State |
| 32 | California State Auditor | State |
| 33 | Californian State Agencies | State |
| 34 | Colorado General Assembly Legislative Council | State |
| 35 | Commonwealth of Massachusetts | State |
| 36 | Commonwealth of Pennsylvania | State |
| 37 | Commonwealth of Virginia | State |
| 38 | Connecticut Office of Legislative Research | State |
| 39 | District of Columbia | State |
| 40 | Illinois Commission on Government Forecasting and Accountability Research Unit | State |
| 41 | Minnesota House Research Department | State |
| 42 | New York State | State |
| 43 | State of Alabama | State |
| 44 | State of Alaska | State |
| 45 | State of Arizona | State |
| 46 | State of Colorado | State |
| 47 | State of Delaware | State |
| 48 | State of Florida | State |
| 49 | State of Georgia | State |
| 50 | State of Hawaii | State |
| 51 | State of Idaho | State |
| 52 | State of Illinois | State |
| 53 | State of Indiana | State |
| 54 | State of Louisiana | State |
| 55 | State of Maryland | State |
| 56 | State of Michigan | State |
| 57 | State of Minnesota | State |
| 58 | State of Missouri | State |
| 59 | State of New Jersey | State |
| 60 | State of North Carolina | State |
| 61 | State of South Carolina | State |
| 62 | State of Tennessee | State |
| 63 | State of Texas | State |
| 64 | State of Washington | State |
| 65 | State of Wisconsin | State |
| 66 | Baltimore City Council | City |
| 67 | City of Albuquerque | City |
| 68 | City of Atlanta | City |
| 69 | City of Austin | City |
| 70 | City of Baltimore | City |
| 71 | City of Boston | City |
| 72 | City of Charlotte | City |
| 73 | City of Chicago | City |
| 74 | City of Columbus | City |
| 75 | City of Dallas | City |
| 76 | City of Denver | City |
| 77 | City of Detroit | City |
| 78 | City of El Paso | City |
| 79 | City of Fresno | City |
| 80 | City of Grand Rapids | City |
| 81 | City of Houston | City |
| 82 | City of Indianapolis | City |
| 83 | City of Jacksonville | City |
| 84 | City of Las Vegas | City |
| 85 | City of Los Angeles | City |
| 86 | City of Louisville | City |
| 87 | City of Memphis | City |
| 88 | City of Mesa | City |
| 89 | City of Miami | City |
| 90 | City of Minneapolis | City |
| 91 | City of New York | City |
| 92 | City of Philadelphia | City |
| 93 | City of Phoenix | City |
| 94 | City of Pittsburgh | City |
| 95 | City of Portland | City |
| 96 | City of Sacramento | City |
| 97 | City of San Antonio | City |
| 98 | City of San Diego | City |
| 99 | City of San Francisco | City |
| 100 | City of San Jose | City |
| 101 | City of Seattle | City |
| 102 | City of South Bend | City |
| 103 | City of Tucson | City |
| 104 | Oklahoma City | City |
| 105 | Philadelphia City Council | City |

Table S20: List of government organizations

| Index | IGO |
|---|---|
| 1 | African Development Bank |
| 2 | African Union |
| 3 | Arctic Council |
| 4 | Asian Development Bank |
| 5 | Association of Southeast Asian Nations |
| 6 | BIS |
| 7 | CGAP |
| 8 | Council of Europe |
| 9 | ECHR |
| 10 | Egmont Group |
| 11 | Food and Agriculture Organization of the United Nations |
| 12 | Geneva Centre for Security Sector Governance |
| 13 | IARC |
| 14 | IPBES |
| 15 | IPCC |
| 16 | Inter-American Development Bank |
| 17 | Inter-Parliamentary Union |
| 18 | International Civil Aviation Organization |
| 19 | International Committee of the Red Cross |
| 20 | International Energy Agency |
| 21 | International IDEA |
| 22 | International Labour Organization |
| 23 | International Monetary Fund |
| 24 | International Organization for Migration |
| 25 | International Renewable Energy Agency |
| 26 | International Transport Forum |
| 27 | Interpol |
| 28 | Mekong River Commission |
| 29 | NATO Stratcom Centre of Excellence |
| 30 | Nordic Council |
| 31 | OECD |
| 32 | OPEC |
| 33 | Organisation for the Prohibition of Chemical Weapons |
| 34 | Organization for Security and Co-operation in Europe |
| 35 | The Global Commission on the Economy and Climate |
| 36 | UN Department of Economic and Social Affairs |
| 37 | UN Office for Disaster Risk Reduction |
| 38 | UN Women |
| 39 | UNAIDS |
| 40 | UNCTAD |
| 41 | UNESCO |
| 42 | UNICEF |
| 43 | UNICEF Office of Research - Innocenti |
| 44 | UNU-CRIS |
| 45 | United Nations |
| 46 | United Nations Development Programme |
| 47 | United Nations ECE |
| 48 | United Nations ESCWA |
| 49 | United Nations Economic Commission for Africa |
| 50 | United Nations Environment Programme |
| 51 | United Nations Human Rights |
| 52 | United Nations Office on Drugs and Crime |
| 53 | United Nations Population Fund |
| 54 | World Bank |
| 55 | World Food Programme |
| 56 | World Health Organization |
| 57 | World Intellectual Property Organization |
| 58 | World Meterological Organization |
| 59 | World Tourism Organization |
| 60 | World Trade Organization |

Table S21: List of intergovernmental organizations